\documentclass[pdflatex,sn-mathphys-num]{sn-jnl}
\usepackage{graphicx}
\usepackage{multirow}%
\usepackage{amsmath,amssymb,amsfonts}%
\usepackage{amsthm}%
\usepackage{mathrsfs}%
\usepackage[title]{appendix}%
\usepackage{xcolor}%
\usepackage{textcomp}%
\usepackage{manyfoot}%
\usepackage{booktabs}%
\usepackage{algorithm}%
\usepackage{algorithmicx}%
\usepackage{algpseudocode}%
\usepackage{listings}%
\usepackage{url}
\usepackage{hyperref}
\usepackage{lscape}
\usepackage{rotating}
\usepackage{float} 
\usepackage{threeparttable}
\usepackage{subcaption}
\usepackage{lipsum} 
\usepackage[normalem]{ulem}
\useunder{\uline}{\ul}{}

\theoremstyle{thmstyleone}%
\theoremstyle{thmstyletwo}%

\theoremstyle{thmstylethree}%

\begin{document}

\title[Deep Generative Models for Discrete Genotype Simulation]{Deep Generative Models for Discrete Genotype Simulation}

\author*[1]{\fnm{Sihan Xie}} \email{sihan.xie@inrae.fr}
\author[1]{\fnm{Thierry Tribout}}
\author[1]{\fnm{Didier Boichard}}
\author[2]{Blaise Hanczar}\equalcont{These authors contributed equally to this work.}
\author[3]{Julien Chiquet}\equalcont{These authors contributed equally to this work.}
\author[1]{\fnm{Eric Barrey}}
\affil[1]{Université Paris-Saclay, INRAE, AgroParisTech, GABI, 78350, Jouy-en-Josas, France} \equalcont{These authors contributed equally to this work.}
\affil[2]{Université Paris-Saclay, Univ Evry, IBISC, 91020, Evry-Courcouronnes, France}
\affil[3]{Université Paris-Saclay, AgroParisTech, INRAE, UMR MIA Paris-Saclay, 91120, Palaiseau, France}

\abstract{Deep generative models open new avenues for simulating realistic genomic data while preserving privacy and addressing data accessibility constraints. While previous studies have primarily focused on generating gene expression or haplotype data, this study explores generating genotype data in both unconditioned and phenotype-conditioned settings, which is inherently more challenging due to the discrete nature of genotype data. In this work, we developed and evaluated commonly used generative models, including Variational Autoencoders (VAEs), Diffusion Models, and Generative Adversarial Networks (GANs), and proposed adaptation tailored to discrete genotype data. We conducted extensive experiments on large-scale datasets, including all chromosomes from cow and multiple chromosomes from human. Model performance was assessed using a well-established set of metrics drawn from both deep learning and quantitative genetics literature. Our results show that these models can effectively capture genetic patterns and preserve genotype–phenotype association. Our findings provide a comprehensive comparison of these models and offer practical guidelines for future research in genotype simulation. We have made our code publicly available at \href{https://github.com/SihanXXX/DiscreteGenoGen}{https://github.com/SihanXXX/DiscreteGenoGen}.}


\maketitle

\section{Introduction}\label{sec:intro}
The development of dense genotyping platforms and high-throughput sequencing technologies has significantly advanced genetic analysis \cite{seq_tech, seq_tech_cardio}. Today, genomic studies rely on large biobanks that contain vast amounts of genomic data. However, working with such datasets presents several challenges, including high sequencing costs, substantial storage requirements, privacy concerns, and access restrictions that limit data sharing. To address these issues, simulation tools and synthetic data are commonly used. Traditional statistical simulation methods are based on evolutionary models like Wright-Fisher model \cite{WF} and coalescent theory \cite{coalescent1, coalescent2, coalescent_sim}, where users need to specify evolutionary parameters or provide ancestral population. While these simulation tools \cite{ms,mbs,msprime, SLiM3, MSMS, simuPOP} are powerful, they often simplify various aspects of population genetics, which may not fully capture the complexities of real-world datasets.

Recently, data-driven simulation methods based on deep generative models have gained attention in genomics. These approaches eliminate the need to explicitly specify genetic parameters by learning directly from data, enabling the reproduction of fine-scale genomic characteristics presented in the given population. By shifting from explicit genomic sequences to generative models, the genome-wide data remains private, while the trained models can be shared publicly without directly exposing individual-level genetic information.

Previous studies have applied generative models to various genomic modalities: \cite{metric_corr, lacan_gan, lacan_dm} focused on gene expression data, \cite{latent_dm_dna,evo2} focused on DNA sequence, and there is a substantial body of literature on haplotype data \cite{GMMN_hap, VAE_hap, VAE_GAN_hap, deep_hap, yelmen2021, yelmen2023, szatkownik2024latent, szatkownik2024dm}. In this work, we propose a new study on genotype data, which represents genetic variation at specific positions in the genome known as Single Nucleotide Polymorphisms (SNPs). Unlike contiguous DNA sequences, which may include both coding and non-coding regions that do not directly reflect individual genetic variability, genotype data focuses on selected variant sites, making them particularly valuable for studying population-level traits and disease association. Unlike binary-valued haplotype, genotype for diploid organisms includes three possible values (0, 1, 2), representing the number of alternative alleles inherited from both parents, which introduces specific modeling challenges. Importantly, directly simulating genotype data offers several advantages over haplotype-based approaches: Haplotype-based simulation methods lack rich conditioning capabilities, whereas our method allows conditioning on phenotypic traits, enabling more flexible and application-oriented use cases. Furthermore, traditional genotype–phenotype simulation workflows via haplotypes involve a multi-step pipeline and our approach consolidates this into a single generative step, reducing complexity and avoiding additional modeling assumptions that may introduce bias. Finally, haplotype simulators are limited to genomic regions with strong linkage disequilibrium (LD) on a single chromosome, whereas our study supports genome-wide simulation, demonstrated in our experiments on cow dataset where we modeled genotypes across all 29 autosomes simultaneously, significantly extending the scale of genomic data simulation.

This paper investigates the use of deep generative models for simulating genotype data, potentially conditioned on phenotype. Specifically, we adapt models such as Variational Autoencoders (VAEs) \cite{VAE}, Generative Adversarial Networks (GANs) \cite{cGAN}, and Diffusion Models \cite{DDPM} to accommodate the discrete nature of genotype. Properly evaluating synthetic genotype is a critical aspect of our study, as the evaluation metrics commonly used in the Generative AI community, such as precision and recall \cite{precision_recall}, have never been applied in previous haplotype generation studies. We propose a comprehensive evaluation framework that integrates both deep learning and quantitative genetics approaches, providing a rigorous comparison of the reviewed models. Section \ref{sec:model} presents the models adapted for genotype data and Section \ref{sec:metric} presents our evaluation framework. Section~\ref{sec:exp} describes the experimental setup and Section~\ref{sec:result} presents the main results. Finally, Section \ref{sec:dis_conclu} discusses how the proposed models can be practically implemented, along with potential challenges and future research directions.

\section{Generative Models for Genotype Data}\label{sec:model}
Building on recent advances in haplotype generation \cite{VAE_hap, yelmen2021, yelmen2023, szatkownik2024latent, szatkownik2024dm}, we adopt generative models such as Variational Autoencoders (VAEs) \cite{VAE}, Generative Adversarial Networks (GANs) \cite{GAN}, and diffusion models \cite{DDPM}. These models are well-suited for capturing global dependencies across all SNPs, as opposed to relying on sequential autoregressive decomposition, which may not align with the underlying biological structure. Since genotype is a discrete sequence represented as $\mathbf{x} \in \{0,1,2\}^n$, we propose adaptation to better handle this structure.

\subsection{Variational Autoencoders}
\label{sec:model:vae}
Variational Autoencoders (VAEs) \cite{VAE} learn to approximate the underlying data distribution by introducing a latent variable. A VAE consists of two neural networks, parameterized by $\phi$ and $\theta$: an encoder that maps the input data $x$ to a latent representation $z$ via the approximate posterior $q_\phi(z \mid x)$, and a decoder that reconstructs $x$ from $z$ via the likelihood $p_\theta(x \mid z)$. The model is trained by maximizing the Evidence Lower BOund (ELBO) on the marginal likelihood $\log p(x)$, using the reparameterization trick to enable efficient gradient-based optimization. The ELBO is given by
\begin{equation}
\label{eq:vae}
\mathcal{L}(\theta, \phi; x) = 
\underbrace{\mathbb{E}_{q_\phi(z \mid x)}\left[\log p_\theta(x \mid z)\right]}_{\text{decoder for reconstruction}} 
- 
\underbrace{D_{\mathrm{KL}}\left(q_\phi(z \mid x) \,\Vert\, p(z)\right)}_{\text{encoder for prior matching}},
\end{equation}
where $p(z)$ is the prior on the latent variable and $D_{\mathrm{KL}}$ denotes the Kullback-Leibler divergence. Optimizing this objective encourages the model to learn a meaningful, structured latent space that can be sampled to generate new, realistic data. For generation purpose, once the VAE is trained, the encoder can be discarded. New samples are generated by drawing random latent vector $z$ from the prior and passing it through the decoder.

\subsection{Diffusion Models}\label{sec:model:dm}
Diffusion models (DMs), and in particular Denoising Diffusion Probabilistic Models (DDPMs) \cite{DDPM}, can be viewed as a Markovian hierarchical VAE where each latent $x_t$ has the same dimension as the data $x_0$ and the encoder is not learned but is a fixed Gaussian noising process. During the encoding phase, also called the forward diffusion process, we gradually add Gaussian noise to input $x_0$ until it becomes pure noise $x_{T}$ over $T$ steps via a Markov chain:
\begin{equation}
\label{eq:dm_forward_step}
q(x_t \!\mid\! x_{t-1})
= \mathcal{N}\!\bigl(\sqrt{\alpha_t}\,x_{t-1},\,\beta_t\,\mathbf I\bigr),
\quad \alpha_t = 1-\beta_t
\quad \text{for t} = 1, \dots, T.
\end{equation}

Because of the Markov property, the Gaussian transition, and the independence of the noise at every step, one can collapse all \(t\) steps into a single closed‑form marginal:
\begin{equation}
\label{eq:dm_marginal}
q(x_t \!\mid\! x_0)
= \mathcal{N}\!\Bigl(x_t;\,\sqrt{\bar\alpha_t}\,x_0,\,(1-\bar\alpha_t)\,\mathbf I\Bigr),
\quad
\bar\alpha_t \;=\;\prod_{s=1}^t \alpha_s.
\end{equation}

Intuitively, the hyperparameter \(\beta_t\) controls the amount of noise injected at step \(t\), and \(\alpha_t=1-\beta_t\) is the fraction of signal retained. Our goal is to undo the added noise by learning $p_\theta(x_{t-1}\!\mid\!x_t)$, so that starting from \(x_T\sim\mathcal N(0,I)\) we can step-by-step recover $x_0$. The true reverse \(\;q(x_{t-1}\!\mid\!x_t)\) is intractable, but during training we know \(x_0\). Hence we can write down the exact one‑step posterior as a Gaussian distribution with a closed‑form mean $\mu_t(x_t,x_0)$ and variance $\sigma_t^2$:
\begin{equation}
\label{eq:dm_one_step_posterior}
q\bigl(x_{t-1}\mid x_t, x_0\bigr)
= \mathcal{N}\!\Bigl(\underbrace{\frac{\sqrt{\alpha_t}\,(1-\bar\alpha_{t-1})}{1-\bar\alpha_t}\,x_t
\;+\;
\frac{\beta_t\,\sqrt{\bar\alpha_{t-1}}}{1-\bar\alpha_t}\,x_0,}_{\mu_t(x_t,x_0)}\;\underbrace{\beta_t\,\frac{1-\bar\alpha_{t-1}}{1-\bar\alpha_t}\,\mathbf I}_{\sigma_t^2} \Bigr).
\end{equation}

For the reverse process, we learn to approximate the posterior in Equation \ref{eq:dm_one_step_posterior}. For the variance part, many implementations simply set $\sigma_{t}^2 = \beta_{t}$, which has a negligible loss on quality. For the mean part, since $\mu_t(x_t,x_0)$ requires the true $x_{0}$ which is unavailable at inference, the direct training objective is to predict $x_{0}$ given $x_{t}$ and $t$. In practice, however, it is more common and empirically more stable to train a network $\epsilon_{\theta}(x_t, t)$ to predict the injected noise at each timestep optimized via mean‑squared error loss. Then using the predicted noise $\epsilon_{\theta} (x_t,t)$, we first infer an estimator of $x_0$, given by $
\hat{x}_0 = \left( x_t - \sqrt{1 - \bar\alpha_t}\;\epsilon_{\theta}(x_t, t) \right) \big/ \sqrt{\bar\alpha_t}$. Substituting \(\hat x_0\) for $x_0$ in Equation \ref{eq:dm_one_step_posterior} gives the familiar reverse‑step update:
\begin{equation}
\label{eq:dm_reverse_step}
x_{t-1}
\;=\;
\frac{1}{\sqrt{\alpha_t}}
\Bigl(
  x_t \;-\; \tfrac{\beta_t}{\sqrt{1-\bar\alpha_t}}\,\epsilon_\theta(x_t,t)
\Bigr)
\;+\;
\sigma_t\,z,
\quad
z\sim\mathcal{N}(0,I).
\end{equation}

Despite their success, DMs are not compatible with discrete data. Two main strategies have been proposed to address this limitation: (1) defining a diffusion-like process that operates in discrete space, or (2) projecting the discrete input into a continuous latent space. Additionally, DMs can be computationally demanding during inference. To address both issues, we adopt the second strategy. Various methods can be used to construct a suitable latent space. For example, \cite{latent_dm_dna} employed a VAE to embed DNA sequences into a continuous representation. We follow the PCA-based approach originally developed for haplotype \cite{szatkownik2024latent, szatkownik2024dm}. Specifically, we project genotype into a lower-dimensional PCA space \cite{PCA} and train the DMs in this continuous latent space. This single transformation yields three major benefits in one shot: it greatly reduces dimensionality and speeds up both training and inference; it transforms the discrete genotype into a continuous representation that matches the assumptions of DMs; And it allows precise reconstruction via a simple linear multiplication (Supplementary Section 1).

\subsection{Generative Adversarial Networks}\label{sec:model:gan}
While VAEs and DMs learn the data distribution by explicitly maximizing likelihood, Generative Adversarial Networks (GANs) \cite{GAN} adopt a fundamentally different strategy. They avoid explicit density estimation by framing generation as a two-player game: a generator $G$ transforms a latent vector $z\sim p_z$ (typically Gaussian) into a synthetic sample $G(z)$, and a discriminator $D$ attempts to distinguish real data from generated samples. Training proceeds by solving the minimax problem:
\begin{equation}
\label{eq:gan}
\min_{G}\,\max_{D}\quad
\underbrace{
  \mathbb{E}_{x \sim p_{\mathrm{data}}}\bigl[\log D(x)\bigr]
  \;+\;
  \mathbb{E}_{z \sim p(z)}\bigl[\log\bigl(1 - D\bigl(G(z)\bigr)\bigr)\bigr]
}_{\text{binary cross‐entropy loss}}.
\end{equation}

Here, \(D\) is trained as a binary classifier to assign high probability to real sample \(x\) and low probability to generated sample \(G(z)\), while \(G\) is trained to fool \(D\) by producing ever more realistic outputs. 

From Equation \ref{eq:gan}, $G$ is trained by backpropagating gradients from $D$ through its outputs $G(z)$. This works when $G(z)$ is continuous but fails for discrete outputs, which break differentiability. In previous GAN-based haplotype generation studies~\cite{yelmen2021,yelmen2023,deep_hap}, no specific treatment was proposed for this issue. $G$ output continuous values between $0$ and $1$, which were passed directly to $D$ during training. At inference, discrete values were recovered using a binarization threshold of $0.5$. This binary setting can be interpreted in two ways. First, from a probabilistic perspective, $G$ outputs the probability of class $1$, and inference chooses the most likely class. Motivated by this view, we tested a probabilistic approach: using a Softmax final layer in $G$ to predict class probabilities per SNP, and training $D$ to distinguish these from one-hot encoded real genotype sequences. However, this method produced unsatisfactory results. The second interpretation views binarization as a quantization operation that maps continuous outputs to a discrete set. We therefore explored several threshold-based strategies for our ternary genotype data, but observed suboptimal results (Supplementary Section 2). To enable end-to-end differentiable training on discrete outputs, we instead integrate a Gumbel-Softmax \cite{GumbelSoftmax,GumbelSoftmaxGAN} layer into $G$. The Gumbel-Softmax distribution provides a continuous approximation to categorical sampling by applying a temperature-controlled softmax to perturbed logits. Concretely, for each SNP, if the final layer of $G$ produces logits $\ell = (\ell_0,\ell_1,\ell_2)$, we then sample Gumbel noise $g_i \sim \mathrm{Gumbel}(0,1),$ and compute the relaxed one-hot vector
\begin{equation}
\label{eq:gumbel_softmax}
\tilde p_i
= \frac{\exp\!\bigl((\ell_i + g_i)/\tau\bigr)}
       {\sum_{j=0}^{2} \exp\!\bigl((\ell_j + g_j)/\tau\bigr)},
\end{equation}
where $\tau$ is a temperature parameter. As $\tau \to 0$, $\tilde p_i$ becomes exactly one-hot vector. During training we anneal $\tau$ from a high initial value down toward $0$ to balance exploration and discretization. At inference we take $\arg\max_i \tilde p_i$ to recover a discrete value in $\{0,1,2\}$.  

\subsection{Wasserstein GANs with Gradient Penalty}\label{sec:model:wgan}
GANs lack an explicit likelihood measure and can suffer from training instabilities such as mode collapse \cite{mode_collapse}. Subsequent refinements like Wasserstein GAN (WGAN) \cite{WGAN} was developed to address these issues. The original WGAN replaces the Jensen-Shannon divergence with the Earth-Mover (Wasserstein-1) distance by solving:
\begin{equation}
\label{eq:wgan}   
\min_{G}\;\max_{D\in\mathcal{D}}\quad
\mathbb{E}_{x\sim p_{\mathrm{data}}}\bigl[D(x)\bigr]
- \mathbb{E}_{z\sim p_{z}}\bigl[D\bigl(G(z)\bigr)\bigr],
\end{equation}
where \(\mathcal{D}\) is the set of 1-Lipschitz functions. While weight clipping was initially proposed to enforce the Lipschitz constraint, this approach is proved unstable in practice. The Wasserstein GAN with gradient penalty (WGAN-GP)~\cite{WGAN-GP} instead introduces a gradient penalty term that penalizes the deviation of the gradient norm from $1$, leading to more stable training dynamics:
\begin{equation}
\label{eq:wgan_gp}    
\min_G\max_D \quad
\underbrace{
 \mathbb{E}_{x \sim p_{data}}[D(x)] - \mathbb{E}_{z \sim p_z}[D(G(z))]
}_\text{Wasserstein distance between real and synthetic}
- \underbrace{
 \lambda\, \mathbb{E}_{\hat{x} \sim p_{\hat{x}}}\left( \|\nabla_{\hat{x}} D(\hat{x})\|_2 - 1 \right)^2
}_\text{Gradient penalty},
\end{equation}
where $p_{\hat x}$ is the distribution of points interpolated between real and generated samples, and $\lambda$ controls the penalty strength. 

\subsection{Conditional Generative Modeling}\label{sec:model:cond}
So far, we have focused on modeling the marginal distribution $p(x)$. In practice, it is more interesting to learn the conditional distribution $p(x \mid y)$, which provides control over the generated data through conditioning variable $y$. A straightforward approach is to append \( y \) to \( x \) as input during training \cite{cGAN}. At inference, we can sample from $p(x \mid y)$ by specifying a desired value of $y$ to guide generation.

In previous work on haplotype generation \cite{VAE_hap, VAE_GAN_hap, szatkownik2024dm}, models have frequently used ancestry group as conditioning variable to reflect population structure. For genotype, a natural choice is phenotype, particularly quantitative traits, which are often associated with genetic variation. This type of conditioning enables the generation of synthetic population that mirrors specific trait distribution, making the approach well suited for phenotype driven genetics research.

\section{Evaluation Metrics for Synthetic Genotype Data}\label{sec:metric}
Genotype lacks the intuitive visual or semantic cues of images and text. We therefore use a diverse set of metrics, offering a multi-angle assessment of synthetic genotype.

\subsection{PCA and UMAP}\label{sec:metric:visu}
The PCA~\cite{PCA} and UMAP~\cite{UMAP} projections provide an initial visual assessment of how well the synthetic population resembles the real one. These dimensionality reduction techniques highlight global structure and potential clustering patterns, offering a qualitative sense of alignment between the two distributions. However, they do not quantitatively measure distributional similarity and should be interpreted as complementary to more rigorous evaluation metrics.

\subsection{Genetic Parameters}\label{sec:metric:ge_param}
\subsubsection{Allele Frequency and Genotype Frequency}\label{sec:metric:ge_param:freq}
We compare allele and genotype frequencies \cite{al_geno_freq} between real and synthetic cohorts as a basic sanity check. Let $N$ be the number of individuals, for a given SNP $i$, let $n_{2,i}$, $n_{1,i}$, and $n_{0,i}$ denote the counts of individuals with genotype 2, 1, and 0, respectively. The allele frequency at locus $i$ is $p_i \;=\; \left(2\,n_{2,i} + n_{1,i}\right) \big/ (2N)\,$.
The genotype frequency is the proportion of each genotype class, given by 
$f_i(2) = n_{2,i} \big/ N\quad\text{(homozygosity for the alternative allele)}, 
\quad
f_i(1) = n_{1,i} \big/ N\quad\text{(heterozygosity)}, 
\quad
f_i(0) = n_{0,i} \big/ N\quad\text{(homozygosity for the reference allele)}.$ A strong concordance indicates that the generative model has accurately reproduced the per-locus marginal distribution, which is a prerequisite before assessing the higher order structure.
 
\subsubsection{Aggregated Fixation Index}\label{sec:metric:ge_param:fst}
The fixation index $F_{ST}$ \cite{fst1,fst2} is a widely used population genetic statistic that quantifies the degree of genetic differentiation among populations. It normalizes the difference between the total heterozygosity and the average heterozygosity within populations, yielding a value between 0 (no genetic differentiation) and 1 (complete genetic differentiation). For SNP $i$, let $p_{\text{real},i}$ and $p_{\text{syn},i}$ denote the allele frequencies in the real and synthetic cohorts respectively, assuming both cohorts are of the same size. Thus, the combined allele frequency for the total population is $p_{T,i} = \left(p_{real,i} + p_{syn,i}\right) \big/ 2$. Recall that for a given SNP $i$, the expected heterozygosity is given by $H = 1 - p^2 - (1-p)^2$. Thus, for the two subpopulations we have $H_{real,i} = 1 - p_{real,i}^2 - (1-p_{real,i})^2$ and $H_{syn,i} = 1 -p_{syn,i}^2 - (1-p_{syn,i})^2$. The average within-subpopulation heterozygosity is $H_{S,i} = (H_{real,i} + H_{syn,i}) / 2$. The heterozygosity in the combined population is $H_{T,i} = 1 - p_{T,i}^2 - (1-p_{T,i})^2$.
The per-SNP fixation index is given by
\begin{equation}
\label{eq:fst}
F_{ST}(i) = \frac{H_{T,i} - H_{S,i}}{H_{T,i}}
\end{equation}
Recognizing that not all SNPs are equally informative, with those exhibiting higher total heterozygosity providing greater insight into genetic diversity, we then aggregate the per-SNP fixation index into a summary metric using a weighted average:
\begin{equation}
\label{eq:fst_ag}
F_{ST}^{\mathrm{aggregated}} = \frac{\sum_i H_{T,i}\,F_{ST}(i)}{\sum_i H_{T,i}}
\end{equation}

\subsubsection{Linkage Disequilibrium and Its Decay with Physical Distance Along Chromosome} \label{sec:metric:ge_param:ld}
Linkage disequilibrium (LD) \cite{ld} measures the non-random association of alleles at different loci. Its decay with increasing physical distance along a chromosome reflects the effect of recombination in reshuffling genetic variation. When working with diploid genotype data, the gametic phase is often unknown, which complicates the accurate computation of LD statistics. To address this, we employ a fast estimator introduced in \cite{ld_geno}, which approximates LD between two loci without relying on the assumption of random mating or requiring iterative computation. This method is implemented in the \textit{scikit-allel} Python library\footnote{\url{https://scikit-allel.readthedocs.io/}}.

\subsection{Unsupervised Metrics for Genotype Structural Similarity}\label{sec:metric:unsuper}
\subsubsection{Precision and Recall}\label{sec:metric:unsuper:precision_recall}
Precision and recall, originally developed for classification, have been adapted to assess generative models \cite{precision_recall}. In this context, precision measures the quality of the synthetic data by quantifying the fraction of generated samples that fall within the support of the real data distribution, while recall measures the diversity of the synthetic data by quantifying the fraction of real samples that fall within the support of the synthetic data distribution. The F1 score is the harmonic mean of precision and recall.

To estimate the support of a data distribution, we define, for each sample in this dataset, a threshold $\epsilon$ as the distance to its $k^{th}$ nearest neighbor within the same set. This distance serves as the radius of a hypersphere centered on that sample, and the union of all such hyperspheres provides an estimate of the underlying manifold. Formally, let $R$ denote the set of real samples and $S$ the set of synthetic samples. Precision and recall are defined as follows:
\begin{equation}
\label{eq:precision}
\text{Precision} = \frac{1}{|S|} \sum_{s \in S} \mathbf{1}\Bigl\{ \exists\, r \in R \text{ such that } \| s - r \| < \epsilon_r \Bigr\},
\end{equation}

\begin{equation}
\label{eq:recall}
\text{Recall} = \frac{1}{|R|} \sum_{r \in R} \mathbf{1}\Bigl\{ \exists\, s \in S \text{ such that } \| s - r \| < \epsilon_s \Bigr\}.
\end{equation}

In image-based applications, precision and recall are typically computed on high-level feature vectors extracted from pretrained VGG-16 \cite{vgg16} classifier. However, for genotype, no widely accepted pretrained network exists. We therefore use the original data directly for evaluation. For KNN-based manifold estimation, the $L_2$ distance is conventionally employed. Given that genotype is discrete, we experimented with both $L_1$ and $L_2$ distances and found no significant differences in the resulting metrics (Supplementary Section 3). We adopted $L_2$ distance since it's more computationally efficient. The choice of $k$ is crucial and we selected the value of $k$ that yielded approximately 90\% precision and recall on two real datasets.

\subsubsection{Correlation Score}\label{sec:metric:unsuper:corr}
To compare the moments of the real and synthetic distributions, correlation score is proposed in \cite{metric_corr}, the idea is to compute the Pearson correlation coefficient between the strictly upper-diagonal elements of the SNP-pairwise correlation matrices $M_{real}$ and $M_{syn}$:
\begin{equation}
\label{eq:corr_score}
\rho(M_{\text{real}}, M_{\text{syn}}) = \frac{2}{n(n - 1)} \sum_{i=1}^{n-1} \sum_{j=i+1}^{n} 
\frac{M_{i,j;\text{real}} - \mu(M_{\text{real}})}{\sigma(M_{\text{real}})} \times 
\frac{M_{i,j;\text{syn}} - \mu(M_{\text{syn}})}{\sigma(M_{\text{syn}})},
\end{equation}
where $n$ is the number of SNPs, $\mu(M)$ is the mean, and $\sigma(M)$ is the standard deviation of the strictly upper-diagonal elements of $M$. 

\subsection{Supervised Metrics for Genotype–Phenotype Association}
\label{sec:metric:unsuper}
\subsubsection{Genome-wide association study (GWAS)}
\label{sec:metric:unsuper:gwas}
In quantitative genetics, Genome-Wide Association Study (GWAS) \cite{GWAS} is a fundamental tool for identifying genetic variants associated with specific traits by examining the relationship between genotype and phenotype. In GWAS, a per-SNP regression is performed, and a two-sided t-test is used to determine whether the regression coefficient $\beta$ is significantly different from $0$. The corresponding $p$-value gives us the significance of the association. GWAS can be viewed as a feature‐importance method, since each SNP’s estimated effect size $\beta$ and its $p$-value indicate how strongly that locus contributes to phenotype prediction. By comparing GWAS results obtained from synthetic population with those from real, we can directly evaluate whether our generative model has preserved key biological signals.

\subsubsection{Phenotype Prediction Performance}
\label{sec:metric:unsuper:pred}
We further evaluate synthetic genotype by its ability to predict the conditioning phenotype. Specifically, we train an XGBoost model and a multilayer perceptron (MLP) on synthetic data, then assess their performance on an independent real dataset not used during generative training. If a predictive model trained solely on synthetic data performs comparably to one trained on real data, it suggests that the synthetic population has faithfully preserved the underlying genotype–phenotype relationship.

\subsection{Privacy Leakage Assessment} \label{sec:metric:privacy}
\subsubsection{Nearest Neighbor Adversarial Accuracy ($AA$)}
\label{sec:metric:privacy:AA}
Since genotype data is highly sensitive, our synthetic data must balance utility with privacy protection. To evaluate this balance, we adopt the nearest neighbour adversarial accuracy ($AA$) metric proposed in \cite{AA}. This metric is conceptually similar to precision and recall. Rather than estimating the entire manifold with a full KNN approach, we use a 1NN measure to compare local neighborhood distances. The intuition is that synthetic data should be close enough to real data to preserve utility, yet not so close as to risk privacy leakage. For each real sample, we measure whether its distance to its nearest synthetic neighbour ($d_{RS}$) is larger than its distance to its nearest real neighbour ($d_{RR}$). Likewise, for each synthetic sample, we check whether its distance to its nearest real neighbour ($d_{SR}$) is larger than its distance to its nearest synthetic neighbour ($d_{SS}$). These comparisons yield two values—one for the real dataset ($AA_{real}$) and one for the synthetic dataset ($AA_{syn}$). The overall $AA$ score is then defined as the average of these two quantities. Formally, we have:
\begin{equation}
\label{eq:AA}
AA = \frac{1}{2} \left(
\underbrace{ \frac{1}{N} \sum_{i=1}^{N} \mathbf{1}\left( d_{RS}(i) > d_{RR}(i) \right) }_{AA_{\text{real}}}
+ 
\underbrace{ \frac{1}{N} \sum_{i=1}^{N} \mathbf{1}\left( d_{SR}(i) > d_{SS}(i) \right) }_{AA_{\text{syn}}}
\right).
\end{equation}
Same as in the calculation of precision and recall, we use $L_2$ distance. An $AA$ value near $0$ indicates overfitting, while an $AA$ value near $1$ suggests underfitting. Ideally, an $AA$ value around $0.5$ reflects a good tradeoff between utility and privacy.

\section{Experimental Setting}\label{sec:exp}
\begin{figure}[H]  
    \centering
\includegraphics[width=1\textwidth]{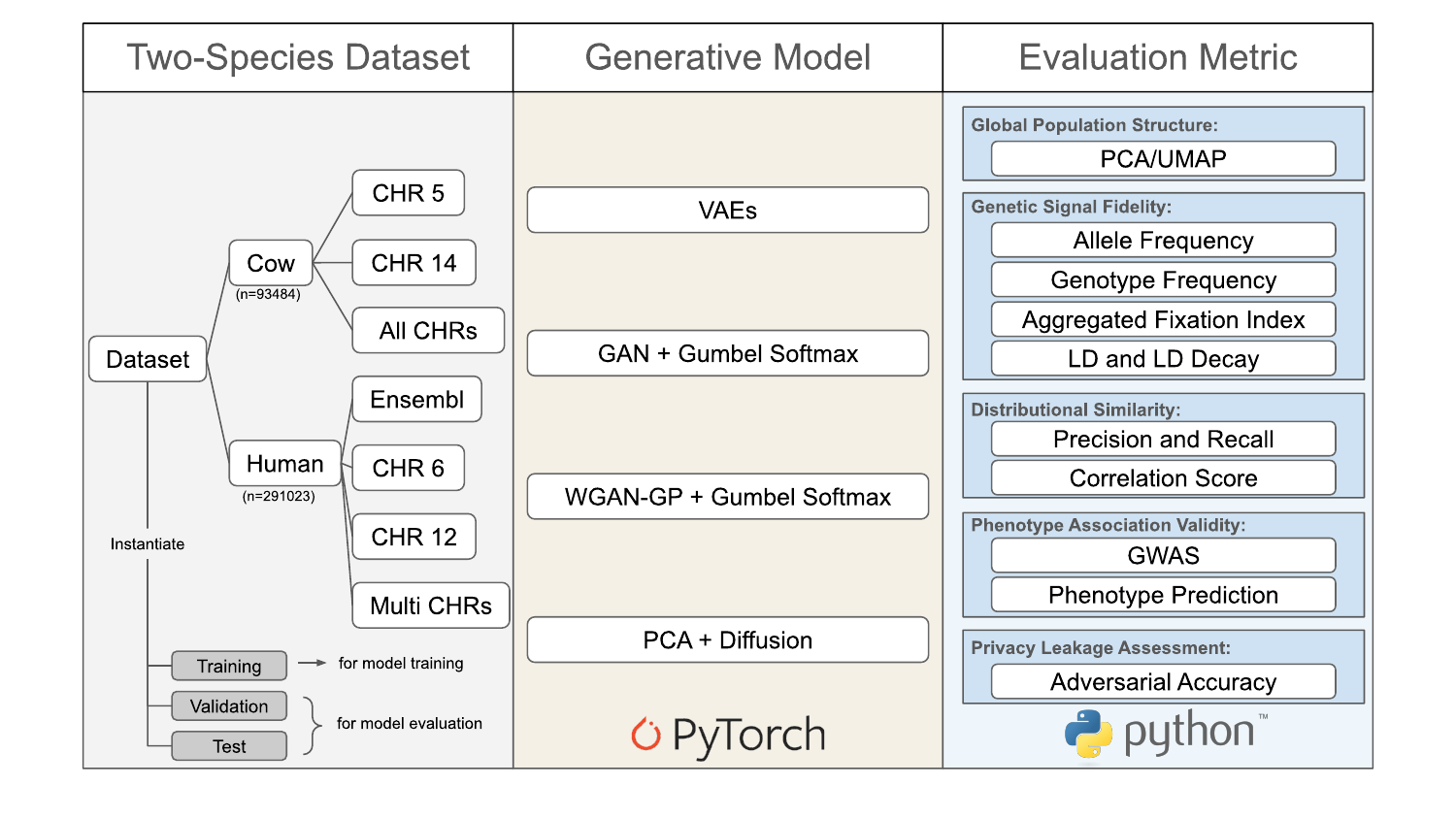} 
    \caption{Schema of our generative modeling task. We used data from two species and constructed sub-datasets at different scales, ranging from single-chromosome to multi-chromosome settings. For the human dataset, the Multi CHRs setting included chromosomes 3, 6, 12, and 17. Four generative models were implemented: VAE, GAN with Gumbel-Softmax, WGAN-GP with Gumbel-Softmax, and PCA combined with diffusion. These models were evaluated across different aspects relevant to performance.}
    \label{fig:schema}
\end{figure}

The following section outlines our experimental setup, including the datasets, model architectures, hyperparameter choices, synthetic data simulation and metric computation process. A schematic overview is provided in Figure~\ref{fig:schema}.

\subsection{Datasets}\label{sec:exp:data}
Since the frequency distribution and correlation between SNPs can vary across population groups, techniques developed using data from one group may not generalize well to others. Therefore, we used two large‐scale genomic datasets from different species: a private Holstein cow cohort and the human dataset from UK Biobank\footnote{\url{https://www.ukbiobank.ac.uk/}} \cite{UKBioBank}. Genotype calls for diploid organisms are encoded as 0 for homozygous reference, 1 for heterozygous, and 2 for homozygous alternative alleles.

\textbf{Cow:}  
Our cow dataset comprises 93,484 Holstein individuals genotyped at 50,161 SNPs across all 29 pairs of autosomes. The selected phenotype is fat content (FC), a milk production trait that reflects the proportion of fat in milk. Fat is a key component in dairy products and influences the taste, texture, and richness of milk, making it an important criterion in milk pricing. FC has relatively high heritability, estimated at approximately $0.50$. For selection purpose, FC was analyzed with a mixed model that accounts for various fixed environmental effects, the permanent environmental effect of the cow, and the breeding value (Supplementary Section 4). The so-called Yield Deviations (YD) are therefore by-products of the French Holstein Single Step genomic evaluation \cite{hssgblup, single_step}. The YD of FC for a cow is the mean of its phenotypes that have been adjusted for all non-genetic effects estimated in the genetic evaluation, and serves as the conditioning phenotype in our study. We assessed model performance on two individual chromosomes: Chromosome 5 (2,238 SNPs), where the MGST1 gene \cite{MGST1} is located, and Chromosome 14 (1,771 SNPs), where the DGAT1 gene \cite{DGAT1} is located. In our experiments, these chromosomes exhibited the strongest GWAS signals for the selected trait. We also evaluated the models using the full concatenated genotype across all chromosomes.

\textbf{Human:}  
The UK Biobank provides genotype and phenotype data for 488,377 participants, including 805,426 variants comprising both SNPs and INDELs \cite{INDEL}, across the 22 autosomes, sex chromosomes, and the mitochondrial chromosome. We used sex and height as conditioning phenotypes, as height is a highly heritable and polygenic trait \cite{human_height_1}. Following the pipeline proposed in \cite{Anderson2010} to assemble our study subsets, we performed quality control using PLINK 1.9\footnote{\url{https://www.cog-genomics.org/plink/1.9/}}, including checks for sex discordance, individual and SNP missingness, minor‐allele‐frequency filtering, Hardy–Weinberg equilibrium testing, and LD‐based tag SNP selection. Missing genotypes were imputed using Beagle 5.4\footnote{\url{https://faculty.washington.edu/browning/beagle/beagle.html}}. To recover biologically relevant height loci, we incorporated annotations from Ensembl\footnote{\url{https://www.ensembl.org/index.html}} and extracted 3,493 SNPs associated with height identified in previous studies. In a final cohort of 291,023 individuals, we constructed $4$ genotype datasets: the 3,493 height‐associated variants; Chromosome 6 (12,283 SNPs) where a QTL was detected by GWAS; Chromosome 12 (9,780 SNPs) where IGF-1 gene \cite{IGF-1} is located; a combined set of 42,409 SNPs from Chromosomes 3, 6, 12, and 17. 

For VAE, GAN, and WGAN models, the input genotype sequences were first transformed using one-hot encoding. For DM, we applied PCA and retained the number of principal components that captured 90\% of the total variance in each dataset. Across all experiments, 70\% of the samples were used for training, 15\% for validation, and 15\% for testing.

\subsection{Models and Training}\label{sec:exp:model}
All the models were based on fully connected layers. Since genotype sequence has no inherent spatial or temporal structure, we avoided convolutional or recurrent layers. Instead, each model was composed of a sequence of dense layers, with layer widths heuristically scaled based on the data dimension. To improve training stability and gradient flow, we added residual connections \cite{resnet}. For VAE, the encoder and decoder shared a symmetric architecture. For GAN and WGAN, we used the identical generator and discriminator architectures. For WGAN, the gradient penalty coefficient $\lambda$ was set to $10$ and we performed $5$ discriminator updates for every generator update. For DM, we experimented with multiple noise-addition schedules and found that a linear $\beta$ schedule yields the best performance. 

We performed a grid search over the network architecture and the key training hyperparameters. We provide a full and detailed description of the model architectures and hyperparameter choices for all four models applied to the full cow chromosome dataset (Supplementary Section 5). To determine when to stop training, we monitored the F1 score since it balances precision and recall. Training was terminated once this score no longer improved.

\subsection{Inference and Evaluation}\label{sec:exp:eval}
We generated synthetic population under two scenarios. In the unconditional setting, the only required input was latent noise sampled from the Gaussian prior used during training. In the conditional setting, phenotype values were additionally sampled from the training set and provided as conditioning inputs. All metrics, except for the phenotype-prediction metric, were computed on the validation set. For the phenotype‐prediction metric, we selected the best model using the validation set and reported its performance on the test set. All metrics were averaged over $5$ independent runs and 10,000 synthetic samples were generated per run. For metrics that require hyperparameter tuning, we suggest selecting the values that deliver satisfactory performance on two real datasets.

\section{Results}\label{sec:result}
\subsection{Do Generative Models Capture the Statistical and Genetic Structure of the Real Population?}\label{sec:geno}

\begin{figure}[htbp]
    \centering
    \begin{minipage}{0.50\textwidth}
        \centering
        \begin{minipage}{0.48\textwidth}
            \centering
            \includegraphics[width=\textwidth]{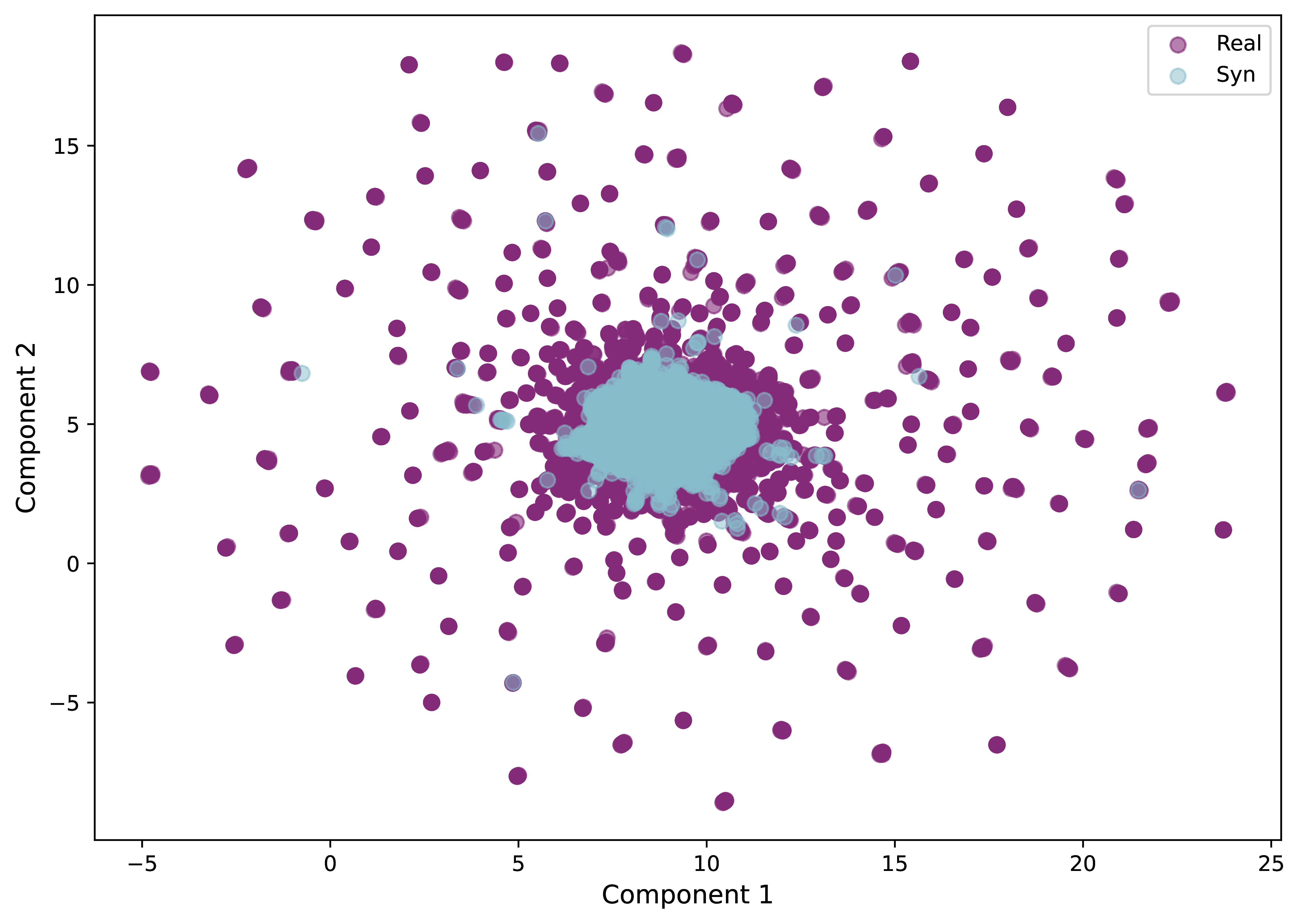}
            \subcaption{VAE}\label{fig:vae_umap}
        \end{minipage}\hfill
        \begin{minipage}{0.48\textwidth}
            \centering
            \includegraphics[width=\textwidth]{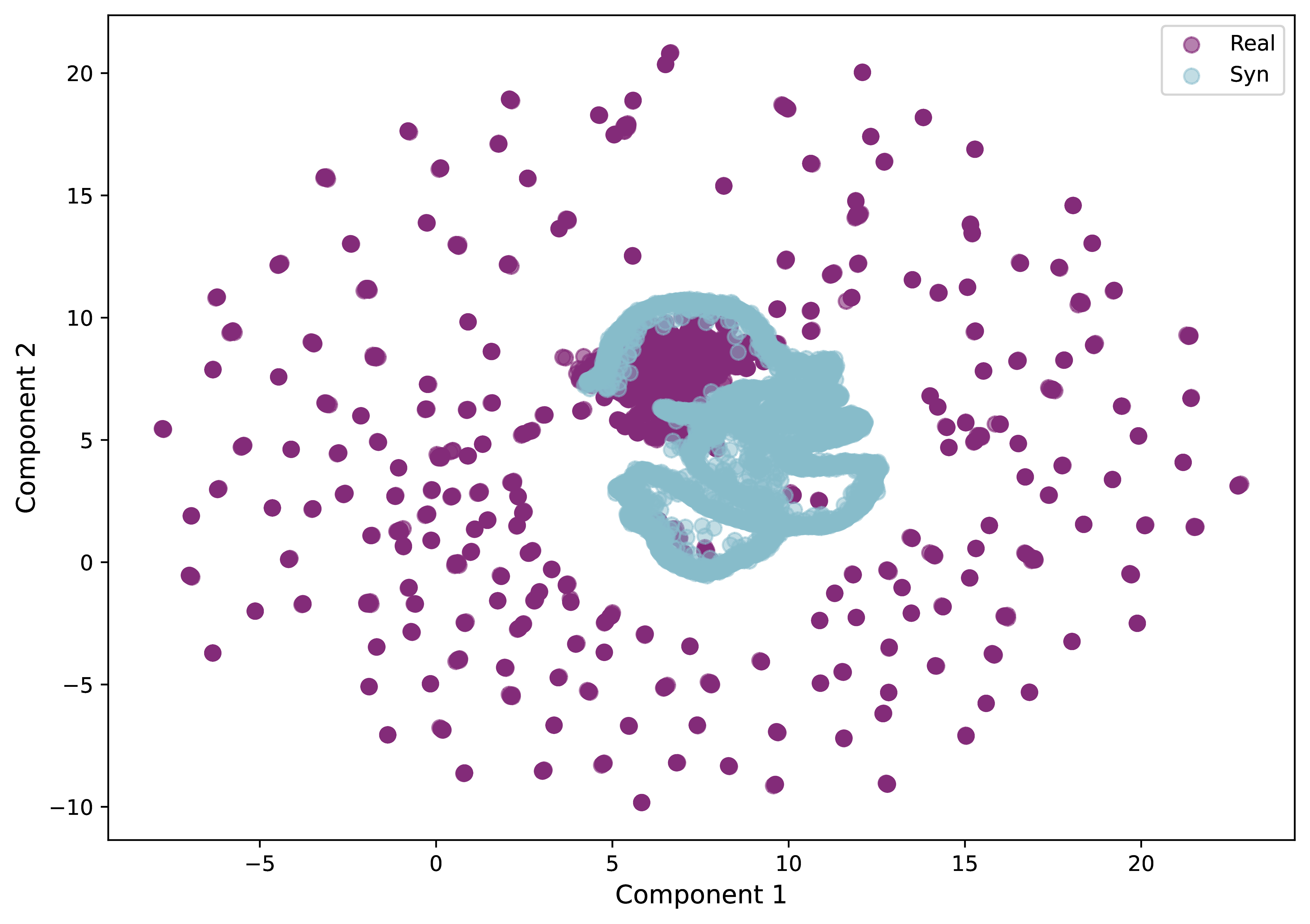}
            \subcaption{GAN}\label{fig:gan_umap}
        \end{minipage}\\[1ex]
        \begin{minipage}{0.48\textwidth}
            \centering
            \includegraphics[width=\textwidth]{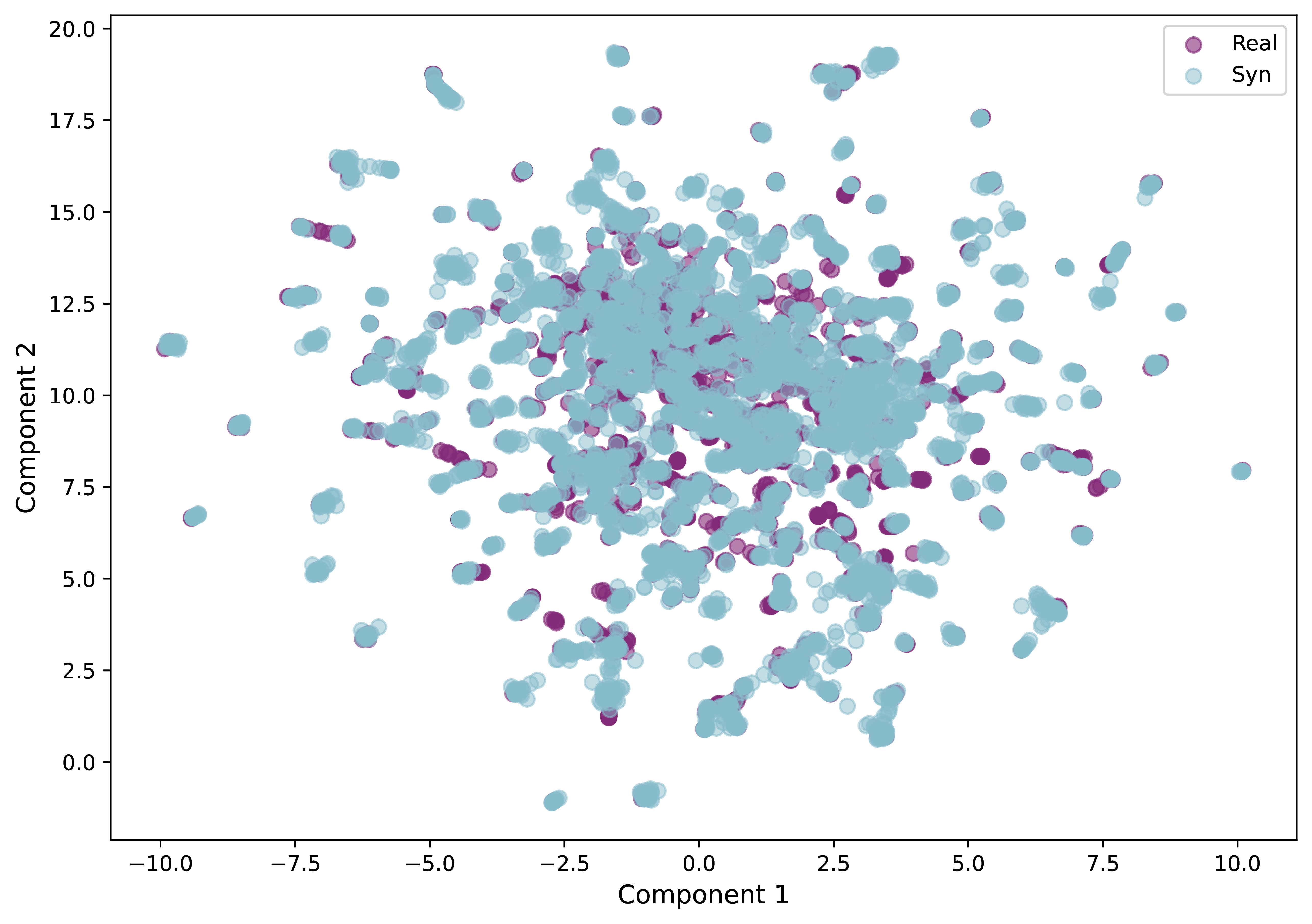}
            \subcaption{WGAN}\label{fig:wgan_umap}
        \end{minipage}\hfill
        \begin{minipage}{0.48\textwidth}
            \centering
            \includegraphics[width=\textwidth]{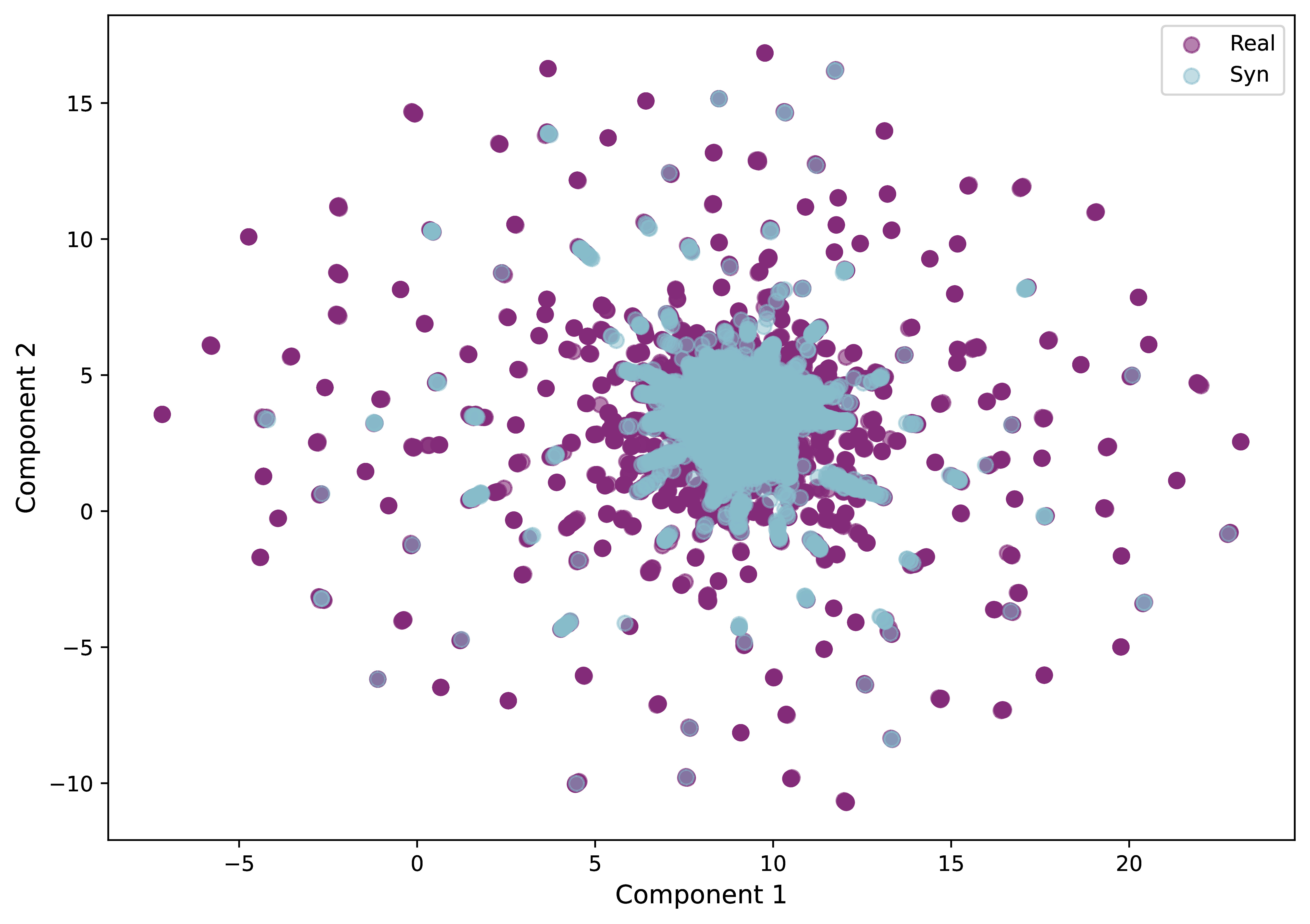}
            \subcaption{DM}\label{fig:dm_umap}
        \end{minipage}
        \label{fig:umap}
    \end{minipage}%
    \hfill
    \begin{minipage}{0.45\textwidth}
        \centering
        \includegraphics[width=\textwidth]{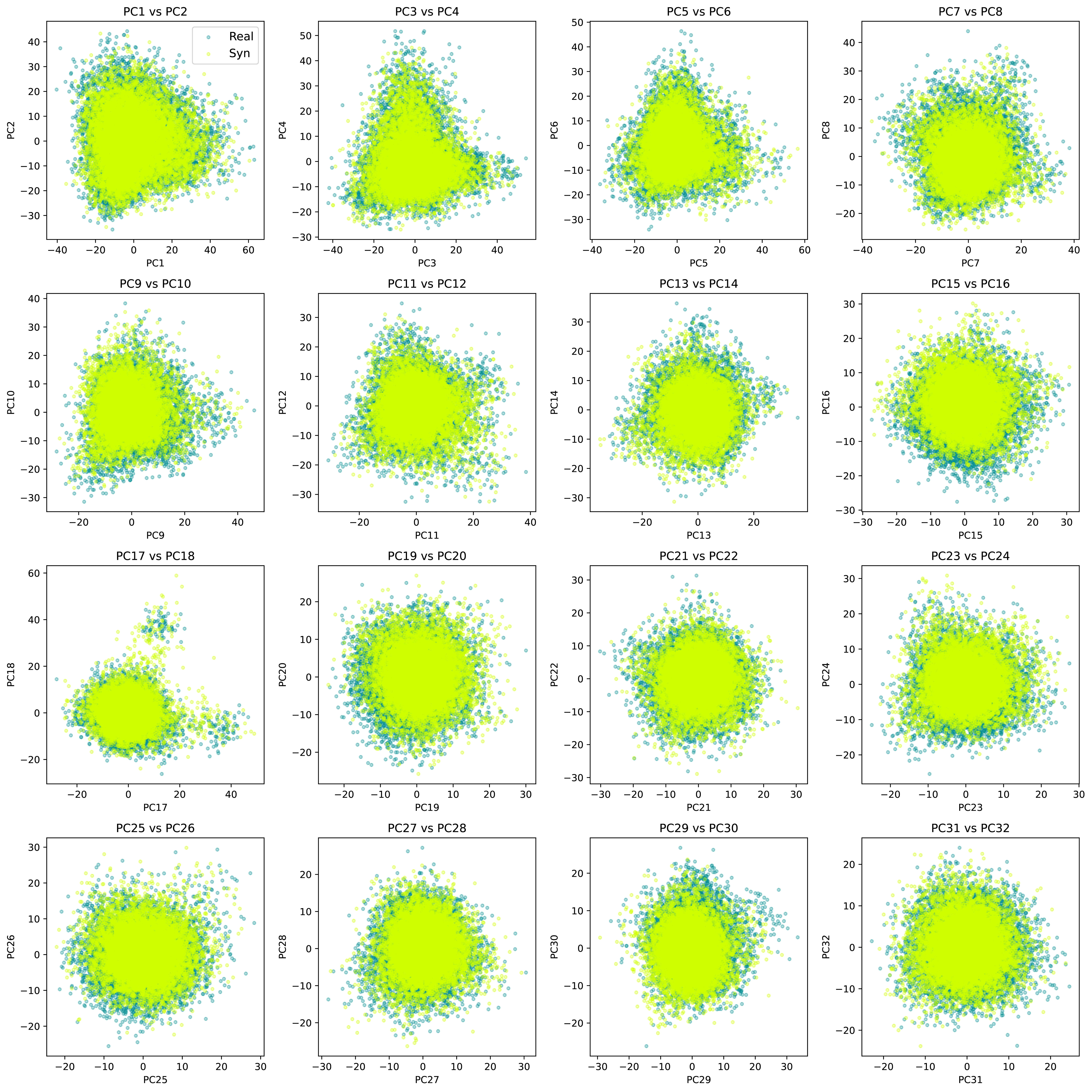}
        \subcaption{WGAN}\label{fig:wgan_pca}
    \end{minipage}
    \caption{PCA and UMAP Visualizations of Real and Model-Generated Synthetic Populations on All Chromosomes of Cow Dataset. (a), (b), (c), (d): UMAP of real data and synthetic genotype generated by VAE, GAN, WGAN, and DM, respectively. (e): First 32 principal components of real and WGAN-generated synthetic genotype, explaining approximately 12\% of the total variance.}
    \label{fig:umap_pca}
\end{figure}

\hspace{2em}\textbf{Global Distribution Resemblance}:
A preliminary UMAP visualization of real and synthetic cow populations (Figure \ref{fig:umap_pca}) shows that all models except GAN can approximate the overall data distribution. VAE and DM, both likelihood-based generative models, capture the global central structure well, with synthetic clusters centered around the real data. DM appears slightly better than VAE, as the latter shows more dispersion at the edges. WGAN performs best in this setting, effectively covering both the central structure and the broader population heterogeneity. To further investigate, we compared the first 32 principal components of WGAN-generated and real data and observed near-perfect alignment (Figure \ref{fig:wgan_pca}), suggesting strong distributional fidelity.

\begin{figure}[H]
    \centering
    \begin{minipage}{0.24\textwidth}
        \centering
        \includegraphics[width=\textwidth]{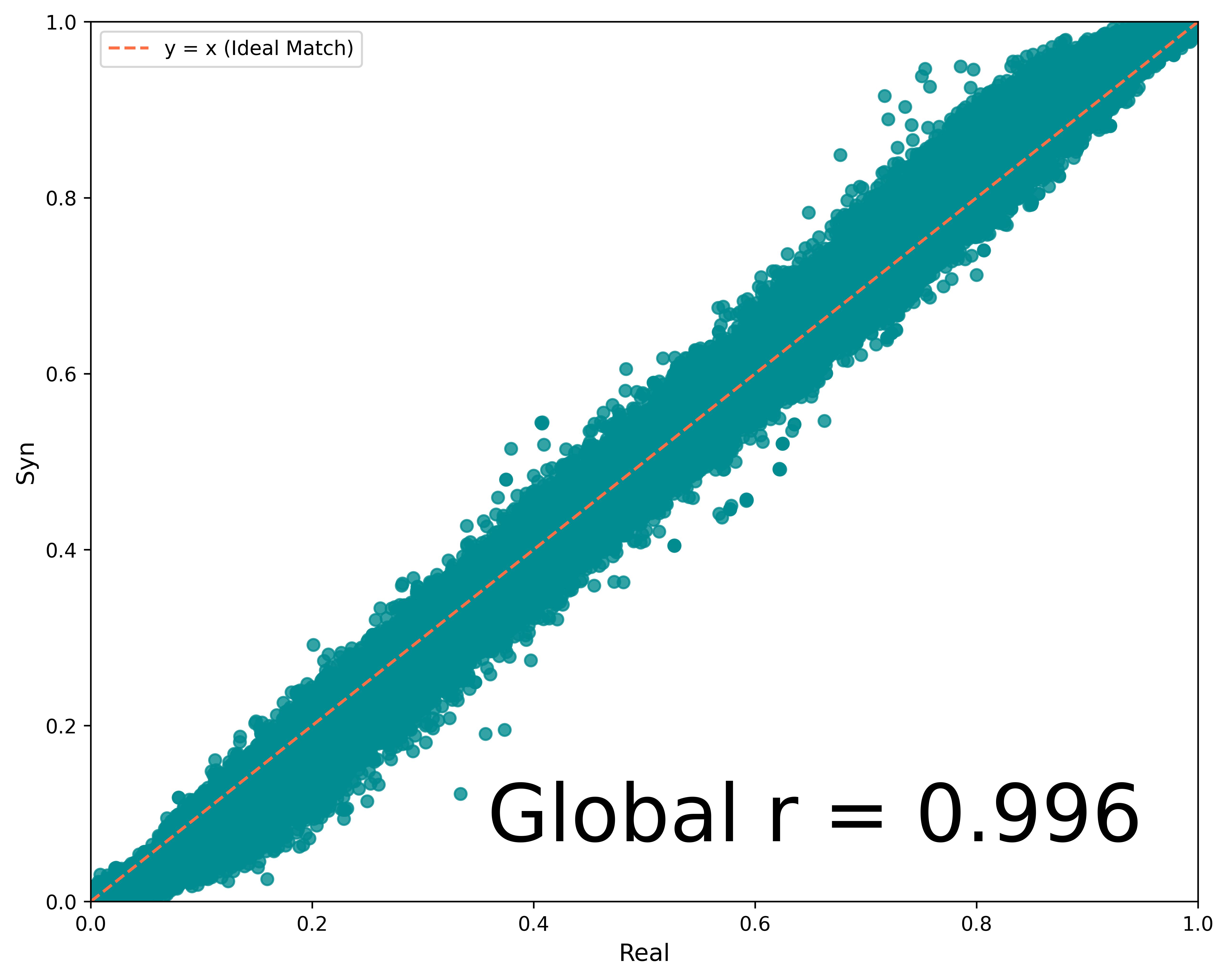}
        \subcaption{VAE}\label{fig:al_freq_vae}
    \end{minipage}%
    \hfill
    \begin{minipage}{0.24\textwidth}
        \centering
        \includegraphics[width=\textwidth]{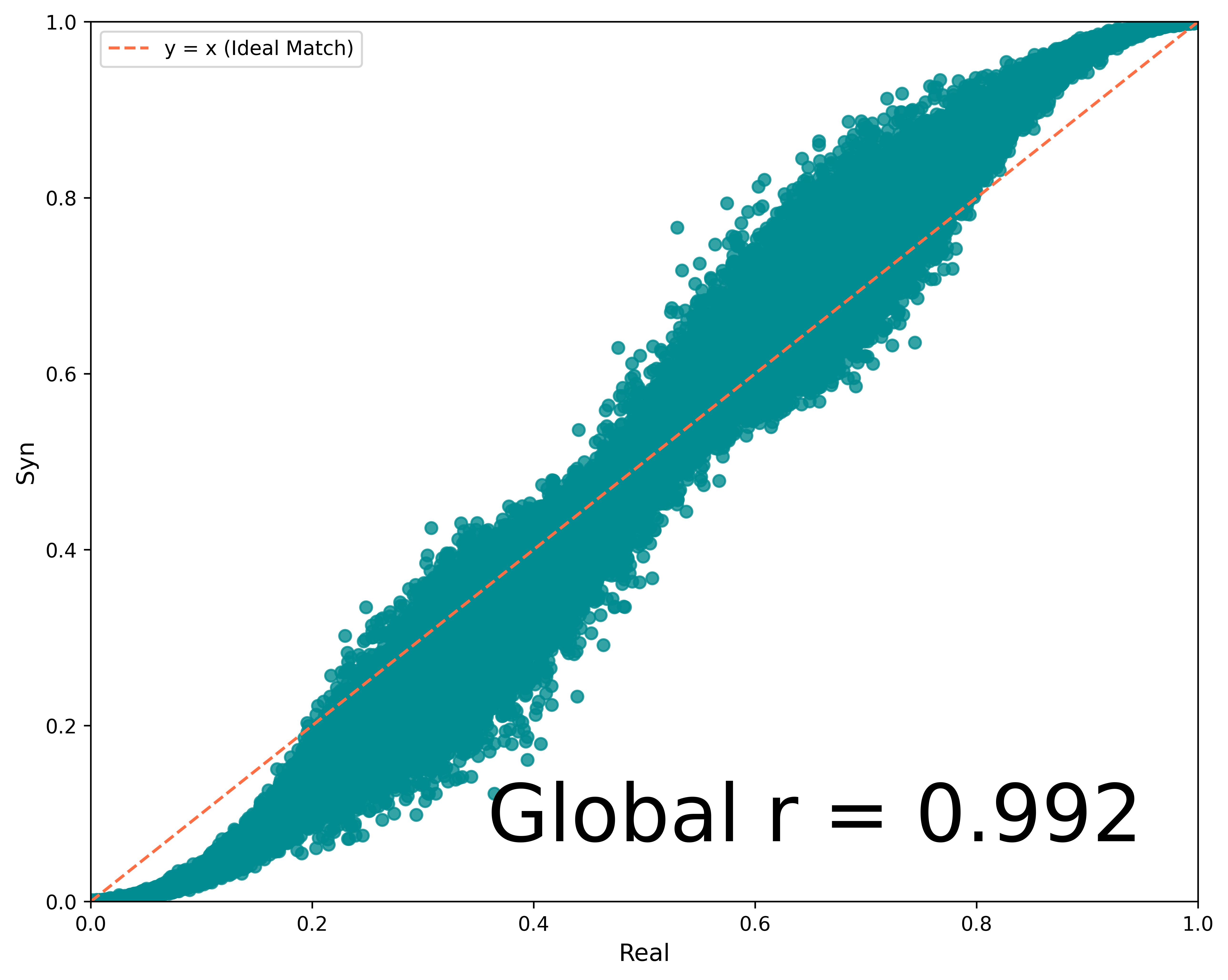}
        \subcaption{GAN}\label{fig:al_freq_gan}
    \end{minipage}%
    \hfill
    \begin{minipage}{0.24\textwidth}
        \centering
        \includegraphics[width=\textwidth]{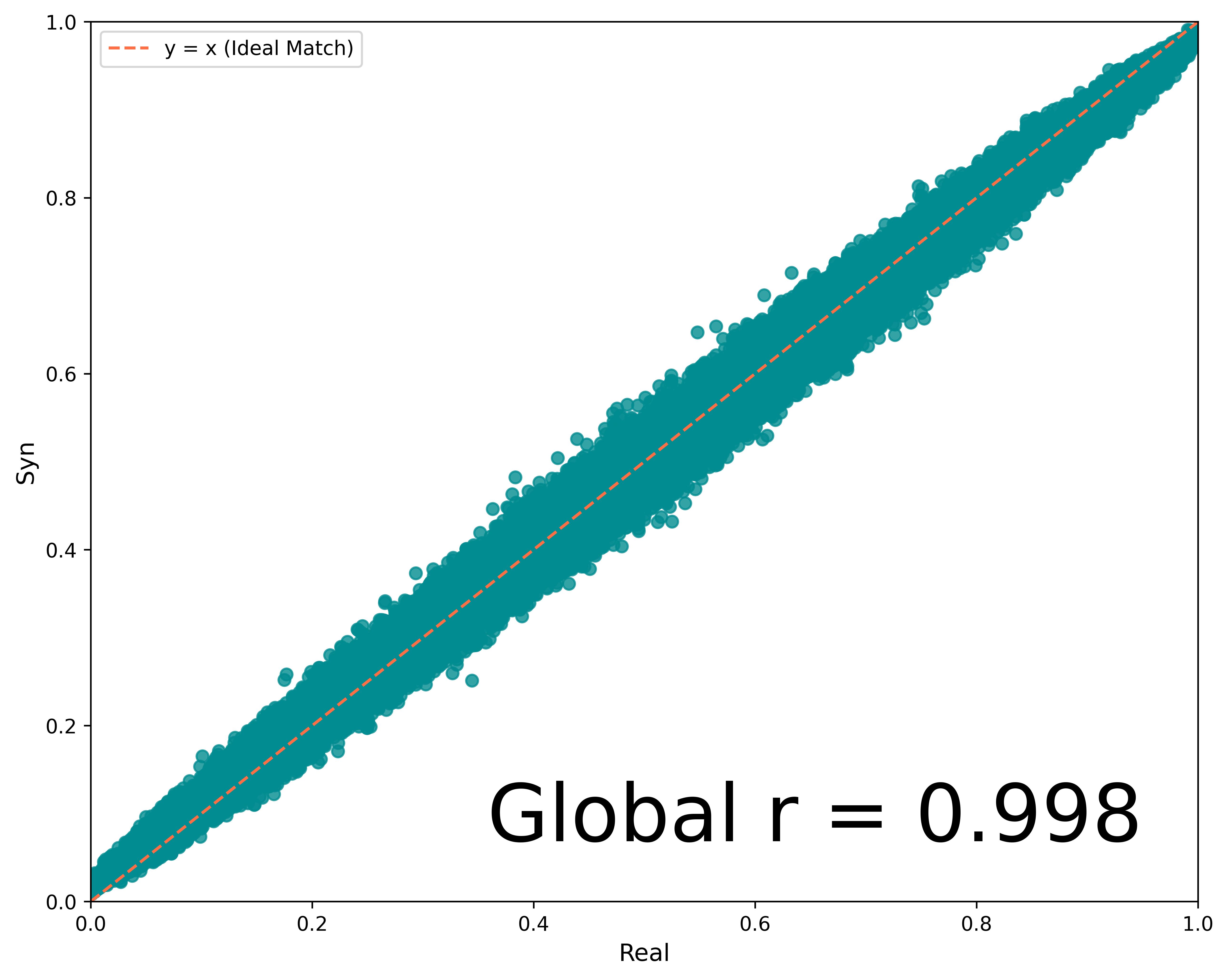}
        \subcaption{WGAN}\label{fig:al_freq_wgan}
    \end{minipage}%
    \hfill
    \begin{minipage}{0.24\textwidth}
        \centering
        \includegraphics[width=\textwidth]{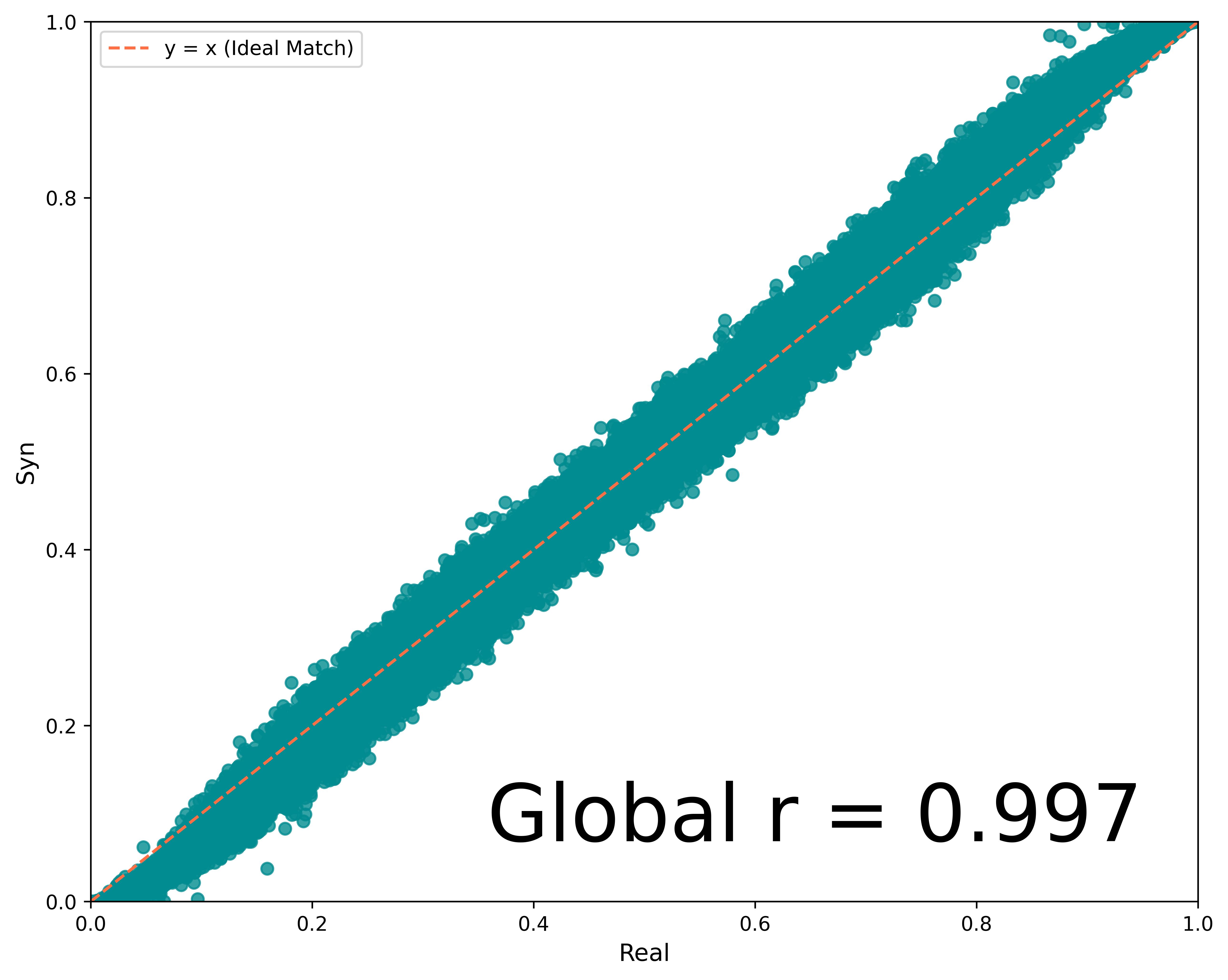}
        \subcaption{DM}\label{fig:al_freq_dm}
    \end{minipage}\\[1ex]
    \begin{minipage}{0.24\textwidth}
        \centering
        \includegraphics[width=\textwidth]{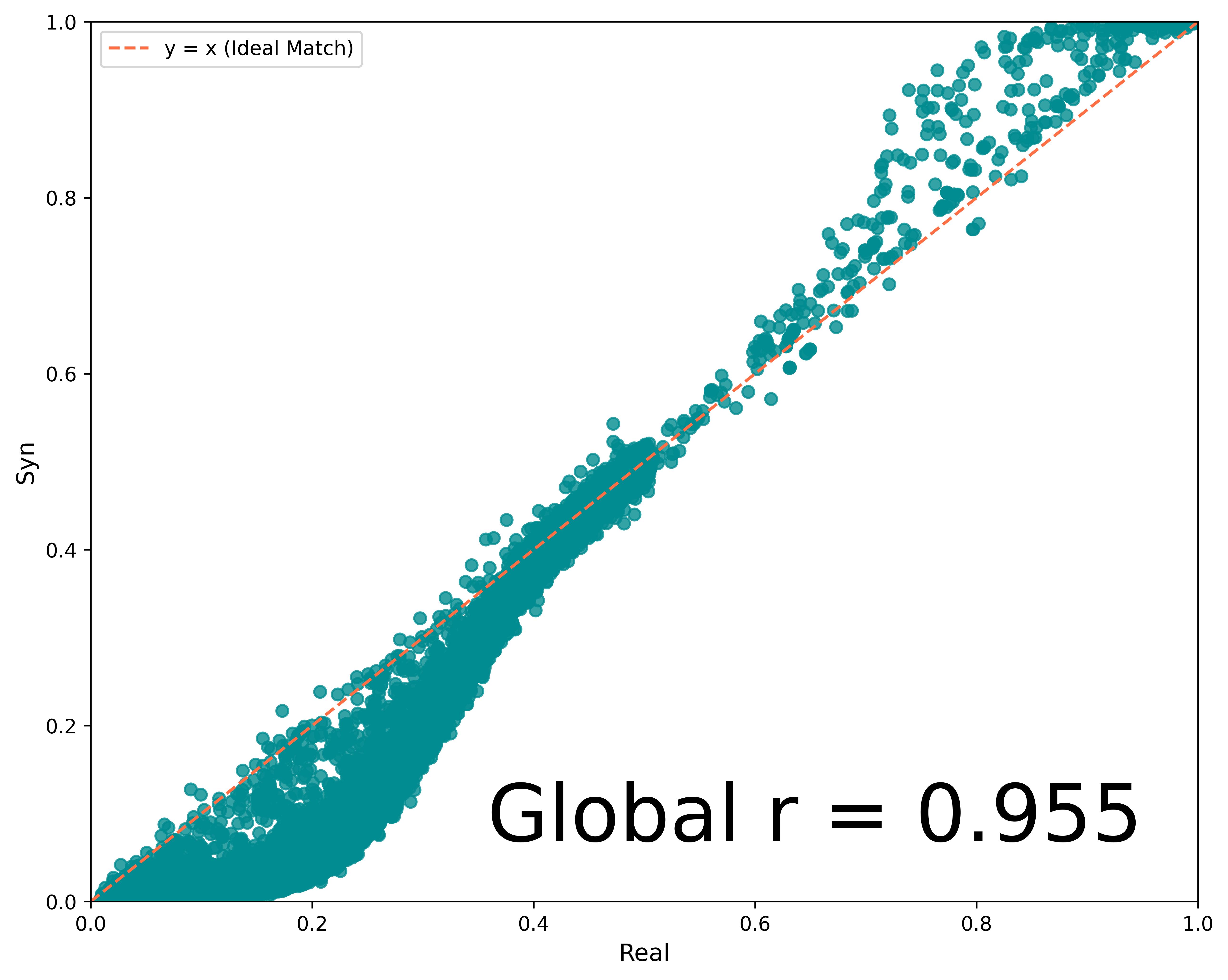}
        \subcaption{VAE}\label{fig:al_freq_vae_ukb}
    \end{minipage}%
    \hfill
    \begin{minipage}{0.24\textwidth}
        \centering
        \includegraphics[width=\textwidth]{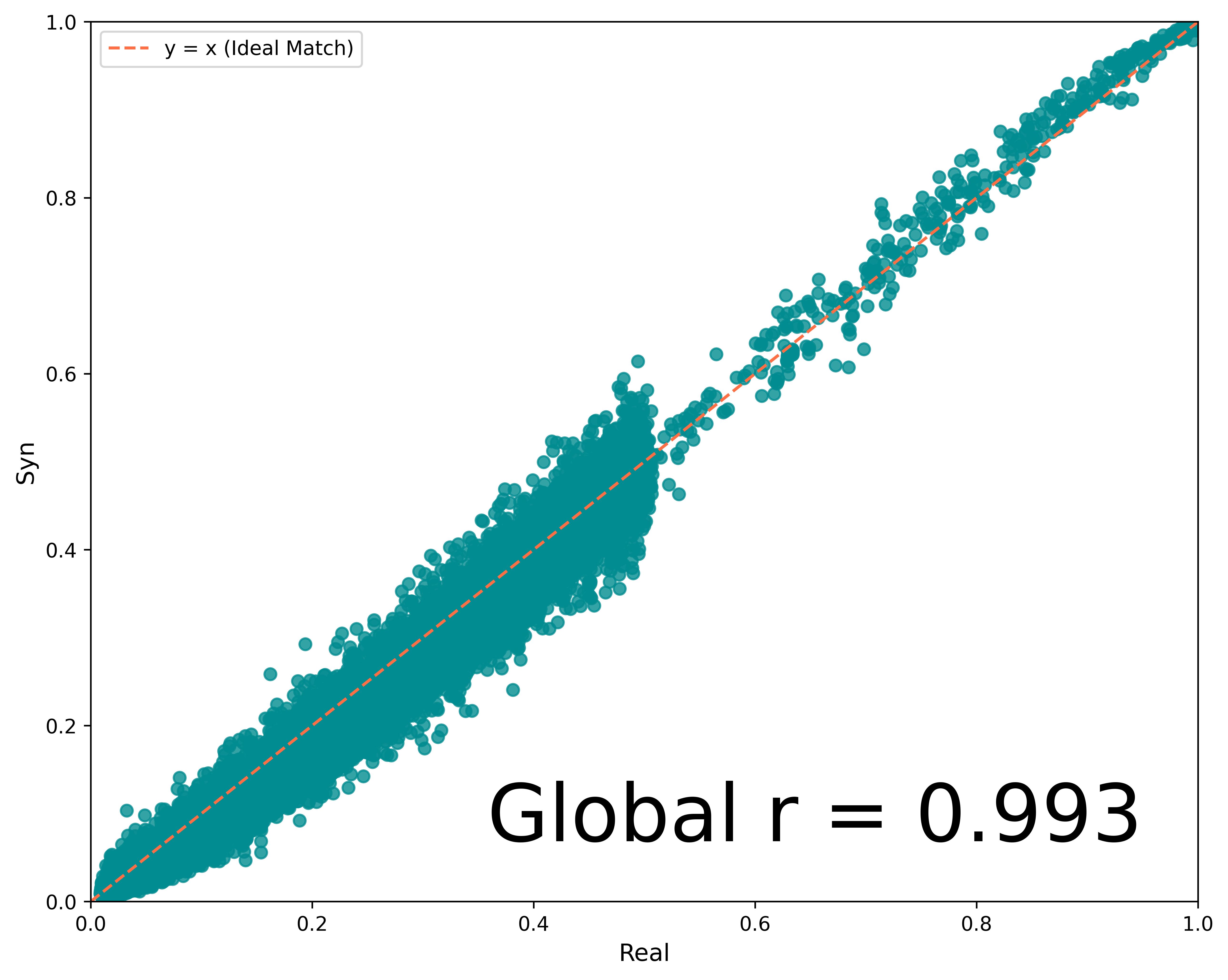}
        \subcaption{GAN}\label{fig:al_freq_gan_ukb}
    \end{minipage}%
    \hfill
    \begin{minipage}{0.24\textwidth}
        \centering
        \includegraphics[width=\textwidth]{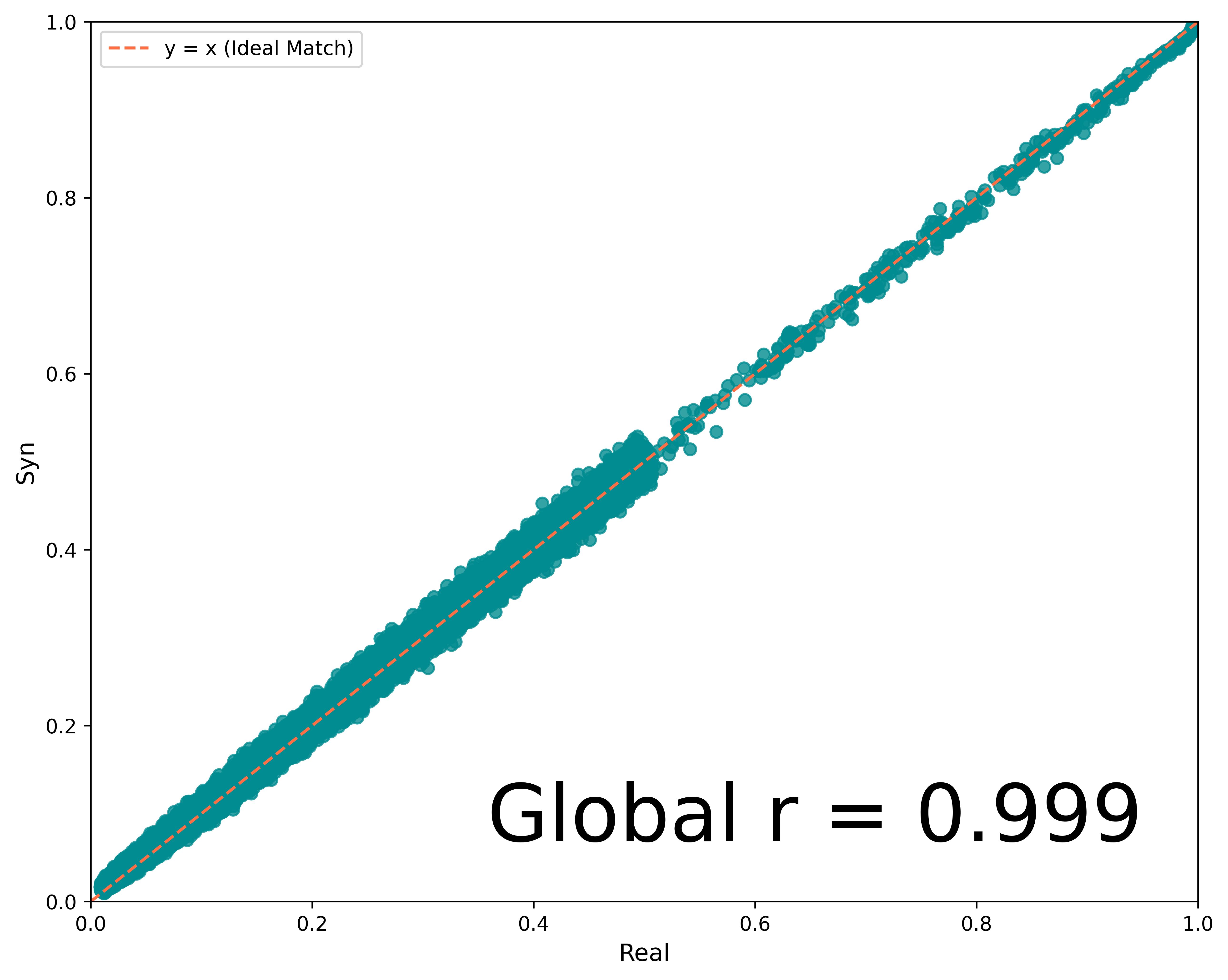}
        \subcaption{WGAN}\label{fig:al_freq_wgan_ukb}
    \end{minipage}%
    \hfill
    \begin{minipage}{0.24\textwidth}
        \centering
        \includegraphics[width=\textwidth]{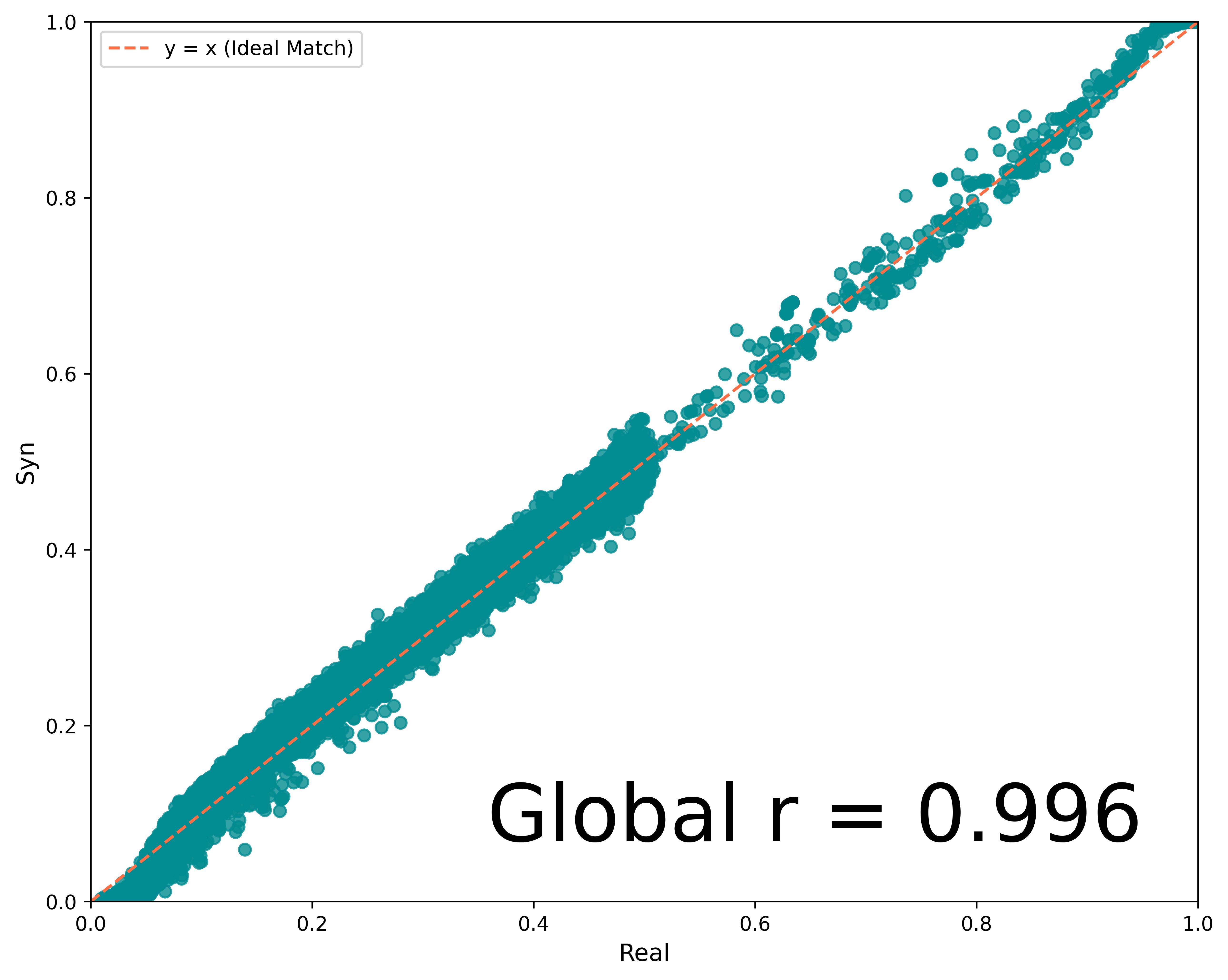}
        \subcaption{DM}\label{fig:al_freq_dm_ukb}
    \end{minipage}\\[1ex]
    \begin{minipage}{0.24\textwidth}
        \centering
        \includegraphics[width=\textwidth]{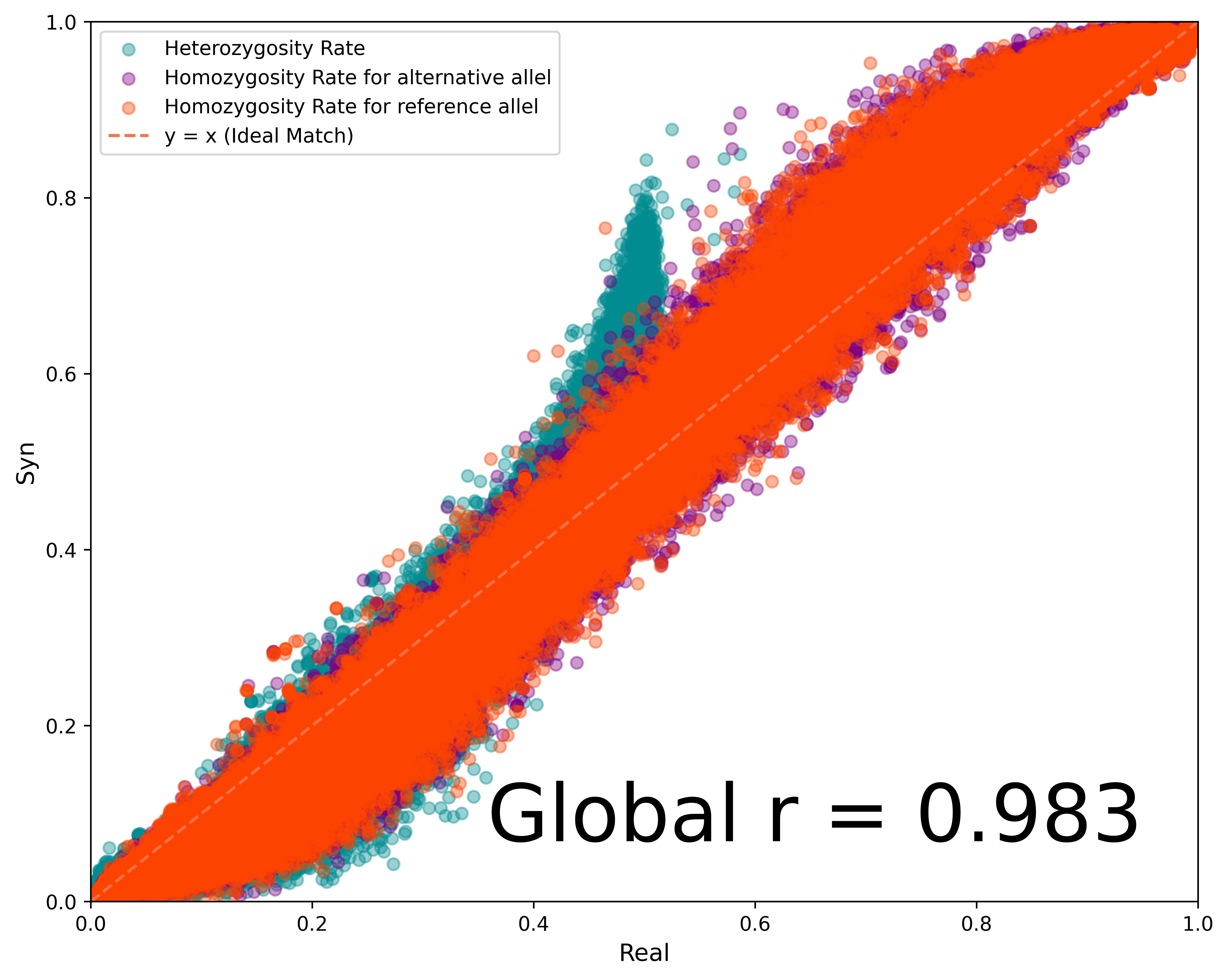}
        \subcaption{VAE}\label{fig:geno_freq_vae}
    \end{minipage}%
    \hfill
    \begin{minipage}{0.24\textwidth}
        \centering
        \includegraphics[width=\textwidth]{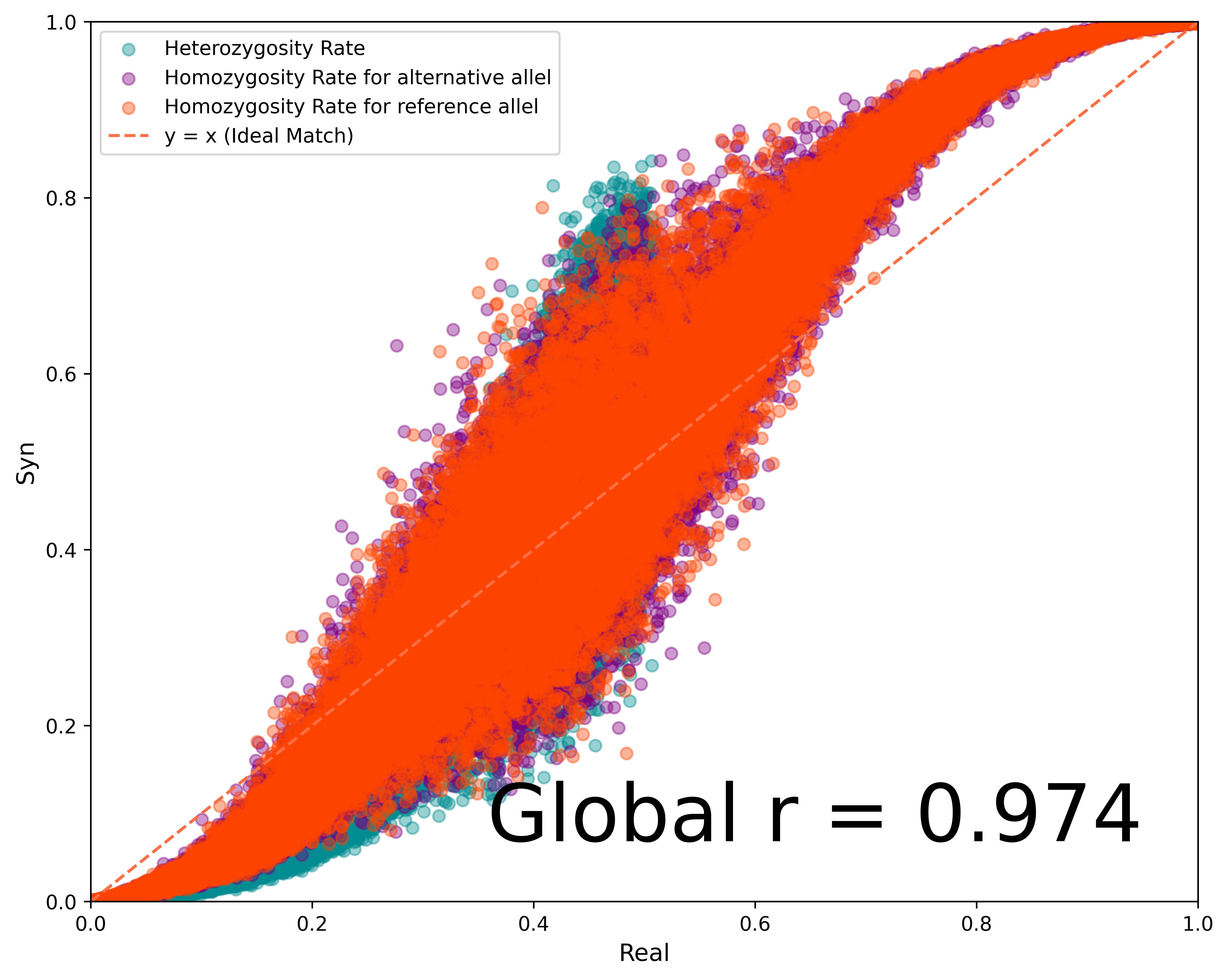}
        \subcaption{GAN}\label{fig:geno_freq_gan}
    \end{minipage}%
    \hfill
    \begin{minipage}{0.24\textwidth}
        \centering
        \includegraphics[width=\textwidth]{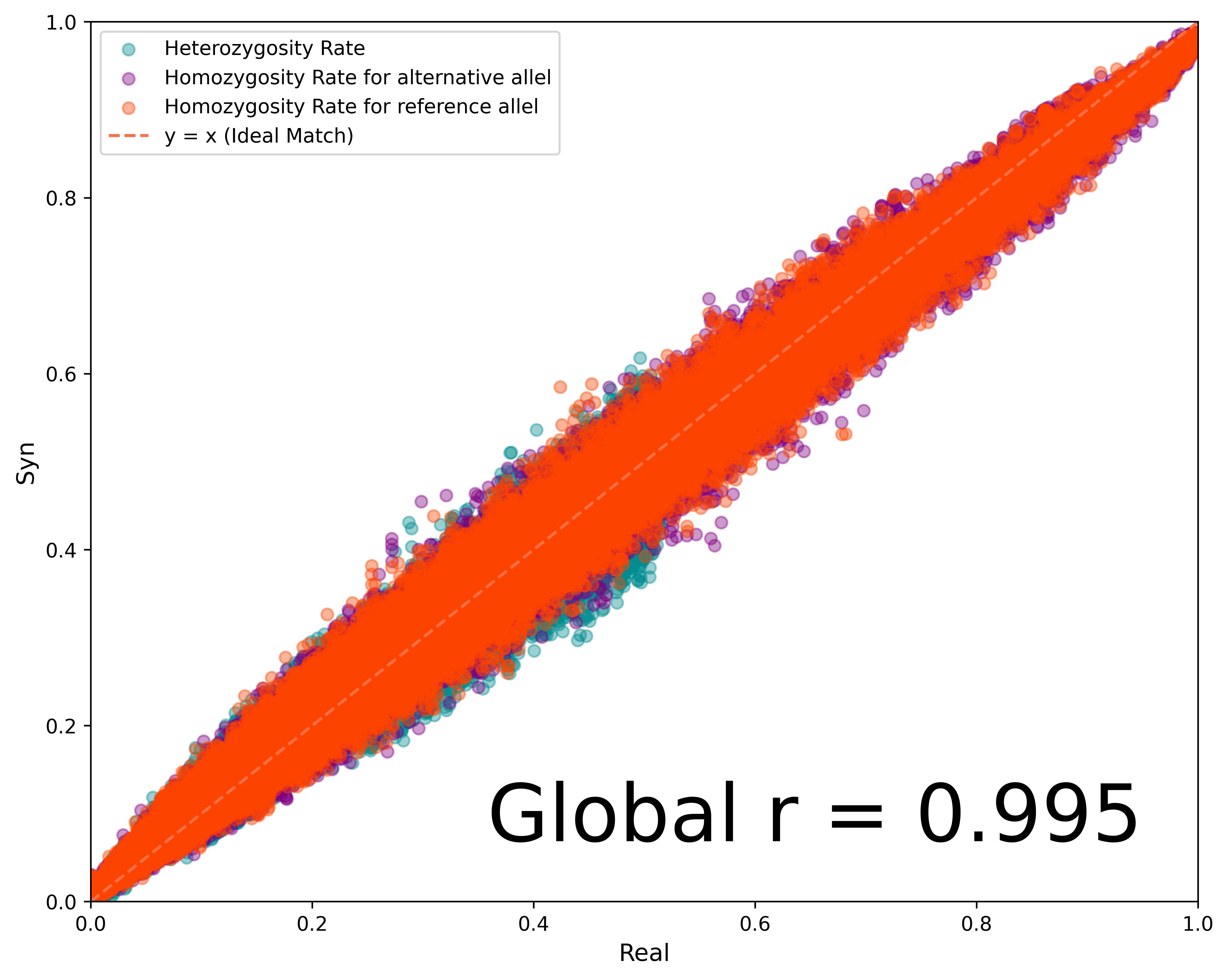}
        \subcaption{WGAN}\label{fig:geno_freq_wgan}
    \end{minipage}%
    \hfill
    \begin{minipage}{0.24\textwidth}
        \centering
        \includegraphics[width=\textwidth]{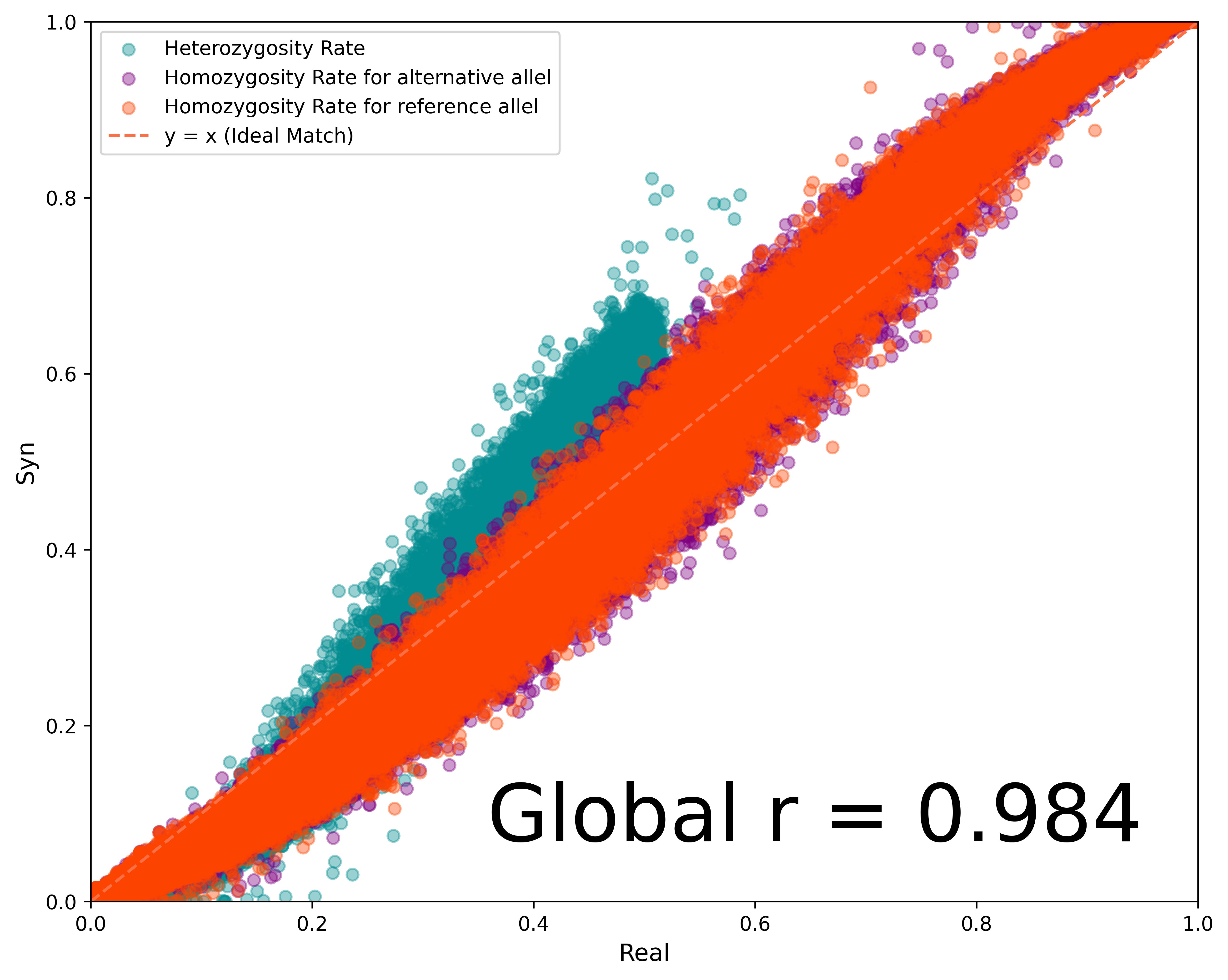}
        \subcaption{DM}\label{fig:geno_freq_dm}
    \end{minipage}
    \begin{minipage}{0.24\textwidth}
        \centering
        \includegraphics[width=\textwidth]{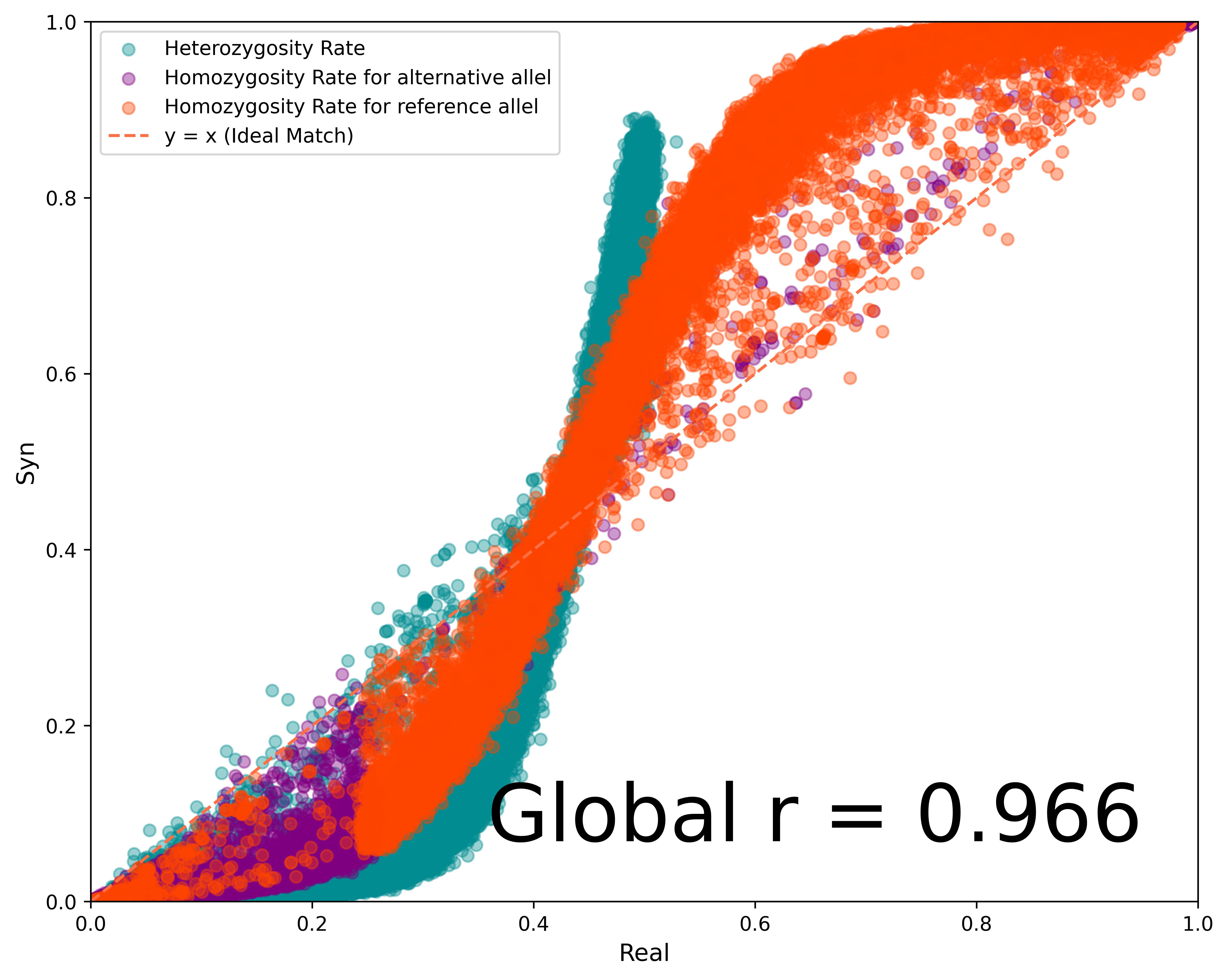}
        \subcaption{VAE}\label{fig:geno_freq_vae_ukb}
    \end{minipage}%
    \hfill
    \begin{minipage}{0.24\textwidth}
        \centering
        \includegraphics[width=\textwidth]{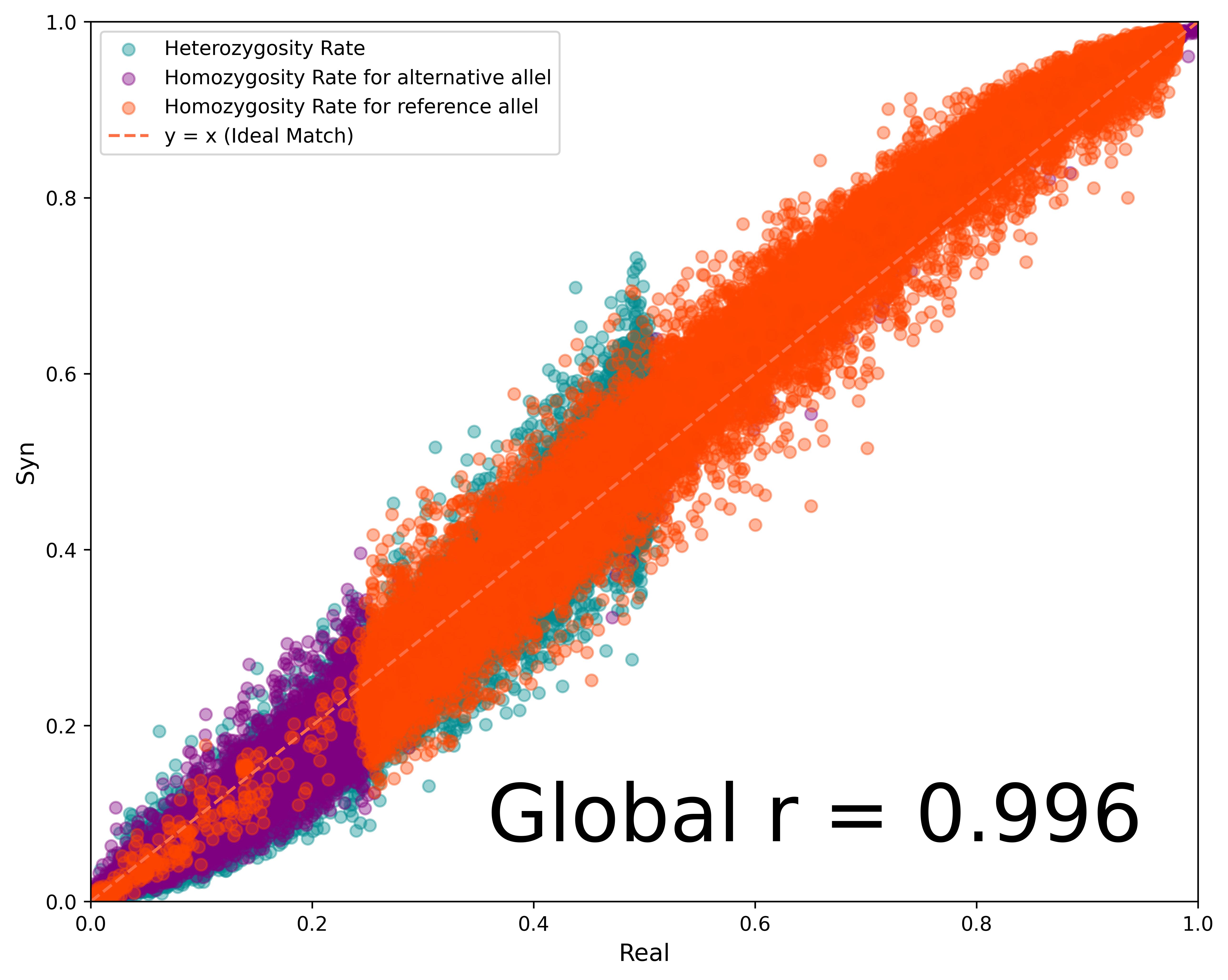}
        \subcaption{GAN}\label{fig:geno_freq_gan_ukb}
    \end{minipage}%
    \hfill
    \begin{minipage}{0.24\textwidth}
        \centering
        \includegraphics[width=\textwidth]{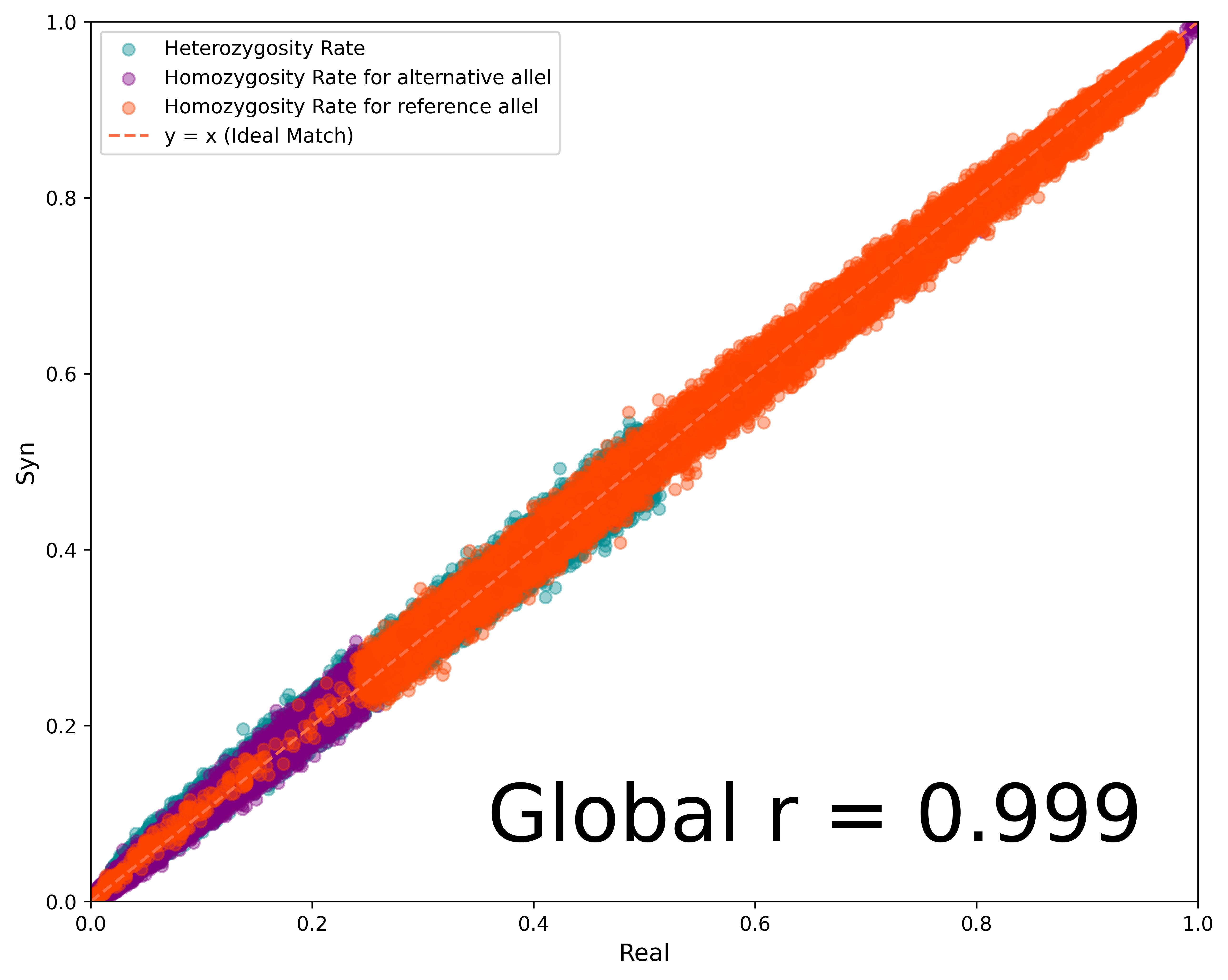}
        \subcaption{WGAN}\label{fig:geno_freq_wgan_ukb}
    \end{minipage}%
    \hfill
    \begin{minipage}{0.24\textwidth}
        \centering
        \includegraphics[width=\textwidth]{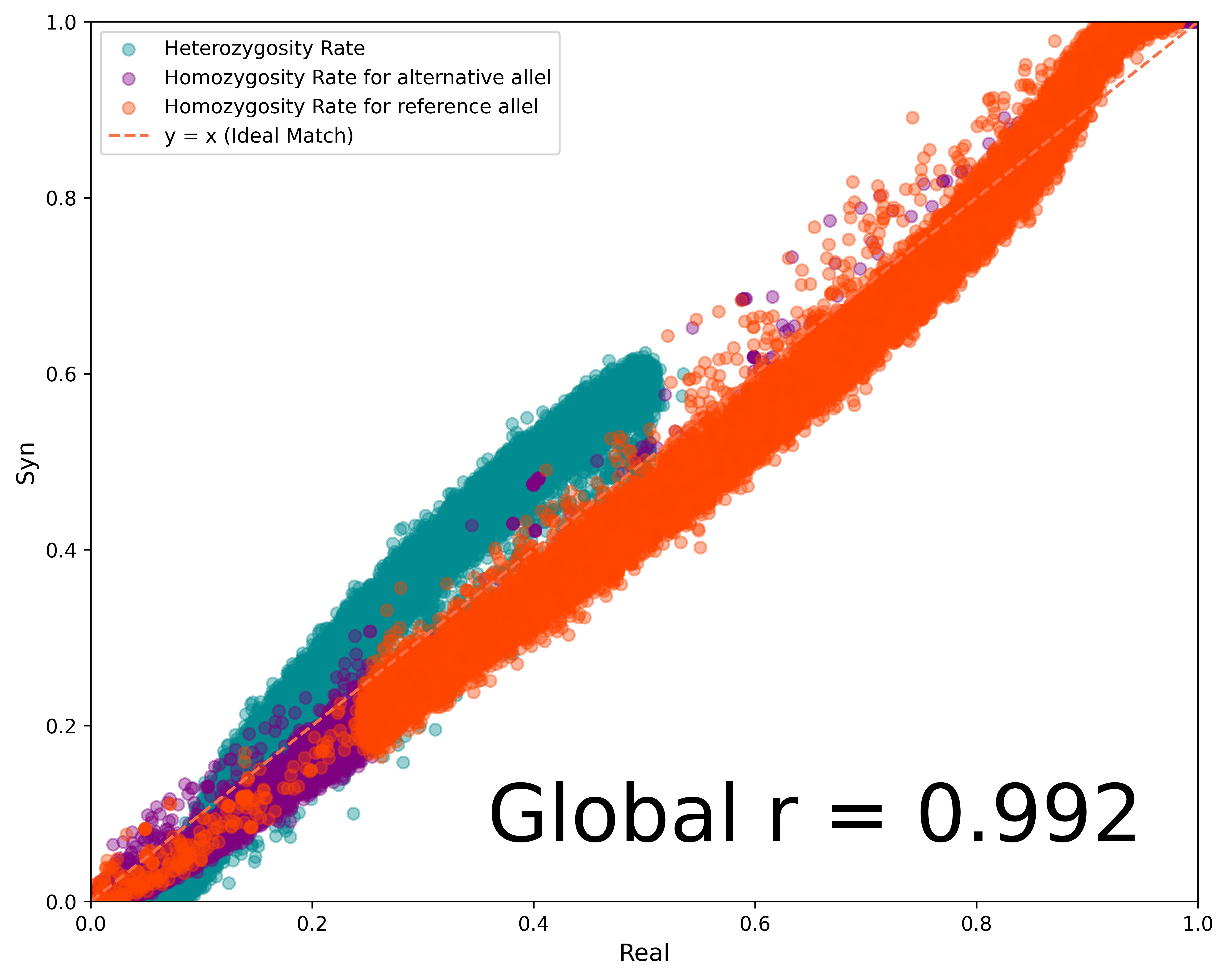}
        \subcaption{DM}\label{fig:geno_freq_dm_ukb}
    \end{minipage}
    \caption{Comparison of Genetic Parameters Between Real and Synthetic Populations Across All Chromosomes of the Cow Dataset and Multiple Chromosomes of the Human Dataset. (a) to (d): Allele frequency comparison between real and synthetic genotype in cow dataset. (e) to (h): Allele frequency comparison between real and synthetic genotype in human dataset. (i) to (l): Genotype frequency comparison between real and synthetic genotype in cow dataset. (m) to (p): Genotype frequency comparison between real and synthetic genotype in human dataset.}
    \label{fig:al_geno_freq}
\end{figure}

\begin{figure}[H]
    \centering
    \begin{minipage}{0.24\textwidth}
        \centering
        \includegraphics[width=\textwidth]{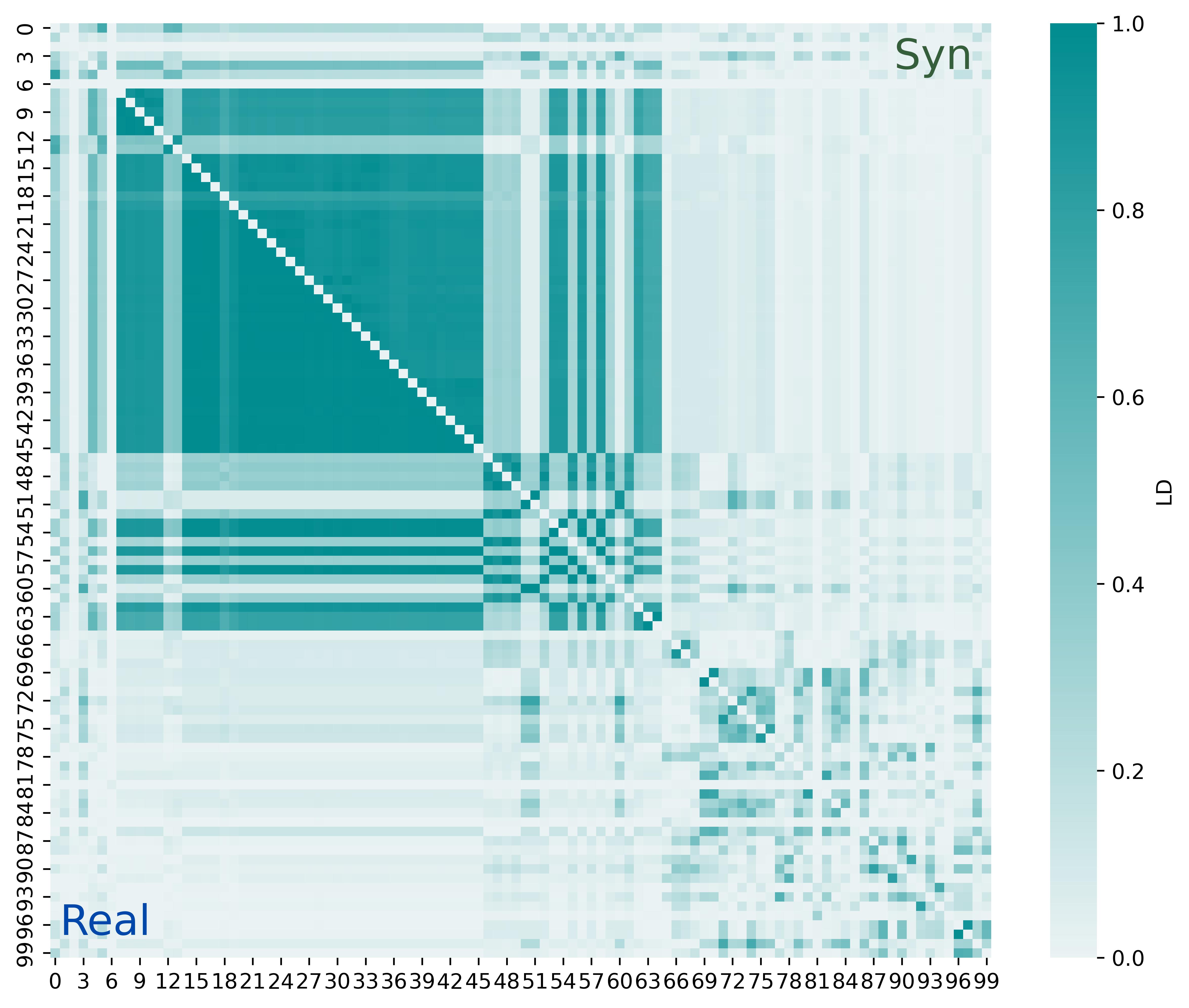}
        \subcaption{VAE}\label{fig:ld_vae}
    \end{minipage}%
    \hfill
    \begin{minipage}{0.24\textwidth}
        \centering
        \includegraphics[width=\textwidth]{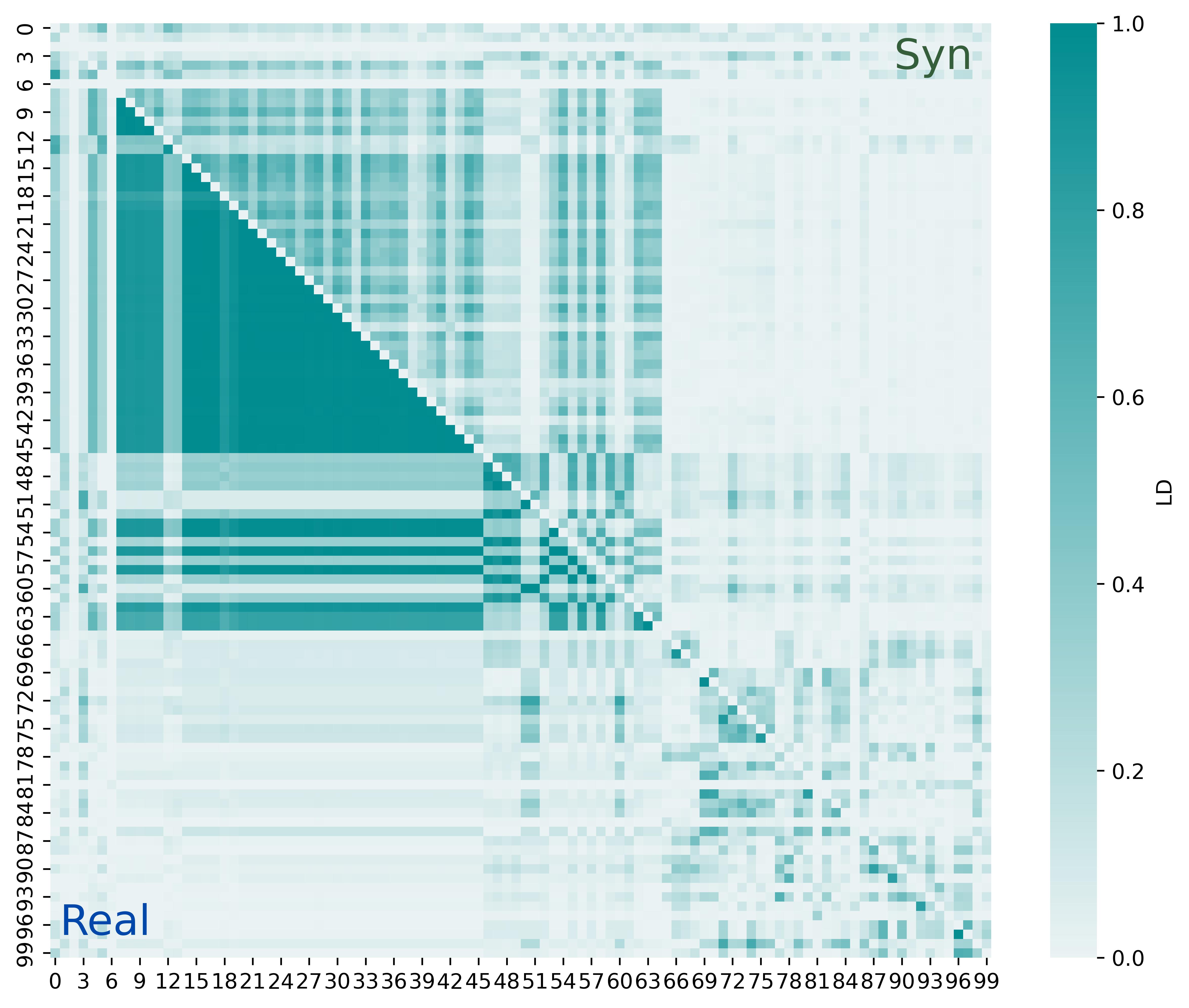}
        \subcaption{GAN}\label{fig:ld_gan}
    \end{minipage}%
    \hfill
    \begin{minipage}{0.24\textwidth}
        \centering
        \includegraphics[width=\textwidth]{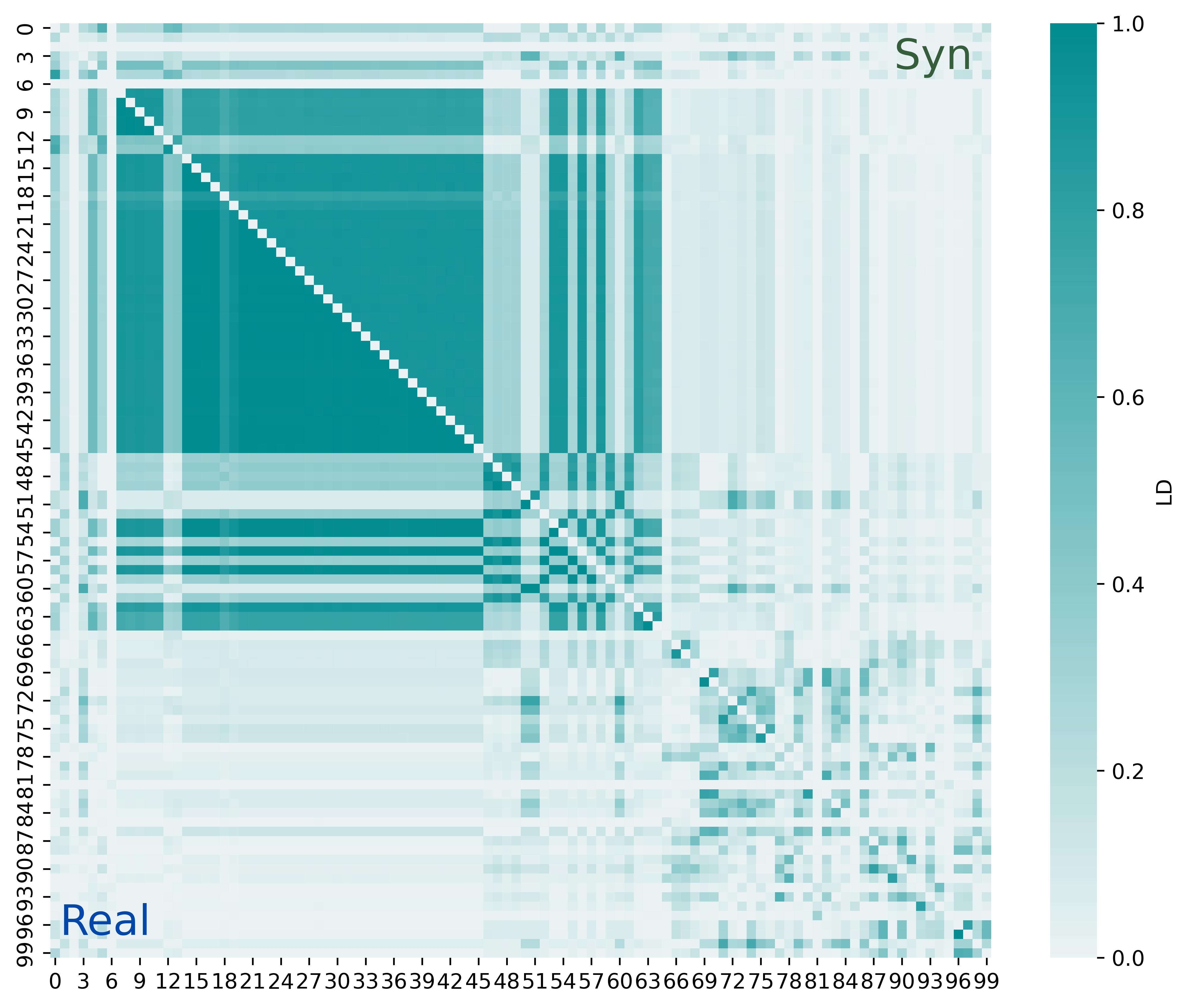}
        \subcaption{WGAN}\label{fig:ld_wgan}
    \end{minipage}%
    \hfill
    \begin{minipage}{0.24\textwidth}
        \centering
        \includegraphics[width=\textwidth]{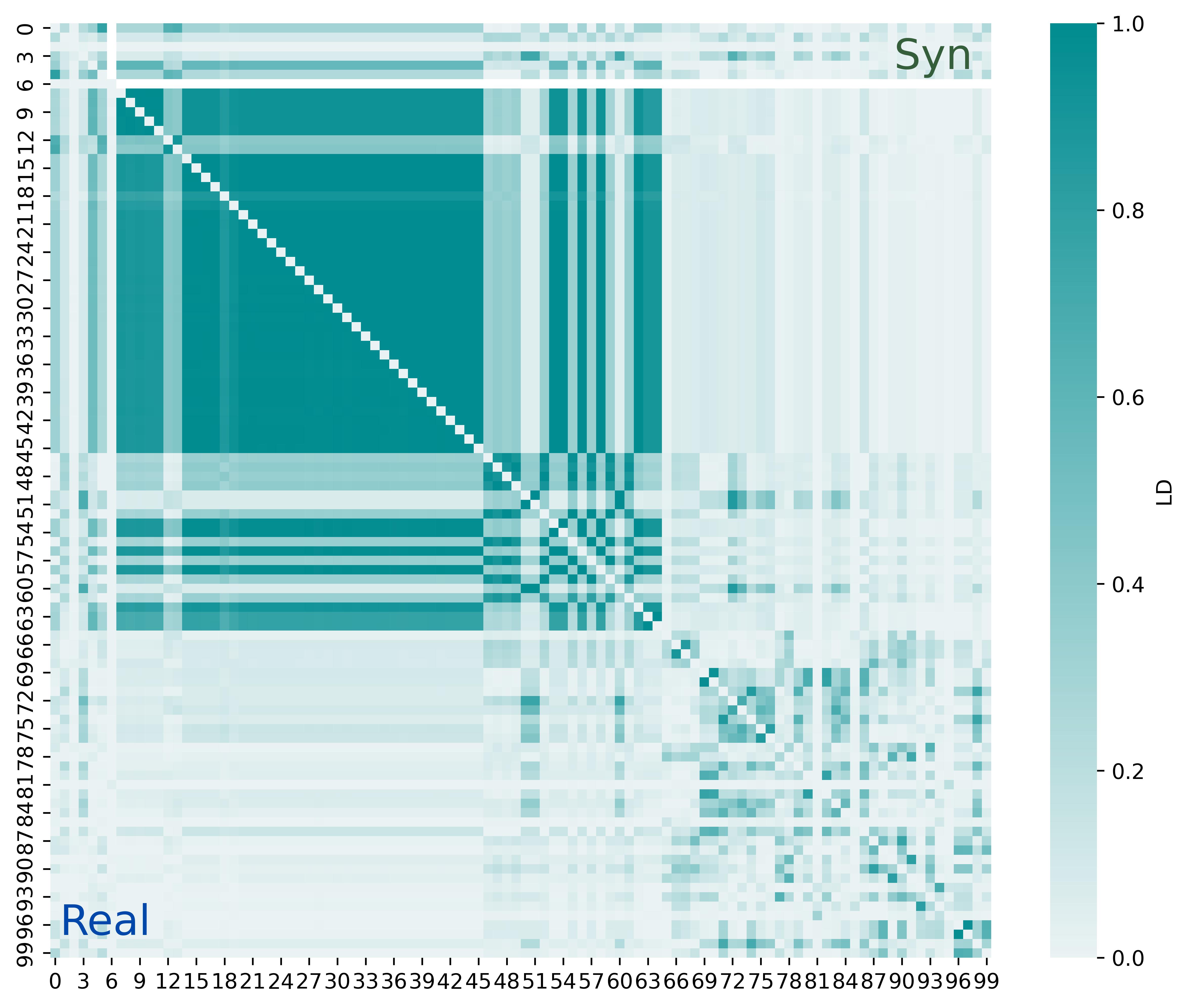}
        \subcaption{DM}\label{fig:ld_dm}
    \end{minipage}\\[1ex]
    \begin{minipage}{0.24\textwidth}
        \centering
        \includegraphics[width=\textwidth]{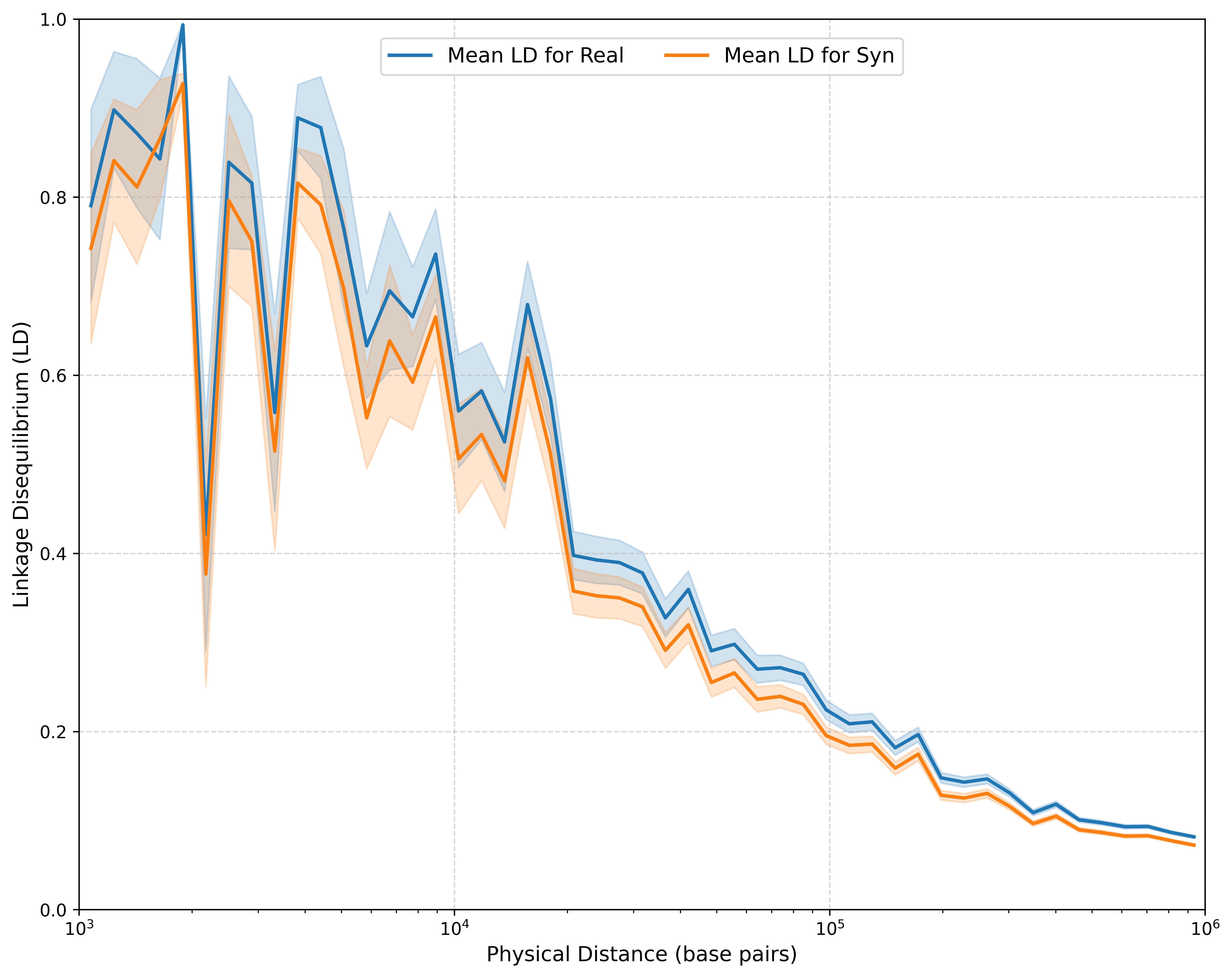}
        \subcaption{VAE}\label{fig:ld_decay_vae}
    \end{minipage}%
    \hfill
    \begin{minipage}{0.24\textwidth}
        \centering
        \includegraphics[width=\textwidth]{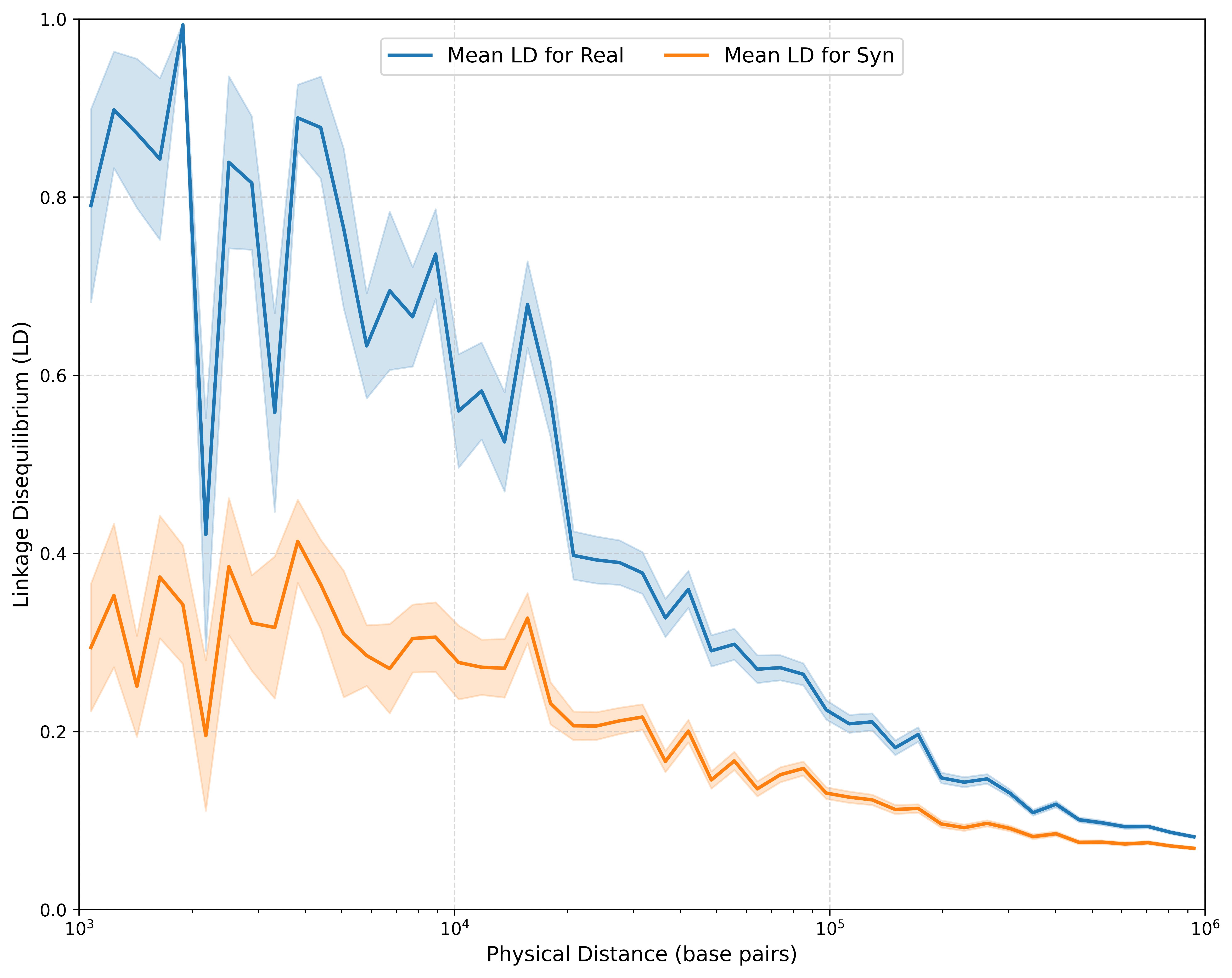}
        \subcaption{GAN}\label{fig:ld_decay_gan}
    \end{minipage}%
    \hfill
    \begin{minipage}{0.24\textwidth}
        \centering
        \includegraphics[width=\textwidth]{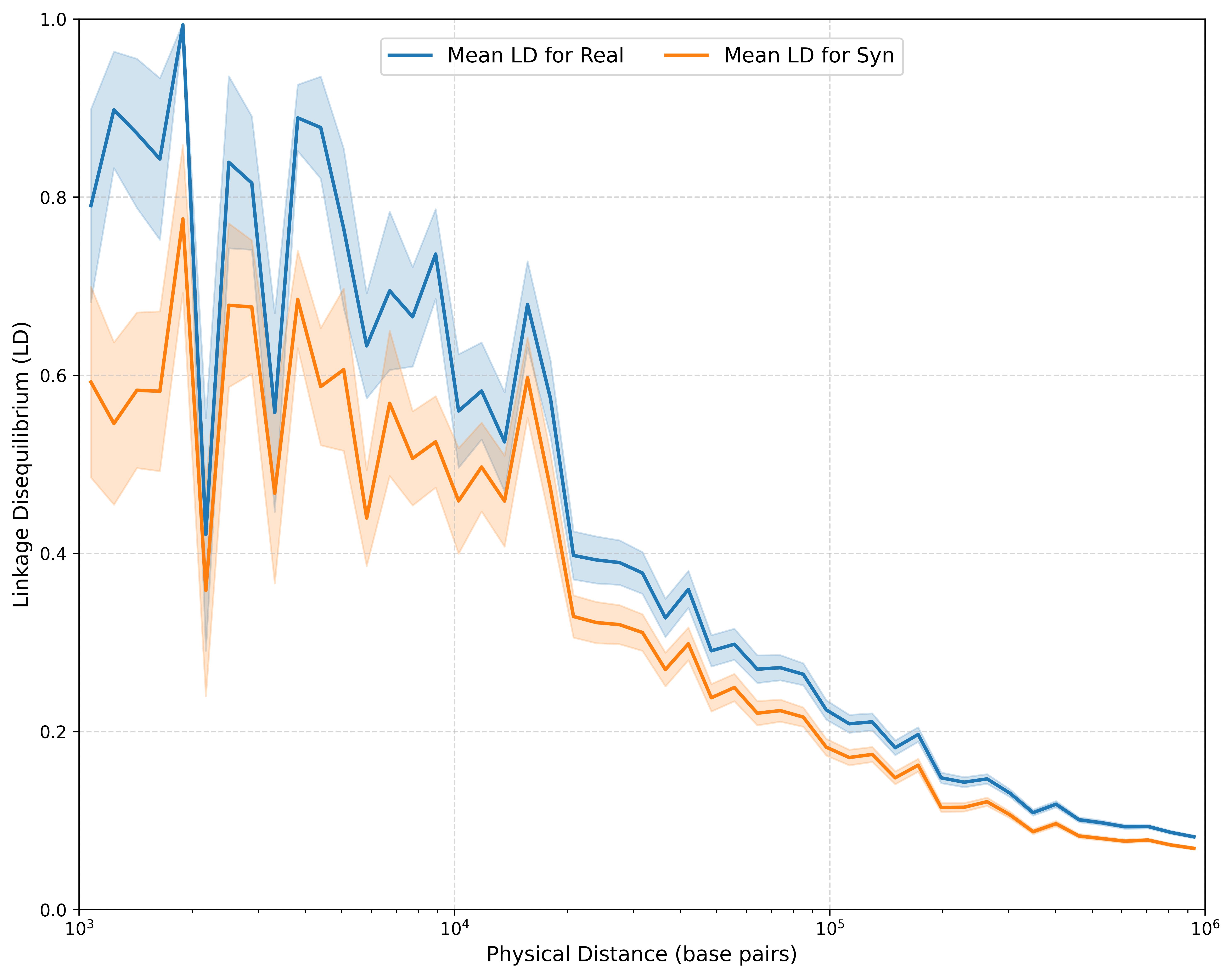}
        \subcaption{WGAN}\label{fig:ld_decay_wgan}
    \end{minipage}%
    \hfill
    \begin{minipage}{0.24\textwidth}
        \centering
        \includegraphics[width=\textwidth]{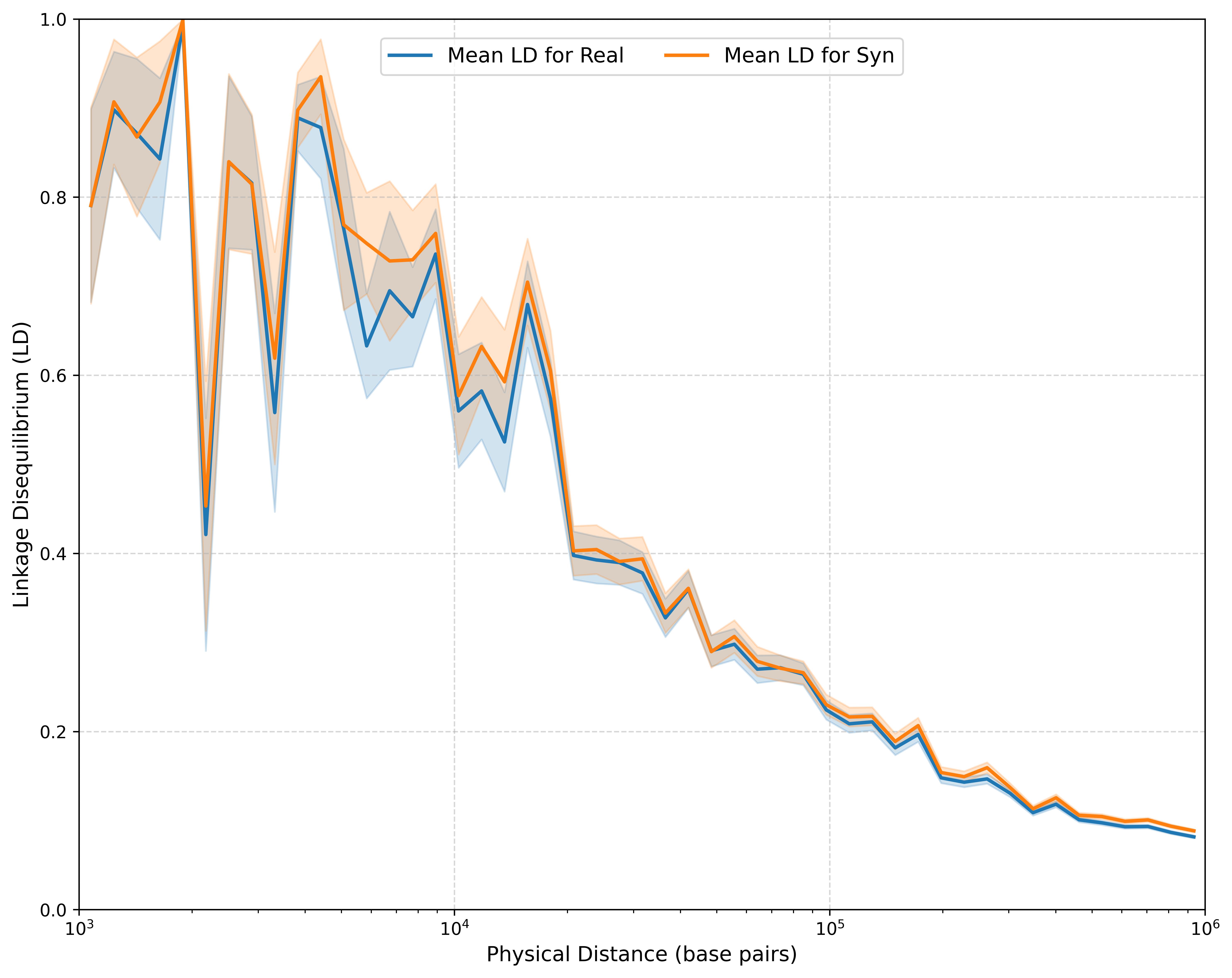}
        \subcaption{DM}\label{fig:ld_decay_dm}
    \end{minipage}
    \caption{Comparison of Linkage Disequilibrium Between Real and Synthetic Populations on Chromosome 14 of Cow Dataset. (a) to (d): LD block heatmaps, where the upper diagonal represents pairwise LD in synthetic population and the lower diagonal represents pairwise LD in real population. (e) to (h): LD decay with respect to physical distance in real and synthetic populations.}
    \label{fig:ld}
\end{figure}

\textbf{Genetic Parameters and Linkage Structure Comparison:} 
Figure~\ref{fig:al_geno_freq} shows the comparison of allele and genotype frequencies between real and synthetic populations. Among all models, WGAN clearly outperforms the others, achieving nearly perfect correlation score between real and synthetic frequency distributions. While the other models can partially capture the frequency profiles, we observed a consistent pattern of deviation: the frequency plots exhibit a sigmoid-like distortion. This phenomenon reflects a known Matthew effect \cite{matthew_effect_1, matthew_effect_2}, where the model tends to overestimate high-frequency variants and underestimate rare ones, amplifying existing disparities in the data. For VAE, this phenomenon may be linked to its likelihood-based objective, which encourages prioritizing frequent patterns to maximize likelihood. DM performs better in this regard, possibly due to its hierarchical noise removal mechanism. GAN, which is known to suffer from mode collapse \cite{mode_collapse}, exhibits this effect more severely. WGAN appears to be the only model that successfully mitigates this bias and accurately preserves the full frequency spectrum. In terms of LD, shown in Figure~\ref{fig:ld}, all models except GAN manage to reproduce the original LD block structure and show a similar decay pattern with increasing distance. However, both VAE and WGAN tend to underestimate the strength of LD, while DM most closely matches the LD structure observed in the real population.

\textbf{Quantitative Evaluation Metrics:}
Table \ref{tab:unsuper} summarizes the results across all quantitative metrics. For relatively small datasets (e.g., single chromosome in cow dataset with around a thousand SNPs), VAE, WGAN, and DM perform well across most metrics. In contrast, GAN suffers from mode collapse, leading to a recall score close to $0$. For larger-scale datasets (e.g., full chromosomes in cow and multiple chromosomes in human), WGAN consistently outperforms the other models. This is particularly evident for the recall metric: while all models tend to achieve high precision, WGAN is the only model that significantly improves recall. This aligns with the UMAP visualization in Figure~\ref{fig:umap_pca}, which shows WGAN covering the full data distribution more effectively. Overall, WGAN achieves the best results across most metrics, although DM occasionally surpasses it in correlation score on human datasets.

\textbf{Factors Affecting the Complexity of Generative Modeling:}
Table~\ref{tab:unsuper} suggests that the difficulty of generative modeling is related to the input dimensionality: higher dimensions generally make learning more challenging, and we indeed observed this trend. However, we also observed that in the human dataset, CHR 6 has a higher input dimension than CHR 12, yet it is easier for models to learn. This indicates that input dimension alone does not fully explain the complexity. Upon further analysis, we found that SNP dependency also plays an important role. When SNPs exhibit stronger dependency, generative models can more easily capture the underlying distribution (Supplementary Section 6). Comparing across datasets, models consistently perform better on cow dataset than on human dataset. This may be attributed to differences in genetic structure: the cow dataset exhibits stronger LD, implying higher SNP dependency, while the human dataset shows greater genetic variability, which increases learning complexity.

\textbf{On the Robustness of Evaluation Metrics:}
When assessing the robustness of the evaluation metrics, we found that $F_{ST}^{\mathrm{aggregated}}$, F1 and correlation score are highly correlated: good performance in one metric typically results in good performance across the others. A clear trade-off exists between precision and recall: models can achieve high precision by capturing only the core of the real distribution, whereas recall reflects how well the model covers the full diversity. A more detailed examination of the $AA$ score reveals an important nuance, consistent with findings reported in \cite{yelmen2023}: extreme scenarios may yield a favorable global $AA$ score while masking poor generative behavior. Ideally, both $AA_{real}$ and $AA_{syn}$ should be close to $0.50$. However, when applying DM to human dataset, we observed a global $AA$ score of $0.50$ resulting from an imbalanced case where $AA_{real} \approx 0$ and $AA_{syn} \approx 1$. This means that real samples are closer to synthetic ones than to other real samples, while synthetic samples are only close to each other and fail to reflect the diversity of the real distribution. This is also consistent with the observation that precision is close to $1$ while recall is near $0$, indicating that the model captures only a subset of the true distribution (Supplementary Section 7).

\begin{sidewaystable}
\centering
\caption{Quantitative performance indicators for all generative models on Cow and Human datasets.}
\label{tab:unsuper}
\begin{tabular}{|c|c|c|c|c|c|c|c|c|}
\toprule
Dataset & Chromosome  & Model  & $F_{ST}^{\mathrm{aggregated}}$ ↓ & Precision (\%) ↑ & Recall (\%) ↑ & F1 (\%) ↑ & Corr(\%) ↑ & AA \\ \hline
\multirow{12}{*}{Cow} & \multirow{4}{*}{CHR 14 (1771 SNPs)} & VAE & $1.81\mathrm{e}{-4}\pm7\mathrm{e}{-6}$
 & 99.06 ± 0.04 & 99.70 ± 0.04 & 99.38 ± 0.02 & 96.87 ± 0.04 & 0.63 ± $2\mathrm{e}{-3}$ \\ 
 &  & GAN & $1.88\mathrm{e}{-4}\pm4\mathrm{e}{-6}$ & 80.88 ± 0.23 & 57.97 ± 0.45 & 67.53 ± 0.24 & 72.60 ± 0.13 & 0.99 ± $3\mathrm{e}{-4}$ \\ 
 &  & WGAN & $\mathbf{1.19\mathrm{e}{-4}\pm2\mathrm{e}{-6}}$ & 99.64 ± 0.04 & \textbf{99.88 ± 0.02} & \textbf{99.76 ± 0.02} & \textbf{98.65 ± 0.02} & $\mathbf{0.55 \pm 3\mathrm{e}{-3}}$ \\ 
 &  & DM & $3.07\mathrm{e}{-4}\pm6\mathrm{e}{-6}$ & \textbf{99.92 ± 0.01} & 99.13 ± 0.03 & 99.52 ± 0.02 & 98.53 ± 0.01 & 0.63 ± $2\mathrm{e}{-3}$ \\  \cmidrule{2-9}
 & \multirow{4}{*}{CHR 5 (2238 SNPs)} & VAE & $4.05\mathrm{e}{-4}\pm5\mathrm{e}{-6}$ & \textbf{99.87 ± 0.02} & 99.51 ± 0.03 & \textbf{99.69 ± 0.02} & 97.21 ± 0.04 & 0.68 ± $2\mathrm{e}{-3}$ \\ 
 &  & GAN & $3.99\mathrm{e}{-3}\pm4\mathrm{e}{-5}$ & 88.40 ± 0.21 & 0.01 ± 0.00 & 0.01 ± 0.01 & 55.40 ± 0.05 & 1.00 ± $7\mathrm{e}{-5}$ \\ 
 &  & WGAN & $\mathbf{1.22\mathrm{e}{-4}\pm4\mathrm{e}{-6}}$ & 98.98 ± 0.07 & \textbf{99.86 ± 0.03} & 99.42 ± 0.04 & \textbf{98.74 ± 0.01} & $\mathbf{0.63 \pm 2\mathrm{e}{-3}}$ \\  
 &  & DM & $3.10\mathrm{e}{-4}\pm4\mathrm{e}{-6}$ & 99.86 ± 0.02 & 99.26 ± 0.07 & 99.56 ± 0.04 & 98.17 ± 0.01 & 0.65 ± $3\mathrm{e}{-3}$ \\  \cmidrule{2-9}
 & \multirow{4}{*}{All CHRs (50161 SNPs)} & VAE & $1.80\mathrm{e}{-3}\pm1\mathrm{e}{-5}$ & 99.99 ± 0.01 & 11.65 ± 1.03 & 20.85 ± 1.65 & 73.03 ± 0.11 & 0.96 ± $1\mathrm{e}{-3}$ \\ 
 &  & GAN & $5.58\mathrm{e}{-3}\pm1\mathrm{e}{-5}$ & \textbf{100 ± 0.00} & 0.00 ± 0.00 & 0.00 ± 0.00 & 0.52 ± 0.01 & 0.98 ± $2\mathrm{e}{-3}$ \\  
 &  & WGAN & $\mathbf{6.21\mathrm{e}{-4}\pm5\mathrm{e}{-6}}$ & 92.00 ± 0.16 & \textbf{99.93 ± 0.01} & \textbf{95.80 ± 0.09} & \textbf{83.32 ± 0.06} & $\mathbf{0.74 \pm 7\mathrm{e}{-3}}$ \\ 
 &  & DM & $1.10\mathrm{e}{-3}\pm1\mathrm{e}{-6}$ & \textbf{100 ± 0.00} & 40.59 ± 0.63 & 57.74 ± 0.64 & 76.56 ± 0.10 & 0.94 ± $1\mathrm{e}{-3}$ \\ \cmidrule{2-9} \midrule
\multirow{16}{*}{Human} & \multirow{4}{*}{Ensembl (3493 SNPs)} & VAE & $2.88\mathrm{e}{-2}\pm3\mathrm{e}{-5}$ & \textbf{100 ± 0.00} & 0.29 ± 0.24 & 0.57 ± 0.44 & 39.74 ± 1.35 & $\mathbf{0.50 \pm 1\mathrm{e}{-5}}$ \\
 &  & GAN & $5.00\mathrm{e}{-3}\pm9\mathrm{e}{-6}$ & 99.98 ± 0.01 & 0.00 ± 0.00 & 0.00 ± 0.00 & 34.03 ± 0.09 & 0.52 ± $5\mathrm{e}{-4}$ \\ 
 &  & WGAN & $\mathbf{1.31\mathrm{e}{-4}\pm2\mathrm{e}{-6}}$ & 71.84 ± 0.11 & \textbf{97.86 ± 0.11} & \textbf{82.85 ± 0.11} & \textbf{83.74 ± 0.03} & 0.76 ± $1\mathrm{e}{-2}$ \\ 
 &  & DM & $1.53\mathrm{e}{-3}\pm5\mathrm{e}{-6}$ & \textbf{100 ± 0.00} & 13.96 ± 0.07 & 24.49 ± 0.11 & 61.73 ± 0.72 & $\mathbf{0.50 \pm 3\mathrm{e}{-4}}$ \\  \cmidrule{2-9}
 & \multirow{4}{*}{CHR 6 (12283 SNPs)} & VAE & $6.08\mathrm{e}{-3}\pm3\mathrm{e}{-5}$ & 99.99 ± 0.01 & 0.05 ± 0.07 & 0.10 ± 0.13 & \textbf{64.93 ± 0.03} & $\mathbf{0.50 \pm 6\mathrm{e}{-5}}$ \\ 
 &  & GAN & $1.62\mathrm{e}{-3}\pm6\mathrm{e}{-6}$ & 99.02 ± 0.12 & 0.17 ± 0.06 & 0.34 ± 0.11 & 20.51 ± 0.46 & 0.52 ± $4\mathrm{e}{-4}$ \\ 
 &  & WGAN & $\mathbf{2.24\mathrm{e}{-4}\pm1\mathrm{e}{-6}}$ & 57.76 ± 0.33 & \textbf{97.83 ± 0.11} & \textbf{72.63 ± 0.23} & 53.97 ± 0.10 & 0.73 ± $2\mathrm{e}{-2}$ \\
 &  & DM & $9.54\mathrm{e}{-4}\pm4\mathrm{e}{-6}$ & \textbf{100 ± 0.00} & 1.20 ± 0.05 & 2.36 ± 0.09 & 54.65 ± 0.77 & $\mathbf{0.50 \pm 9\mathrm{e}{-5}}$ \\ \cmidrule{2-9}
 & \multirow{4}{*}{CHR 12 (9780 SNPs)} & VAE & $1.40\mathrm{e}{-2}\pm5\mathrm{e}{-5}$ & 99.99 ± 0.01 & 0.04 ± 0.01 & 0.08 ± 0.03 & 26.91 ± 0.21 & $\mathbf{0.50 \pm 6\mathrm{e}{-4}}$ \\ 
 &  & GAN & $9.20\mathrm{e}{-4}\pm6\mathrm{e}{-6}$ & 21.15 ± 0.25 & 0.95 ± 0.09 & 1.82 ± 0.17 & 8.30 ± 0.08 & 0.99 ± $6\mathrm{e}{-3}$ \\ 
 &  & WGAN & $\mathbf{1.13\mathrm{e}{-4}\pm1\mathrm{e}{-6}}$ & 55.28 ± 0.72 & \textbf{75.19 ± 0.53} & \textbf{63.71 ± 0.57} & 40.16 ± 0.19 & 0.55 ± $4\mathrm{e}{-3}$ \\
 &  & DM & $9.23\mathrm{e}{-4}\pm3\mathrm{e}{-6}$ & \textbf{100 ± 0.00} & 1.20 ± 0.03 & 2.38 ± 0.06 & \textbf{40.35 ± 0.30} & $\mathbf{0.50 \pm 1\mathrm{e}{-4}}$ \\ \cmidrule{2-9}
 & \multirow{4}{*}{Multi CHRs (42409 SNPs)} & VAE & $1.77\mathrm{e}{-2}\pm5\mathrm{e}{-5}$ & \textbf{100 ± 0.00} & 0.00 ± 0.00 & 0.00 ± 0.00 & 5.57 ± 0.07 & $\mathbf{0.50 \pm 2\mathrm{e}{-3}}$ \\
 &  & GAN & $1.46\mathrm{e}{-3}\pm5\mathrm{e}{-6}$ & \textbf{100 ± 0.01} & 0.65 ± 0.03 & 1.30 ± 0.05 & 5.10 ± 0.17 & 0.51 ± $7\mathrm{e}{-5}$ \\  &  & WGAN & $\mathbf{1.63\mathrm{e}{-4}\pm1\mathrm{e}{-6}}$ & 45.80 ± 0.42 & \textbf{64.35 ± 1.28} & \textbf{53.50 ± 0.46} & 19.58 ± 0.50 & 0.52 ± $1\mathrm{e}{-2}$ \\ 
 &  & DM & $9.56\mathrm{e}{-4}\pm3\mathrm{e}{-6}$ & \textbf{100 ± 0.00} & 1.00 ± 0.01 & 1.98 ± 0.02 & \textbf{20.04 ± 0.23} & $\mathbf{0.50 \pm 8\mathrm{e}{-5}}$ \\ \cmidrule{2-9} \botrule
\end{tabular}
\begin{tablenotes}
\footnotesize
\item Note: Scientific notation is used, for example, \texttt{a\textnormal{e}-b} denotes \( a \times 10^{-b} \), where \texttt{e} represents base-10 exponentiation.
\end{tablenotes}
\end{sidewaystable}

\subsection{Do Generative Models Preserve Genotype–Phenotype Association?}\label{sec:geno_pheno}
\begin{figure}[htbp]
    \centering
    \begin{minipage}{0.49\textwidth}
        \centering
        \includegraphics[width=\textwidth]{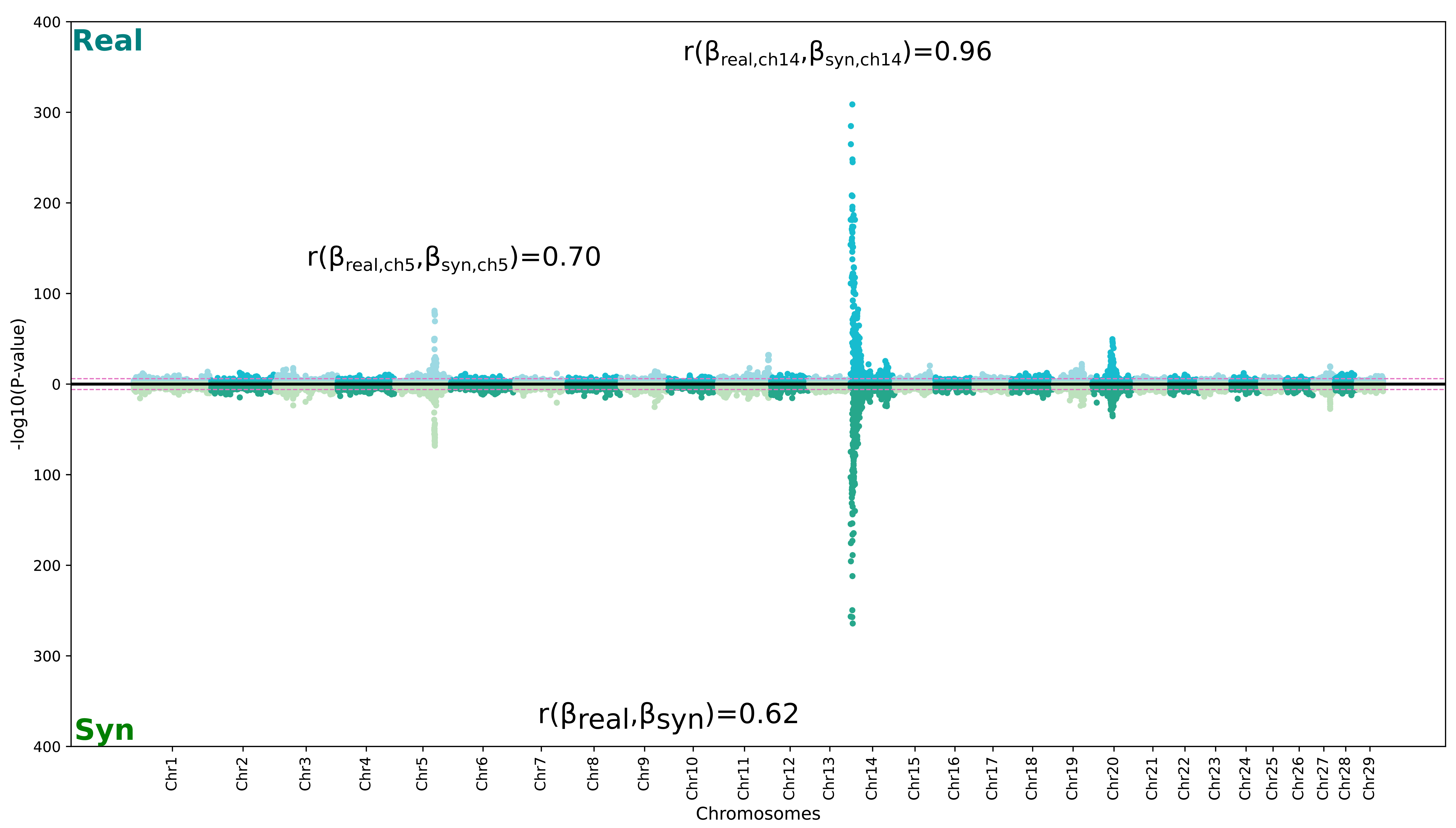}
        \subcaption{WGAN}\label{fig:gwas_wgan}
    \end{minipage}%
    \hfill
    \begin{minipage}{0.49\textwidth}
        \centering
        \includegraphics[width=\textwidth]{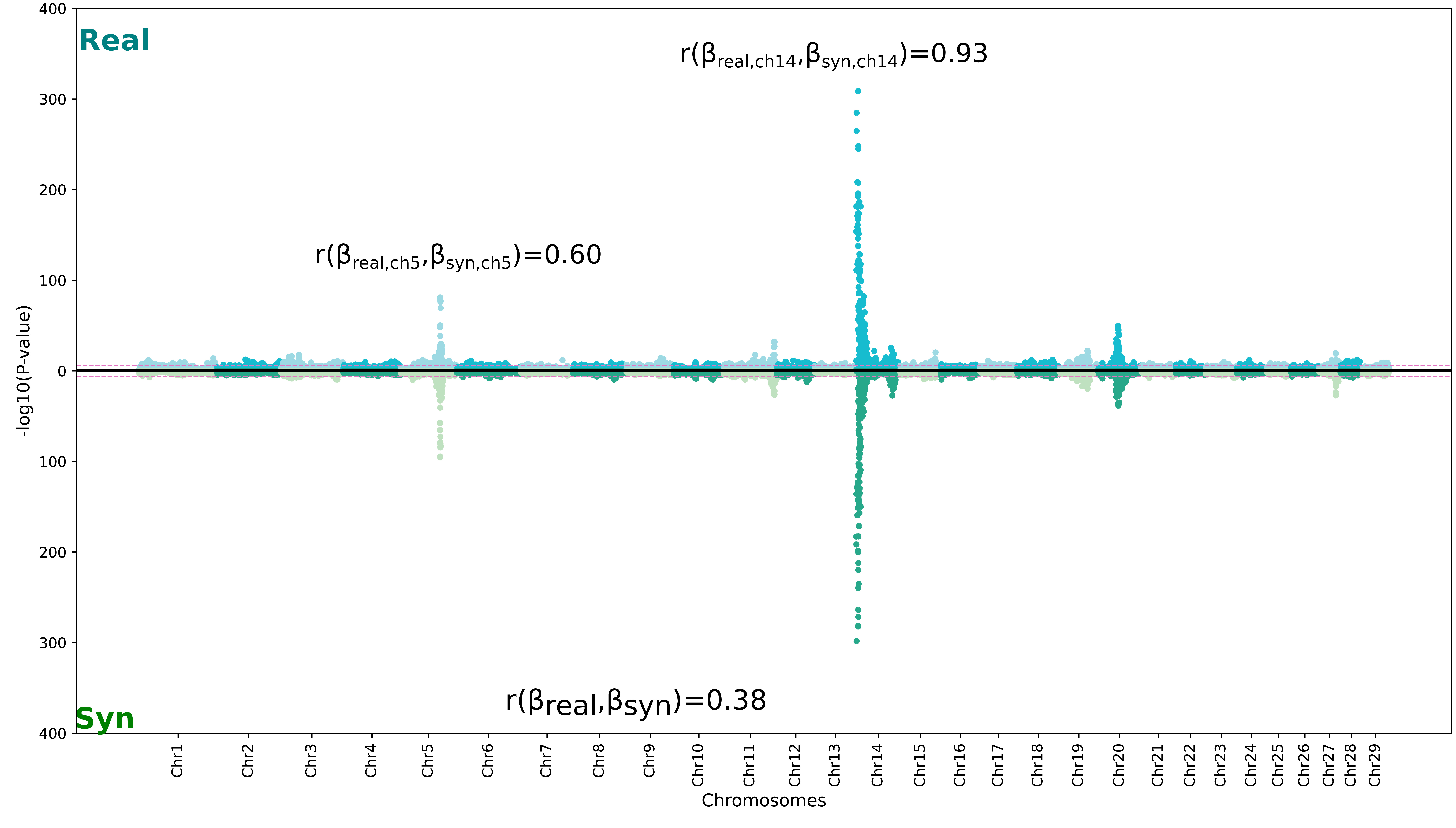}
        \subcaption{DM}\label{fig:gwas_dm}
    \end{minipage}
    \caption{GWAS Comparison of Real and Synthetic Populations Across All Chromosomes of Cow Dataset. (a) Compared with WGAN-generated genotype. (b) Compared with DM-generated genotype.}
    \label{fig:gwas}
\end{figure}

In the previous unconditional setting, WGAN and DM demonstrated superior performance, especially for large-dimensional datasets. We then evaluated their performance under the conditional setting by using phenotype as conditioning variable to investigate whether the models could also capture the genotype-phenotype association. Figure \ref{fig:gwas} presents the results of GWAS analysis on the full set of chromosomes in cow dataset, comparing real and synthetic populations. Both WGAN and DM are able to recover the $3$ main quantitative trait locus (QTL) regions. When examining the regression coefficients $\beta$ in GWAS, we observed that the WGAN-generated synthetic population shows a higher correlation with the real population's $\beta$ values compared to the DM-generated population. 

Table \ref{tab:super} summarizes the predictive performance of machine learning and deep learning models on synthetic genotype to predict the conditioning phenotype, compared to the results from the real population. Similarly to the previous unconditional setting, for relatively small datasets (e.g., single chromosome in cow dataset), both models achieve comparable performance to that obtained with real datasets. However, for more complex datasets, WGAN-generated synthetic genotype appears to better preserve the complex genotype-phenotype relationship, especially when using MLP as the prediction model. This aligns with WGAN's ability to improve the recall metric and more fully capture the distribution of real data. Overall, these results suggest that WGAN is able to generate synthetic population with genotype–phenotype association that closely mirror those observed in real data, as reflected by its consistently strong predictive performance across datasets and predictive model types.

\begin{table}[h]
\centering
\caption{Comparison of phenotype‐prediction performance using real and synthetic genotype data.}
\label{tab:super}
\begin{tabular}{|c|c|c|c|c|c|c|}
\toprule
\multicolumn{3}{|c|}{\multirow{2}{*}{Dataset}} & \multicolumn{2}{c|}{XGBoost} & \multicolumn{2}{c|}{MLP} \\ \cmidrule{4-7} 
\multicolumn{3}{|c|}{} & MSE ↓ & r ↑ & MSE ↓ & r ↑ \\ \hline
\multirow{9}{*}{Cow} & \multirow{3}{*}{CHR 14} & {\ul Real} & 0.60 ± 0.0054 & 0.65 ± 0.0034 & 0.67± 0.0088 & 0.61 ± 0.0066 \\ \cmidrule{3-7} 
 &  & WGAN & 0.62 ± 0.0050 & 0.63 ± 0.0012 & \textbf{0.72 ± 0.0323} & \textbf{0.59 ± 0.0041} \\ \cmidrule{3-7} 
 &  & DM & \textbf{0.61 ± 0.0056} & \textbf{0.64 ± 0.0023} & 0.74 ± 0.0188 & 0.55 ± 0.0033 \\ \cmidrule{2-7} 
 & \multirow{3}{*}{CHR 5} & {\ul Real} & 0.95 ± 0.0004 & 0.23 ± 0.0031 & 1.12 ± 0.0096 & 0.16 ± 0.0082 \\ \cmidrule{3-7} 
 &  & WGAN & \textbf{0.95 ± 0.0014} & \textbf{0.23 ± 0.0021} & \textbf{1.14 ± 0.0168} & \textbf{0.15 ± 0.0151} \\ \cmidrule{3-7} 
 &  & DM & 0.96 ± 0.0019 & 0.21 ± 0.0045 & 1.14 ± 0.0320 & 0.11± 0.0305 \\ \cmidrule{2-7} 
 & \multirow{3}{*}{All CHRs} & {\ul Real} & 0.46 ± 0.0053 & 0.76 ± 0.0023 & 0.40 ± 0.0135 & 0.81 ± 0.0042 \\ \cmidrule{3-7} 
 &  & WGAN & \textbf{0.52 ± 0.0069} & \textbf{0.72 ± 0.0027} & 0.49 ± 0.0261 & 0.75 ± 0.0067 \\ \cmidrule{3-7} 
 &  & DM & 0.53 ± 0.0145 & 0.71 ± 0.0084 & \textbf{0.47 ± 0.0248} & \textbf{0.76 ± 0.0025} \\ \midrule
\multirow{12}{*}{Human} & \multirow{3}{*}{Ensembl} & Real & 40.87± 0.1936 & 0.72 ± 0.0016 & 68.33 ± 6.1033 & 0.64 ± 0.0473 \\ \cmidrule{3-7} 
 &  & WGAN & 43.44 ± 0.1392 & \textbf{0.70 ± 0.0011} & \textbf{69.87 ± 9.3012} & \textbf{0.59 ± 0.0105} \\ \cmidrule{3-7} 
 &  & DM & \textbf{43.28 ± 0.6887} & \textbf{0.70 ± 0.0047} & 114.62 ± 31.7599 & 0.49± 0.0320 \\ \cmidrule{2-7} 
 & \multirow{3}{*}{CHR 6} & Real & 42.81 ± 0.0791 & 0.71 ± 0.0006 & 88.53 ± 4.7244 & 0.44 ± 0.0384  \\ \cmidrule{3-7} 
 &  & WGAN & \textbf{43.30 ± 0.1146} & \textbf{0.70 ± 0.0010} & \textbf{94.62 ± 5.2922}   & \textbf{0.40 ± 0.0221} \\ \cmidrule{3-7} 
 &  & DM & 44.65 ± 0.9555 & 0.69 ± 0.0079 & 218.95 ± 27.5067 & 0.24 ± 0.0099 \\ \cmidrule{2-7} 
 & \multirow{3}{*}{CHR 12} & Real & 42.96 ± 0.0912 & 0.70 ± 0.0007 & 94.33 ± 5.6021  & 0.46 ± 0.0097 \\ \cmidrule{3-7} 
 &  & WGAN & \textbf{43.41 ± 0.0684} & \textbf{0.70 ± 0.0006} & \textbf{96.52 ± 6.2831} & \textbf{0.42 ± 0.0158} \\ \cmidrule{3-7} 
 &  & DM & 45.15 ± 1.4170 & 0.69 ± 0.0123 & 237.82 ± 14.8840 & 0.26 ± 0.0036 \\ \cmidrule{2-7} 
 & \multirow{3}{*}{Multi CHRs} & Real & 42.63 ± 0.1547 & 0.71 ± 0.0013 & 92.91 ± 3.2856  & 0.26 ± 0.0327\\ \cmidrule{3-7} 
 &  & WGAN & \textbf{43.44 ± 0.0800} & \textbf{0.70 ± 0.0006} & \textbf{96.51 ± 1.4907}  & \textbf{0.23 ± 0.0334} \\ \cmidrule{3-7} 
 &  & DM & 44.65 ± 0.9555 & 0.69 ± 0.0079 & 149.87 ± 5.7670  & 0.18 ± 0.0096 \\ \botrule
\end{tabular}
\end{table}

\section{Discussion and Conclusion}\label{sec:dis_conclu}
The primary objective of this study was to investigate the effectiveness of widely used deep generative models for simulating genotype. We proposed specific adaptation for VAE, GAN, WGAN and diffusion models to better handle the discrete nature of genotype representation. Our experiments revealed that no single model performs best across all evaluation metrics and datasets. Each dataset exhibits distinct genetic properties, and we found that model performance is influenced by both the dimensionality of genotype sequence and the degree of SNP dependence. For relatively small and simple datasets (e.g., with a few thousand SNPs), we recommend using VAE due to its computational efficiency, training stability, and minimal hyperparameter tuning. For larger and more complex datasets with higher genetic diversity, WGAN-based model consistently outperforms the other models, particularly in capturing the overall distribution and the genotype-phenotype association.

We also proposed a comprehensive evaluation framework that combines multiple metrics to assess synthetic genotype quality from different angles. Since each metric captures a specific aspect, using them in combination provides a more complete evaluation. We found that not all previously developed metrics are robust. The $AA$ score can produce misleading results in certain edge cases. Other metrics, such as the correlation score, are reliable but computationally intensive. Among all the metrics, recall stands out as a particularly valuable supervision signal during training, as it is more difficult to optimize and indicative of a model's ability to capture diversity. We recommend using PCA, $F_{ST}^{\mathrm{aggregated}}$, precision, and recall during training to decide when to stop and only compute the more costly metrics afterward.

Our results align with previous studies on haplotype generation that suggest generative models can accurately capture the genetic structure. Moreover, to our knowledge, this is the first work to show that conditioning on phenotype allows generative models to produce synthetic population that preserve genotype–phenotype association. The resulting synthetic population can be effectively used in downstream applications, such as GWAS, highlighting the potential of generative models to support genetics research.

Several future research directions can be envisioned. In this study, we focused on models that learn the joint distribution of the entire genotype sequence, favoring a biologically grounded approach over sequential modeling. However, recent advances inspired by natural language processing, such as transformer-based models applied to DNA sequence \cite{DNABERT, DNAGPT, nucle_transformer}, could also be adapted for genotype and merit further investigation. Another potential direction is the development of post-training refinement algorithms to improve the quality of generated sequences \cite{absorb_escape}. On the data side, future work could aim to better model additional features of genotype data, such as rare variants, population heterogeneity, and multi-phenotype conditioning. Incorporating modules that explicitly capture genotype–phenotype interaction could further enhance biological relevance. Lastly, exploring frugal learning strategies would be valuable, given the high dimensionality of genotype data and the computational demands of generative models.

\bmhead{Supplementary Data}
Supplementary data is available online.

\bmhead{Data Availability Statement}
We provide code for model training, evaluation metrics, and experiments (\href{https://github.com/SihanXXX/DiscreteGenoGen}{https://github.com/SihanXXX/DiscreteGenoGen}). We also provide trained generative models for cows, allowing others to use them directly to generate synthetic genotype. This supports the main motivation of our research, which is to enable genotype data sharing in a compact and privacy-preserving way. Access to the UK Biobank requires a separate application, which can be submitted at: \href{https://www.ukbiobank.ac.uk/enable-your-research/apply-for-access}{https://www.ukbiobank.ac.uk/enable-your-research/apply-for-access}.

\bmhead{Ethics Statement}
This study demonstrates that generative models can effectively capture genetic structure and reproduce genotype–phenotype association within the given population dataset. However, all validations were conducted in a numerical and computational framework. No experimental validation has been conducted to confirm the biological relevance of the synthetic population. For instance, if a SNP appears to be associated with a trait in the synthetic population, this merely reflects statistical pattern learned from the original data and should not be interpreted as a new biological discovery. All data handling procedures in this project comply with the General Data Protection Regulation (GDPR).

\bmhead{Acknowledgements}
This work was supported by the INRAE DigitBio Metaprogram. We thank Jocelyn De-Goër-De-Herve for managing the GPU infrastructure and GPT4 for spelling and grammar checks. This study makes use of data from the UKBiobank under application number 96326.

\bmhead{Author Contributions Statement}
SX curated the human dataset, implemented the methods and wrote the paper. TT and DB curated the cow dataset. EB, JC, and BH supervised the project and reviewed the manuscript. All the authors read and approved the manuscript.

\bibliography{sn-bibliography-et-al}


\begin{thebibliography}{61}
\ifx \bisbn   \undefined \def \bisbn  #1{ISBN #1}\fi
\ifx \binits  \undefined \def \binits#1{#1}\fi
\ifx \bauthor  \undefined \def \bauthor#1{#1}\fi
\ifx \batitle  \undefined \def \batitle#1{#1}\fi
\ifx \bjtitle  \undefined \def \bjtitle#1{#1}\fi
\ifx \bvolume  \undefined \def \bvolume#1{\textbf{#1}}\fi
\ifx \byear  \undefined \def \byear#1{#1}\fi
\ifx \bissue  \undefined \def \bissue#1{#1}\fi
\ifx \bfpage  \undefined \def \bfpage#1{#1}\fi
\ifx \blpage  \undefined \def \blpage #1{#1}\fi
\ifx \burl  \undefined \def \burl#1{\textsf{#1}}\fi
\ifx \doiurl  \undefined \def \doiurl#1{\url{https://doi.org/#1}}\fi
\ifx \betal  \undefined \def \betal{\textit{et al.}}\fi
\ifx \binstitute  \undefined \def \binstitute#1{#1}\fi
\ifx \binstitutionaled  \undefined \def \binstitutionaled#1{#1}\fi
\ifx \bctitle  \undefined \def \bctitle#1{#1}\fi
\ifx \beditor  \undefined \def \beditor#1{#1}\fi
\ifx \bpublisher  \undefined \def \bpublisher#1{#1}\fi
\ifx \bbtitle  \undefined \def \bbtitle#1{#1}\fi
\ifx \bedition  \undefined \def \bedition#1{#1}\fi
\ifx \bseriesno  \undefined \def \bseriesno#1{#1}\fi
\ifx \blocation  \undefined \def \blocation#1{#1}\fi
\ifx \bsertitle  \undefined \def \bsertitle#1{#1}\fi
\ifx \bsnm \undefined \def \bsnm#1{#1}\fi
\ifx \bsuffix \undefined \def \bsuffix#1{#1}\fi
\ifx \bparticle \undefined \def \bparticle#1{#1}\fi
\ifx \barticle \undefined \def \barticle#1{#1}\fi
\bibcommenthead
\ifx \bconfdate \undefined \def \bconfdate #1{#1}\fi
\ifx \botherref \undefined \def \botherref #1{#1}\fi
\ifx \url \undefined \def \url#1{\textsf{#1}}\fi
\ifx \bchapter \undefined \def \bchapter#1{#1}\fi
\ifx \bbook \undefined \def \bbook#1{#1}\fi
\ifx \bcomment \undefined \def \bcomment#1{#1}\fi
\ifx \oauthor \undefined \def \oauthor#1{#1}\fi
\ifx \citeauthoryear \undefined \def \citeauthoryear#1{#1}\fi
\ifx \endbibitem  \undefined \def \endbibitem {}\fi
\ifx \bconflocation  \undefined \def \bconflocation#1{#1}\fi
\ifx \arxivurl  \undefined \def \arxivurl#1{\textsf{#1}}\fi
\csname PreBibitemsHook\endcsname

\bibitem[\protect\citeauthoryear{Reuter et~al.}{2015}]{seq_tech}
\begin{barticle}
\bauthor{\bsnm{Reuter}, \binits{J.}},
\bauthor{\bsnm{Spacek}, \binits{D.V.}},
\bauthor{\bsnm{Snyder}, \binits{M.}}:
\batitle{High-throughput sequencing technologies}.
\bjtitle{Molecular Cell}
\bvolume{58}(\bissue{4}),
\bfpage{586}--\blpage{597}
(\byear{2015})
\end{barticle}
\endbibitem

\bibitem[\protect\citeauthoryear{Churko et~al.}{2013}]{seq_tech_cardio}
\begin{barticle}
\bauthor{\bsnm{Churko}, \binits{J.}},
\bauthor{\bsnm{Mantalas}, \binits{G.}},
\bauthor{\bsnm{Snyder}, \binits{M.}}, \betal:
\batitle{Overview of high throughput sequencing technologies to elucidate molecular pathways in cardiovascular diseases}.
\bjtitle{Circulation research}
\bvolume{112},
\bfpage{1613}--\blpage{23}
(\byear{2013})
\end{barticle}
\endbibitem

\bibitem[\protect\citeauthoryear{Gravel}{2012}]{WF}
\begin{barticle}
\bauthor{\bsnm{Gravel}, \binits{S.}}:
\batitle{Population genetics models of local ancestry}.
\bjtitle{Genetics}
\bvolume{191}(\bissue{2}),
\bfpage{607}--\blpage{619}
(\byear{2012})
\end{barticle}
\endbibitem

\bibitem[\protect\citeauthoryear{Kingman}{1982}]{coalescent1}
\begin{barticle}
\bauthor{\bsnm{Kingman}, \binits{J.F.C.}}:
\batitle{The coalescent}.
\bjtitle{Stochastic Processes and their Applications}
\bvolume{13}(\bissue{3}),
\bfpage{235}--\blpage{248}
(\byear{1982})
\end{barticle}
\endbibitem

\bibitem[\protect\citeauthoryear{Hudson}{1990}]{coalescent2}
\begin{barticle}
\bauthor{\bsnm{Hudson}, \binits{R.R.}}:
\batitle{Gene genealogies and the coalescent process}.
\bjtitle{Oxford surveys in evolutionary biology}
\bvolume{7}(\bissue{1}),
\bfpage{44}
(\byear{1990})
\end{barticle}
\endbibitem

\bibitem[\protect\citeauthoryear{Kelleher et~al.}{2016}]{coalescent_sim}
\begin{barticle}
\bauthor{\bsnm{Kelleher}, \binits{J.}},
\bauthor{\bsnm{Etheridge}, \binits{A.M.}},
\bauthor{\bsnm{McVean}, \binits{G.}}:
\batitle{Efficient coalescent simulation and genealogical analysis for large sample sizes}.
\bjtitle{PLOS Computational Biology}
\bvolume{12}(\bissue{5}),
\bfpage{1}--\blpage{22}
(\byear{2016})
\end{barticle}
\endbibitem

\bibitem[\protect\citeauthoryear{Hudson}{2002}]{ms}
\begin{barticle}
\bauthor{\bsnm{Hudson}, \binits{R.R.}}:
\batitle{Generating samples under a {Wright--Fisher} neutral model of genetic variation}.
\bjtitle{Bioinformatics}
\bvolume{18}(\bissue{2}),
\bfpage{337}--\blpage{338}
(\byear{2002})
\end{barticle}
\endbibitem

\bibitem[\protect\citeauthoryear{Teshima and Innan}{2009}]{mbs}
\begin{barticle}
\bauthor{\bsnm{Teshima}, \binits{K.M.}},
\bauthor{\bsnm{Innan}, \binits{H.}}:
\batitle{mbs: modifying hudson's ms software to generate samples of dna sequences with a biallelic site under selection}.
\bjtitle{BMC Bioinformatics}
\bvolume{10}(\bissue{1}),
\bfpage{166}
(\byear{2009})
\end{barticle}
\endbibitem

\bibitem[\protect\citeauthoryear{Baumdicker et~al.}{2021}]{msprime}
\begin{barticle}
\bauthor{\bsnm{Baumdicker}, \binits{F.}},
\bauthor{\bsnm{Bisschop}, \binits{G.}},
\bauthor{\bsnm{Goldstein}, \binits{D.}}, \betal:
\batitle{Efficient ancestry and mutation simulation with msprime 1.0}.
\bjtitle{Genetics}
\bvolume{220}(\bissue{3}),
\bfpage{229}
(\byear{2021})
\end{barticle}
\endbibitem

\bibitem[\protect\citeauthoryear{Haller and Messer}{2019}]{SLiM3}
\begin{barticle}
\bauthor{\bsnm{Haller}, \binits{B.C.}},
\bauthor{\bsnm{Messer}, \binits{P.W.}}:
\batitle{Slim 3: Forward genetic simulations beyond the {Wright--Fisher} model}.
\bjtitle{Molecular Biology and Evolution}
\bvolume{36}(\bissue{3}),
\bfpage{632}--\blpage{637}
(\byear{2019})
\end{barticle}
\endbibitem

\bibitem[\protect\citeauthoryear{Ewing and Hermisson}{2010}]{MSMS}
\begin{barticle}
\bauthor{\bsnm{Ewing}, \binits{G.}},
\bauthor{\bsnm{Hermisson}, \binits{J.}}:
\batitle{Msms: a coalescent simulation program including recombination, demographic structure and selection at a single locus}.
\bjtitle{Bioinformatics}
\bvolume{26}(\bissue{16}),
\bfpage{2064}--\blpage{2065}
(\byear{2010})
\end{barticle}
\endbibitem

\bibitem[\protect\citeauthoryear{Peng and Kimmel}{2005}]{simuPOP}
\begin{barticle}
\bauthor{\bsnm{Peng}, \binits{B.}},
\bauthor{\bsnm{Kimmel}, \binits{M.}}:
\batitle{simupop: a forward-time population genetics simulation environment}.
\bjtitle{Bioinformatics}
\bvolume{21}(\bissue{18}),
\bfpage{3686}--\blpage{3687}
(\byear{2005})
\end{barticle}
\endbibitem

\bibitem[\protect\citeauthoryear{Viñas et~al.}{2021}]{metric_corr}
\begin{barticle}
\bauthor{\bsnm{Viñas}, \binits{R.}},
\bauthor{\bsnm{Andrés-Terré}, \binits{H.}},
\bauthor{\bsnm{Liò}, \binits{P.}}, \betal:
\batitle{Adversarial generation of gene expression data}.
\bjtitle{Bioinformatics}
\bvolume{38}(\bissue{3}),
\bfpage{730}--\blpage{737}
(\byear{2021})
\end{barticle}
\endbibitem

\bibitem[\protect\citeauthoryear{Lacan et~al.}{2023}]{lacan_gan}
\begin{barticle}
\bauthor{\bsnm{Lacan}, \binits{A.}},
\bauthor{\bsnm{Sebag}, \binits{M.}},
\bauthor{\bsnm{Hanczar}, \binits{B.}}:
\batitle{Gan-based data augmentation for transcriptomics: survey and comparative assessment}.
\bjtitle{Bioinformatics}
\bvolume{39}(\bissue{Supplement\_1}),
\bfpage{111}--\blpage{120}
(\byear{2023})
\end{barticle}
\endbibitem

\bibitem[\protect\citeauthoryear{Lacan et~al.}{2024}]{lacan_dm}
\begin{botherref}
\oauthor{\bsnm{Lacan}, \binits{A.}},
\oauthor{\bsnm{Andr{\'e}}, \binits{R.}},
\oauthor{\bsnm{Sebag}, \binits{M.}}, et al.:
In silico generation of gene expression profiles using diffusion models.
bioRxiv
(2024)
\end{botherref}
\endbibitem

\bibitem[\protect\citeauthoryear{Li et~al.}{2023}]{latent_dm_dna}
\begin{botherref}
\oauthor{\bsnm{Li}, \binits{Z.}},
\oauthor{\bsnm{Ni}, \binits{Y.}},
\oauthor{\bsnm{Huygelen}, \binits{T.A.B.}}, et al.:
Latent Diffusion Model for DNA Sequence Generation
(2023)
\end{botherref}
\endbibitem

\bibitem[\protect\citeauthoryear{Brixi et~al.}{2025}]{evo2}
\begin{botherref}
\oauthor{\bsnm{Brixi}, \binits{G.}},
\oauthor{\bsnm{Durrant}, \binits{M.G.}},
\oauthor{\bsnm{Ku}, \binits{J.}}, et al.:
Genome modeling and design across all domains of life with evo 2.
bioRxiv
(2025)
\end{botherref}
\endbibitem

\bibitem[\protect\citeauthoryear{Perera et~al.}{2022}]{GMMN_hap}
\begin{botherref}
\oauthor{\bsnm{Perera}, \binits{M.}},
\oauthor{\bsnm{Montserrat}, \binits{D.M.}},
\oauthor{\bsnm{Barrab{\'e}s}, \binits{M.}}, et al.:
Generative moment matching networks for genotype simulation.
bioRxiv
(2022)
\end{botherref}
\endbibitem

\bibitem[\protect\citeauthoryear{Geleta et~al.}{2023}]{VAE_hap}
\begin{botherref}
\oauthor{\bsnm{Geleta}, \binits{M.}},
\oauthor{\bsnm{Montserrat}, \binits{D.M.}},
\oauthor{\bsnm{Giro-i-Nieto}, \binits{X.}}, et al.:
Deep variational autoencoders for population genetics.
bioRxiv
(2023)
\end{botherref}
\endbibitem

\bibitem[\protect\citeauthoryear{Montserrat et~al.}{2019}]{VAE_GAN_hap}
\begin{botherref}
\oauthor{\bsnm{Montserrat}, \binits{D.M.}},
\oauthor{\bsnm{Bustamante}, \binits{C.}},
\oauthor{\bsnm{Ioannidis}, \binits{A.}}:
Class-conditional vae-gan for local-ancestry simulation.
arXiv preprint arXiv:1911.13220
(2019)
\end{botherref}
\endbibitem

\bibitem[\protect\citeauthoryear{Nußberger et~al.}{2020}]{deep_hap}
\begin{barticle}
\bauthor{\bsnm{Nußberger}, \binits{J.}},
\bauthor{\bsnm{Boesel}, \binits{F.}},
\bauthor{\bsnm{Lenz}, \binits{S.}}, \betal:
\batitle{Synthetic observations from deep generative models and binary omics data with limited sample size}.
\bjtitle{Briefings in Bioinformatics}
\bvolume{22}(\bissue{4}),
\bfpage{226}
(\byear{2020})
\end{barticle}
\endbibitem

\bibitem[\protect\citeauthoryear{Yelmen et~al.}{2021}]{yelmen2021}
\begin{barticle}
\bauthor{\bsnm{Yelmen}, \binits{B.}},
\bauthor{\bsnm{Decelle}, \binits{A.}},
\bauthor{\bsnm{Ongaro}, \binits{L.}}, \betal:
\batitle{Creating artificial human genomes using generative neural networks}.
\bjtitle{PLoS Genetics}
\bvolume{17}(\bissue{2}),
\bfpage{1009303}
(\byear{2021})
\end{barticle}
\endbibitem

\bibitem[\protect\citeauthoryear{Yelmen et~al.}{2023}]{yelmen2023}
\begin{barticle}
\bauthor{\bsnm{Yelmen}, \binits{B.}},
\bauthor{\bsnm{Decelle}, \binits{A.}},
\bauthor{\bsnm{Boulos}, \binits{L.L.}}, \betal:
\batitle{Deep convolutional and conditional neural networks for large-scale genomic data generation}.
\bjtitle{PLOS Computational Biology}
\bvolume{19}(\bissue{10}),
\bfpage{1011584}
(\byear{2023})
\end{barticle}
\endbibitem

\bibitem[\protect\citeauthoryear{Szatkownik et~al.}{2024a}]{szatkownik2024latent}
\begin{botherref}
\oauthor{\bsnm{Szatkownik}, \binits{A.}},
\oauthor{\bsnm{Furtlehner}, \binits{C.}},
\oauthor{\bsnm{Charpiat}, \binits{G.}}, et al.:
Latent generative modeling of long genetic sequences with gans.
bioRxiv,
2024--08
(2024)
\end{botherref}
\endbibitem

\bibitem[\protect\citeauthoryear{Szatkownik et~al.}{2024b}]{szatkownik2024dm}
\begin{botherref}
\oauthor{\bsnm{Szatkownik}, \binits{A.}},
\oauthor{\bsnm{Planche}, \binits{L.}},
\oauthor{\bsnm{Demeulle}, \binits{M.}}, et al.:
Diffusion-based artificial genomes and their usefulness for local ancestry inference.
bioRxiv,
2024--10
(2024)
\end{botherref}
\endbibitem

\bibitem[\protect\citeauthoryear{Kingma and Welling}{2019}]{VAE}
\begin{barticle}
\bauthor{\bsnm{Kingma}, \binits{D.P.}},
\bauthor{\bsnm{Welling}, \binits{M.}}:
\batitle{An introduction to variational autoencoders}.
\bjtitle{Foundations and Trends in Machine Learning}
\bvolume{12}(\bissue{4}),
\bfpage{307}--\blpage{392}
(\byear{2019})
\end{barticle}
\endbibitem

\bibitem[\protect\citeauthoryear{Mirza and Osindero}{2014}]{cGAN}
\begin{botherref}
\oauthor{\bsnm{Mirza}, \binits{M.}},
\oauthor{\bsnm{Osindero}, \binits{S.}}:
Conditional generative adversarial nets.
CoRR
(2014)
\end{botherref}
\endbibitem

\bibitem[\protect\citeauthoryear{Ho et~al.}{2020}]{DDPM}
\begin{bchapter}
\bauthor{\bsnm{Ho}, \binits{J.}},
\bauthor{\bsnm{Jain}, \binits{A.}},
\bauthor{\bsnm{Abbeel}, \binits{P.}}:
\bctitle{Denoising diffusion probabilistic models}.
In: \bbtitle{Advances in Neural Information Processing Systems 33: Annual Conference on Neural Information Processing Systems 2020, NeurIPS 2020, December 6-12, 2020, Virtual}
(\byear{2020})
\end{bchapter}
\endbibitem

\bibitem[\protect\citeauthoryear{Kynk\"{a}\"{a}nniemi et~al.}{2019}]{precision_recall}
\begin{bchapter}
\bauthor{\bsnm{Kynk\"{a}\"{a}nniemi}, \binits{T.}},
\bauthor{\bsnm{Karras}, \binits{T.}},
\bauthor{\bsnm{Laine}, \binits{S.}}, \betal:
\bctitle{Improved precision and recall metric for assessing generative models}.
In: \bbtitle{Advances in Neural Information Processing Systems},
vol. \bseriesno{32}
(\byear{2019})
\end{bchapter}
\endbibitem

\bibitem[\protect\citeauthoryear{Goodfellow et~al.}{2014}]{GAN}
\begin{bchapter}
\bauthor{\bsnm{Goodfellow}, \binits{I.J.}},
\bauthor{\bsnm{Pouget-Abadie}, \binits{J.}},
\bauthor{\bsnm{Mirza}, \binits{M.}}, \betal:
\bctitle{Generative adversarial nets}.
In: \bbtitle{Advances in Neural Information Processing Systems},
vol. \bseriesno{27}
(\byear{2014})
\end{bchapter}
\endbibitem

\bibitem[\protect\citeauthoryear{Wold et~al.}{1987}]{PCA}
\begin{barticle}
\bauthor{\bsnm{Wold}, \binits{S.}},
\bauthor{\bsnm{Esbensen}, \binits{K.}},
\bauthor{\bsnm{Geladi}, \binits{P.}}:
\batitle{Principal component analysis}.
\bjtitle{Chemometrics and Intelligent Laboratory Systems}
\bvolume{2}(\bissue{1}),
\bfpage{37}--\blpage{52}
(\byear{1987}).
\bcomment{Proceedings of the Multivariate Statistical Workshop for Geologists and Geochemists}
\end{barticle}
\endbibitem

\bibitem[\protect\citeauthoryear{Jang et~al.}{2017}]{GumbelSoftmax}
\begin{bchapter}
\bauthor{\bsnm{Jang}, \binits{E.}},
\bauthor{\bsnm{Gu}, \binits{S.}},
\bauthor{\bsnm{Poole}, \binits{B.}}:
\bctitle{Categorical reparameterization with gumbel-softmax}.
In: \bbtitle{5th International Conference on Learning Representations, {ICLR} 2017, Toulon, France, April 24-26, 2017, Conference Track Proceedings}
(\byear{2017})
\end{bchapter}
\endbibitem

\bibitem[\protect\citeauthoryear{Kusner and Hern{\'a}ndez-Lobato}{2016}]{GumbelSoftmaxGAN}
\begin{botherref}
\oauthor{\bsnm{Kusner}, \binits{M.J.}},
\oauthor{\bsnm{Hern{\'a}ndez-Lobato}, \binits{J.M.}}:
Gans for sequences of discrete elements with the gumbel-softmax distribution.
arXiv preprint arXiv:1611.04051
(2016)
\end{botherref}
\endbibitem

\bibitem[\protect\citeauthoryear{Bau et~al.}{2019}]{mode_collapse}
\begin{bchapter}
\bauthor{\bsnm{Bau}, \binits{D.}},
\bauthor{\bsnm{Zhu}, \binits{J.}},
\bauthor{\bsnm{Wulff}, \binits{J.}}, \betal:
\bctitle{Seeing what a {GAN} cannot generate}.
In: \bbtitle{2019 {IEEE/CVF} International Conference on Computer Vision, {ICCV} 2019, Seoul, Korea (South), October 27 - November 2, 2019},
pp. \bfpage{4501}--\blpage{4510}
(\byear{2019})
\end{bchapter}
\endbibitem

\bibitem[\protect\citeauthoryear{Arjovsky et~al.}{2017}]{WGAN}
\begin{bchapter}
\bauthor{\bsnm{Arjovsky}, \binits{M.}},
\bauthor{\bsnm{Chintala}, \binits{S.}},
\bauthor{\bsnm{Bottou}, \binits{L.}}:
\bctitle{{W}asserstein generative adversarial networks}.
In: \bbtitle{Proceedings of the 34th International Conference on Machine Learning}.
\bsertitle{Proceedings of Machine Learning Research},
vol. \bseriesno{70},
pp. \bfpage{214}--\blpage{223}
(\byear{2017})
\end{bchapter}
\endbibitem

\bibitem[\protect\citeauthoryear{Gulrajani et~al.}{2017}]{WGAN-GP}
\begin{bchapter}
\bauthor{\bsnm{Gulrajani}, \binits{I.}},
\bauthor{\bsnm{Ahmed}, \binits{F.}},
\bauthor{\bsnm{Arjovsky}, \binits{M.}}, \betal:
\bctitle{Improved training of wasserstein gans}.
In: \bbtitle{Advances in Neural Information Processing Systems},
vol. \bseriesno{30}
(\byear{2017})
\end{bchapter}
\endbibitem

\bibitem[\protect\citeauthoryear{McInnes et~al.}{2018}]{UMAP}
\begin{barticle}
\bauthor{\bsnm{McInnes}, \binits{L.}},
\bauthor{\bsnm{Healy}, \binits{J.}},
\bauthor{\bsnm{Saul}, \binits{N.}}, \betal:
\batitle{Umap: Uniform manifold approximation and projection}.
\bjtitle{The Journal of Open Source Software}
\bvolume{3}(\bissue{29}),
\bfpage{861}
(\byear{2018})
\end{barticle}
\endbibitem

\bibitem[\protect\citeauthoryear{Laland et~al.}{2015}]{al_geno_freq}
\begin{barticle}
\bauthor{\bsnm{Laland}, \binits{K.N.}},
\bauthor{\bsnm{Uller}, \binits{T.}},
\bauthor{\bsnm{Feldman}, \binits{M.W.}}, \betal:
\batitle{The extended evolutionary synthesis: its structure, assumptions and predictions}.
\bjtitle{Proceedings of the Royal Society B: Biological Sciences}
\bvolume{282}(\bissue{1813}),
\bfpage{20151019}
(\byear{2015})
\end{barticle}
\endbibitem

\bibitem[\protect\citeauthoryear{Wright}{1949}]{fst1}
\begin{barticle}
\bauthor{\bsnm{Wright}, \binits{S.}}:
\batitle{The genetical structure of populations}.
\bjtitle{Annals of Eugenics}
\bvolume{15}(\bissue{1}),
\bfpage{323}--\blpage{354}
(\byear{1949})
\end{barticle}
\endbibitem

\bibitem[\protect\citeauthoryear{Weir and Cockerham}{1984}]{fst2}
\begin{barticle}
\bauthor{\bsnm{Weir}, \binits{B.S.}},
\bauthor{\bsnm{Cockerham}, \binits{C.C.}}:
\batitle{Estimating f-statistics for the analysis of population structure}.
\bjtitle{Evolution}
\bvolume{38}(\bissue{6}),
\bfpage{1358}--\blpage{1370}
(\byear{1984})
\end{barticle}
\endbibitem

\bibitem[\protect\citeauthoryear{Slatkin}{2008}]{ld}
\begin{barticle}
\bauthor{\bsnm{Slatkin}, \binits{M.}}:
\batitle{Linkage disequilibrium: Understanding the evolutionary past and mapping the medical future}.
\bjtitle{Nature Reviews Genetics}
\bvolume{9}(\bissue{6}),
\bfpage{477}--\blpage{485}
(\byear{2008})
\end{barticle}
\endbibitem

\bibitem[\protect\citeauthoryear{Rogers and Huff}{2009}]{ld_geno}
\begin{barticle}
\bauthor{\bsnm{Rogers}, \binits{A.R.}},
\bauthor{\bsnm{Huff}, \binits{C.}}:
\batitle{Linkage disequilibrium between loci with unknown phase}.
\bjtitle{Genetics}
\bvolume{182}(\bissue{3}),
\bfpage{839}--\blpage{844}
(\byear{2009})
\end{barticle}
\endbibitem

\bibitem[\protect\citeauthoryear{Simonyan and Zisserman}{2015}]{vgg16}
\begin{bchapter}
\bauthor{\bsnm{Simonyan}, \binits{K.}},
\bauthor{\bsnm{Zisserman}, \binits{A.}}:
\bctitle{Very deep convolutional networks for large-scale image recognition}.
In: \bbtitle{3rd International Conference on Learning Representations, {ICLR} 2015, San Diego, CA, USA, May 7-9, 2015, Conference Track Proceedings}
(\byear{2015})
\end{bchapter}
\endbibitem

\bibitem[\protect\citeauthoryear{Uffelmann et~al.}{2021}]{GWAS}
\begin{barticle}
\bauthor{\bsnm{Uffelmann}, \binits{E.}},
\bauthor{\bsnm{Huang}, \binits{Q.Q.}},
\bauthor{\bsnm{Munung}, \binits{N.S.}}, \betal:
\batitle{Genome-wide association studies}.
\bjtitle{Nature Reviews Methods Primers}
\bvolume{1}(\bissue{1}),
\bfpage{59}
(\byear{2021})
\end{barticle}
\endbibitem

\bibitem[\protect\citeauthoryear{Yale et~al.}{2020}]{AA}
\begin{barticle}
\bauthor{\bsnm{Yale}, \binits{A.}},
\bauthor{\bsnm{Dash}, \binits{S.}},
\bauthor{\bsnm{Dutta}, \binits{R.}}, \betal:
\batitle{{Generation and Evaluation of Privacy Preserving Synthetic Health Data}}.
\bjtitle{{Neurocomputing}}
\bvolume{416},
\bfpage{244}--\blpage{255}
(\byear{2020})
\end{barticle}
\endbibitem

\bibitem[\protect\citeauthoryear{Sudlow et~al.}{2015}]{UKBioBank}
\begin{barticle}
\bauthor{\bsnm{Sudlow}, \binits{C.}},
\bauthor{\bsnm{Gallacher}, \binits{J.}},
\bauthor{\bsnm{Allen}, \binits{N.}}, \betal:
\batitle{Uk biobank: an open access resource for identifying the causes of a wide range of complex diseases of middle and old age}.
\bjtitle{PLoS Medicine}
\bvolume{12}(\bissue{3}),
\bfpage{1001779}
(\byear{2015})
\end{barticle}
\endbibitem

\bibitem[\protect\citeauthoryear{Tribout et~al.}{2020}]{hssgblup}
\begin{bchapter}
\bauthor{\bsnm{Tribout}, \binits{T.}},
\bauthor{\bsnm{Ducrocq}, \binits{V.}},
\bauthor{\bsnm{Boichard}, \binits{D.}}:
\bctitle{Hssgblup: a {Single-Step SNP BLUP} genomic evaluation software adapted to large livestock populations}.
In: \bbtitle{Proceedings of the 6th International Conference of Quantitative Genetics},
pp. \bfpage{2}--\blpage{12}
(\byear{2020})
\end{bchapter}
\endbibitem

\bibitem[\protect\citeauthoryear{Fernando et~al.}{2016}]{single_step}
\begin{barticle}
\bauthor{\bsnm{Fernando}, \binits{R.L.}},
\bauthor{\bsnm{Cheng}, \binits{H.}},
\bauthor{\bsnm{Golden}, \binits{B.L.}}, \betal:
\batitle{Computational strategies for alternative single-step bayesian regression models with large numbers of genotyped and non-genotyped animals}.
\bjtitle{Genetics Selection Evolution}
\bvolume{48}(\bissue{1}),
\bfpage{96}
(\byear{2016})
\end{barticle}
\endbibitem

\bibitem[\protect\citeauthoryear{Littlejohn et~al.}{2016}]{MGST1}
\begin{botherref}
\oauthor{\bsnm{Littlejohn}, \binits{M.}},
\oauthor{\bsnm{Tiplady}, \binits{K.}},
\oauthor{\bsnm{Fink}, \binits{T.}}, et al.:
Sequence-based association analysis reveals an mgst1 eqtl with pleiotropic effects on bovine milk composition.
Scientific Reports
\textbf{6}
(2016)
\end{botherref}
\endbibitem

\bibitem[\protect\citeauthoryear{Winter et~al.}{2002}]{DGAT1}
\begin{barticle}
\bauthor{\bsnm{Winter}, \binits{A.}},
\bauthor{\bsnm{Krämer}, \binits{W.}},
\bauthor{\bsnm{Werner}, \binits{F.}}, \betal:
\batitle{Association of a lysine-232/alanine polymorphism in a bovine gene encoding acyl-coa:diacylglycerol acyltransferase (dgat1) with variation at a quantitative trait locus for milk fat content}.
\bjtitle{Proceedings of the National Academy of Sciences of the United States of America}
\bvolume{99},
\bfpage{9300}--\blpage{5}
(\byear{2002})
\end{barticle}
\endbibitem

\bibitem[\protect\citeauthoryear{Mullaney et~al.}{2010}]{INDEL}
\begin{barticle}
\bauthor{\bsnm{Mullaney}, \binits{J.M.}},
\bauthor{\bsnm{Mills}, \binits{R.E.}},
\bauthor{\bsnm{Pittard}, \binits{W.S.}}, \betal:
\batitle{Small insertions and deletions (indels) in human genomes}.
\bjtitle{Human Molecular Genetics}
\bvolume{19}(\bissue{R2}),
\bfpage{131}--\blpage{136}
(\byear{2010})
\end{barticle}
\endbibitem

\bibitem[\protect\citeauthoryear{{Lango Allen} et~al.}{2010}]{human_height_1}
\begin{barticle}
\bauthor{\bsnm{{Lango Allen}}, \binits{H.}},
\bauthor{\bsnm{Estrada}, \binits{K.}},
\bauthor{\bsnm{Lettre}, \binits{G.}}, \betal:
\batitle{Hundreds of variants clustered in genomic loci and biological pathways affect human height}.
\bjtitle{Nature}
\bvolume{467}(\bissue{7317}),
\bfpage{832}--\blpage{838}
(\byear{2010})
\end{barticle}
\endbibitem

\bibitem[\protect\citeauthoryear{Anderson et~al.}{2010}]{Anderson2010}
\begin{barticle}
\bauthor{\bsnm{Anderson}, \binits{C.A.}},
\bauthor{\bsnm{Pettersson}, \binits{F.H.}},
\bauthor{\bsnm{Clarke}, \binits{G.M.}}, \betal:
\batitle{Data quality control in genetic case-control association studies}.
\bjtitle{Nature Protocols}
\bvolume{5}(\bissue{9}),
\bfpage{1564}--\blpage{1573}
(\byear{2010})
\end{barticle}
\endbibitem

\bibitem[\protect\citeauthoryear{Tahimic et~al.}{2013}]{IGF-1}
\begin{barticle}
\bauthor{\bsnm{Tahimic}, \binits{C.}},
\bauthor{\bsnm{Wang}, \binits{Y.}},
\bauthor{\bsnm{Bikle}, \binits{D.}}:
\batitle{Anabolic effects of igf-1 signaling on the skeleton}.
\bjtitle{Frontiers in Endocrinology}
\bvolume{4},
\bfpage{6}
(\byear{2013})
\end{barticle}
\endbibitem

\bibitem[\protect\citeauthoryear{He et~al.}{2016}]{resnet}
\begin{bchapter}
\bauthor{\bsnm{He}, \binits{K.}},
\bauthor{\bsnm{Zhang}, \binits{X.}},
\bauthor{\bsnm{Ren}, \binits{S.}}, \betal:
\bctitle{{Deep Residual Learning for Image Recognition}}.
In: \bbtitle{Proceedings of 2016 IEEE Conference on Computer Vision and Pattern Recognition}.
\bsertitle{CVPR '16},
pp. \bfpage{770}--\blpage{778}
(\byear{2016})
\end{bchapter}
\endbibitem

\bibitem[\protect\citeauthoryear{Gao et~al.}{2023}]{matthew_effect_1}
\begin{bchapter}
\bauthor{\bsnm{Gao}, \binits{C.}},
\bauthor{\bsnm{Huang}, \binits{K.}},
\bauthor{\bsnm{Chen}, \binits{J.}}, \betal:
\bctitle{Alleviating matthew effect of offline reinforcement learning in interactive recommendation}.
In: \bbtitle{Proceedings of the 46th International ACM SIGIR Conference on Research and Development in Information Retrieval},
pp. \bfpage{238}--\blpage{248}
(\byear{2023})
\end{bchapter}
\endbibitem

\bibitem[\protect\citeauthoryear{Ganev et~al.}{2022}]{matthew_effect_2}
\begin{botherref}
\oauthor{\bsnm{Ganev}, \binits{G.}},
\oauthor{\bsnm{Oprisanu}, \binits{B.}},
\oauthor{\bsnm{Cristofaro}, \binits{E.D.}}:
Robin Hood and Matthew Effects: Differential Privacy Has Disparate Impact on Synthetic Data
(2022)
\end{botherref}
\endbibitem

\bibitem[\protect\citeauthoryear{Ji et~al.}{2021}]{DNABERT}
\begin{barticle}
\bauthor{\bsnm{Ji}, \binits{Y.}},
\bauthor{\bsnm{Zhou}, \binits{Z.}},
\bauthor{\bsnm{Liu}, \binits{H.}}, \betal:
\batitle{Dnabert: pre-trained bidirectional encoder representations from transformers model for dna-language in genome}.
\bjtitle{Bioinformatics}
\bvolume{37}(\bissue{15}),
\bfpage{2112}--\blpage{2120}
(\byear{2021})
\end{barticle}
\endbibitem

\bibitem[\protect\citeauthoryear{Zhang et~al.}{2024}]{DNAGPT}
\begin{botherref}
\oauthor{\bsnm{Zhang}, \binits{D.}},
\oauthor{\bsnm{Zhang}, \binits{W.}},
\oauthor{\bsnm{Zhao}, \binits{Y.}}, et al.:
Dnagpt: A generalized pre-trained tool for multiple dna sequence analysis tasks.
bioRxiv
(2024)
\end{botherref}
\endbibitem

\bibitem[\protect\citeauthoryear{Dalla-Torre et~al.}{2025}]{nucle_transformer}
\begin{barticle}
\bauthor{\bsnm{Dalla-Torre}, \binits{H.}},
\bauthor{\bsnm{Gonzalez}, \binits{L.}},
\bauthor{\bsnm{Mendoza-Revilla}, \binits{J.}}, \betal:
\batitle{Nucleotide transformer: building and evaluating robust foundation models for human genomics}.
\bjtitle{Nature Methods}
\bvolume{22},
\bfpage{287}--\blpage{297}
(\byear{2025})
\end{barticle}
\endbibitem

\bibitem[\protect\citeauthoryear{Li et~al.}{2024}]{absorb_escape}
\begin{bchapter}
\bauthor{\bsnm{Li}, \binits{Z.}},
\bauthor{\bsnm{Ni}, \binits{Y.}},
\bauthor{\bsnm{Xia}, \binits{G.}}, \betal:
\bctitle{Absorb \& escape: Overcoming single model limitations in generating heterogeneous genomic sequences}.
In: \bbtitle{Advances in Neural Information Processing Systems},
vol. \bseriesno{37},
pp. \bfpage{21949}--\blpage{21978}
(\byear{2024})
\end{bchapter}
\endbibitem

\end{thebibliography}

\end{document}


\maketitle 

\section{Principal Component Analysis (PCA) Results on Genotype Datasets} \label{sec:pca_analysis}
\begin{table}[htbp]
\centering
\begin{threeparttable}
\begin{tabular}{|c|c|c|c|}
\hline
\textbf{Dataset} & \textbf{Number of PCs} & \textbf{MSE} & \textbf{SNP Matching Accuracy} \\
\hline
Cow CHR 14 (1,771 SNPs)& 155 & 0.0290 & 97.1089\% \\
Cow CHR 5 (2,238 SNPs)& 206 & 0.0281 & 97.1995\% \\
Cow All CHRs (50,161 SNPs)\tnote{*} & 4,819 & 0.0277 & 97.2411\% \\
Human Ensembl (3,493 SNPs) & 2,026 & 0.0339 & 96.7983\% \\
Human CHR 6 (12,283 SNPs)& 5,149 & 0.0143 & 98.5745\% \\
Human CHR 12 (9,780 SNPs)& 4,438 & 0.0143 & 98.5726\% \\
Human Multi CHRs (42,409 SNPs)\tnote{*}& 18,769 & 0.0144 & 98.5688\% \\
\hline
\end{tabular}
\caption{Principal Component Analysis (PCA) performed on each dataset with 90\% of the variance retained from the original data. The continuous values are then reconstructed to discrete values (0, 1, or 2) by assigning the closest discrete value, for example, $(-\infty, 0.5) \to 0$, $[0.5, 1.5] \to 1$, and $(1.5, +\infty) \to 2$.}
\begin{tablenotes}
\footnotesize
\item[*] We apply PCA to each chromosome individually, and then concatenate the results.
\end{tablenotes}
\end{threeparttable}
\end{table}

\section{Failed thresholding strategy when attempting to map generated continuous values to 0/1/2}
\label{sec:failed_thresholding}
\begin{figure}[H]
    \centering
    \begin{minipage}{0.32\textwidth}
        \centering
        \includegraphics[width=\textwidth]{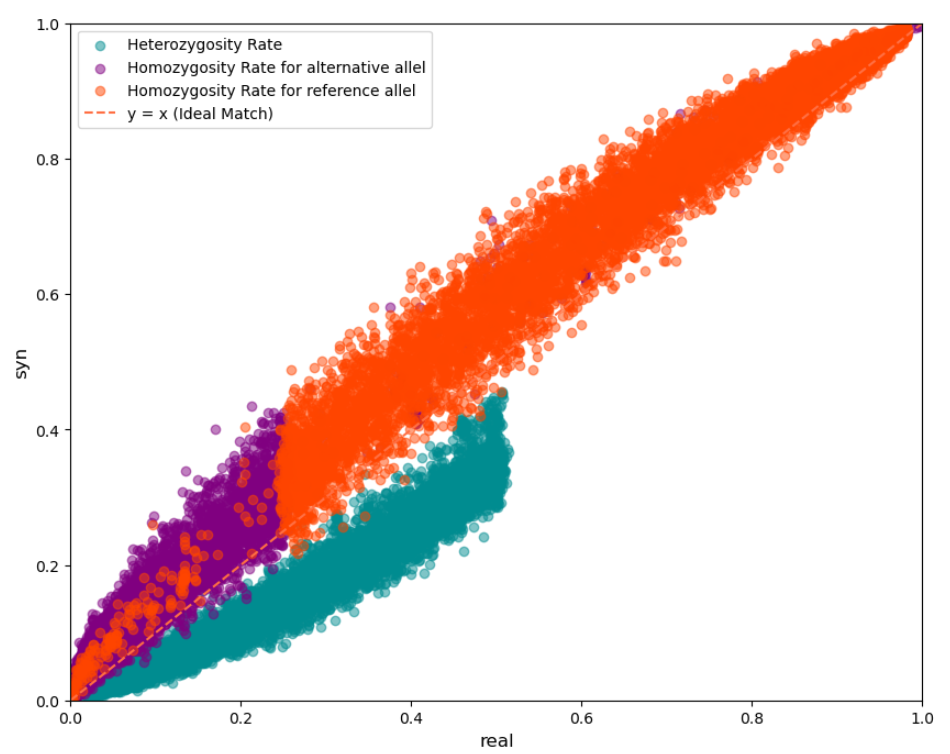}
        \subcaption{}\label{fig:thre1}
    \end{minipage}%
    \hfill
    \begin{minipage}{0.32\textwidth}
        \centering
        \includegraphics[width=\textwidth]{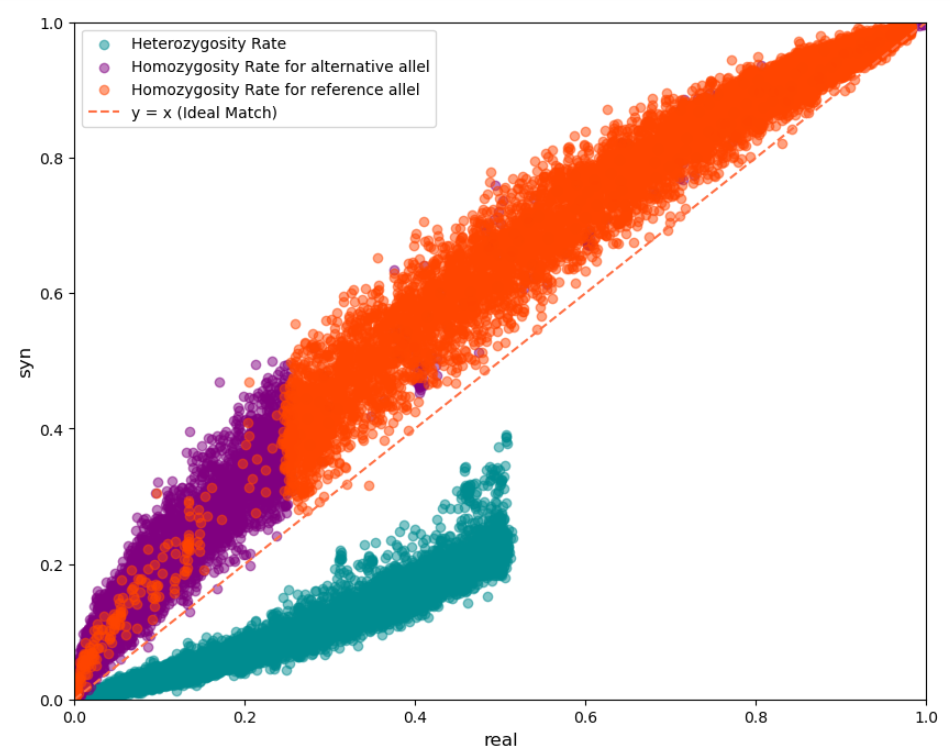}
        \subcaption{}\label{fig:thre2}
    \end{minipage}%
    \hfill
    \begin{minipage}{0.32\textwidth}
        \centering
        \includegraphics[width=\textwidth]{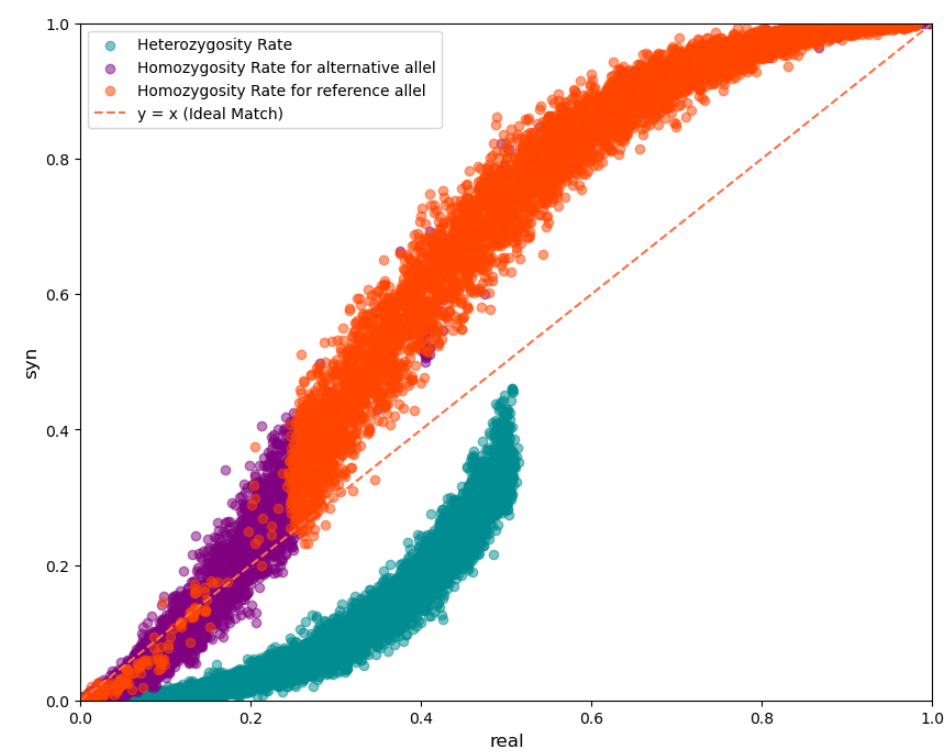}
        \subcaption{}\label{fig:thre3}
    \end{minipage}
    \caption{Comparison of genotype frequencies under different thresholding strategies used to map continuous values to discrete set {0, 1, 2}. (a) Thresholds at [0, $\frac{1}{2}$, $\frac{3}{2}$, 2], (b) Thresholds at [0, $\frac{2}{3}$, $\frac{4}{3}$, 2], and (c) Dynamic thresholding for each SNP based on Hardy–Weinberg equilibrium and allele frequency.}
\end{figure}

Additionally, it is possible to construct a synthetic population with exactly the same genotype frequencies as the original population. This can be achieved by first ranking all the continuous outputs and then mapping them to genotype categories (0/1/2) according to the frequency proportion observed in the original data. While this approach ensures perfect alignment in genotype frequency, it tends to degrade performance on other metrics. In particular, the recall metric typically drops to $0$, indicating poor diversity in the generated samples.

\begin{figure}[H]
  \centering
  \includegraphics[width=0.7\textwidth]{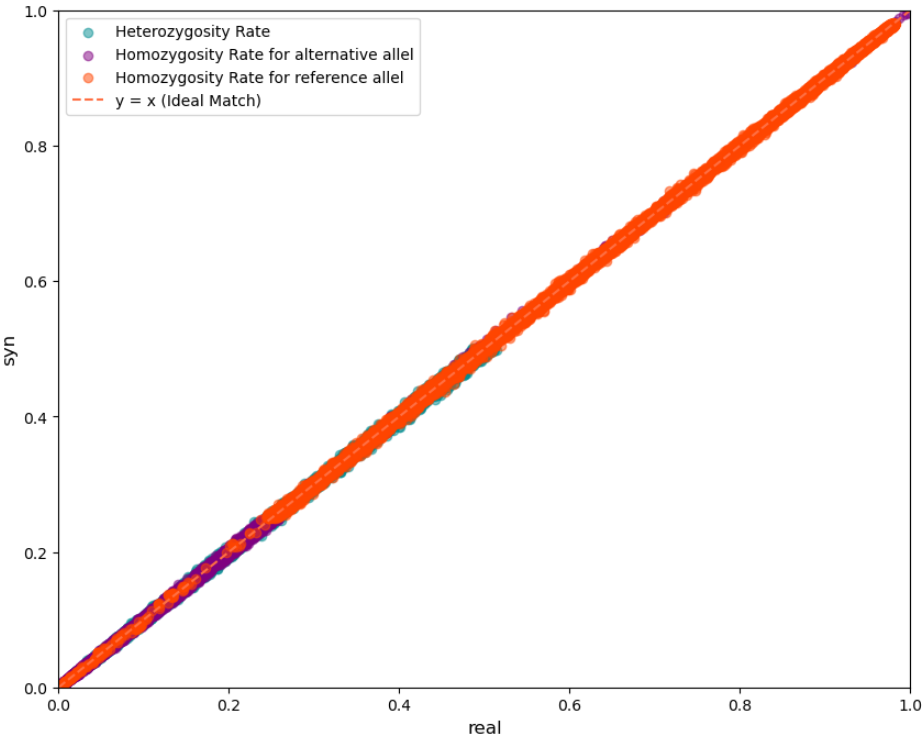}
  \caption{Exact genotype frequency matching between real and synthetic populations using a thresholding strategy based on genotype frequency.}
  \label{fig:thre4}
\end{figure}

\section{Precision and Recall with Different Distance Metrics} \label{sec:precision_recall_dist}
\begin{figure}[H]
    \centering
    \begin{minipage}{0.46\textwidth}
        \centering
        \includegraphics[width=\textwidth]{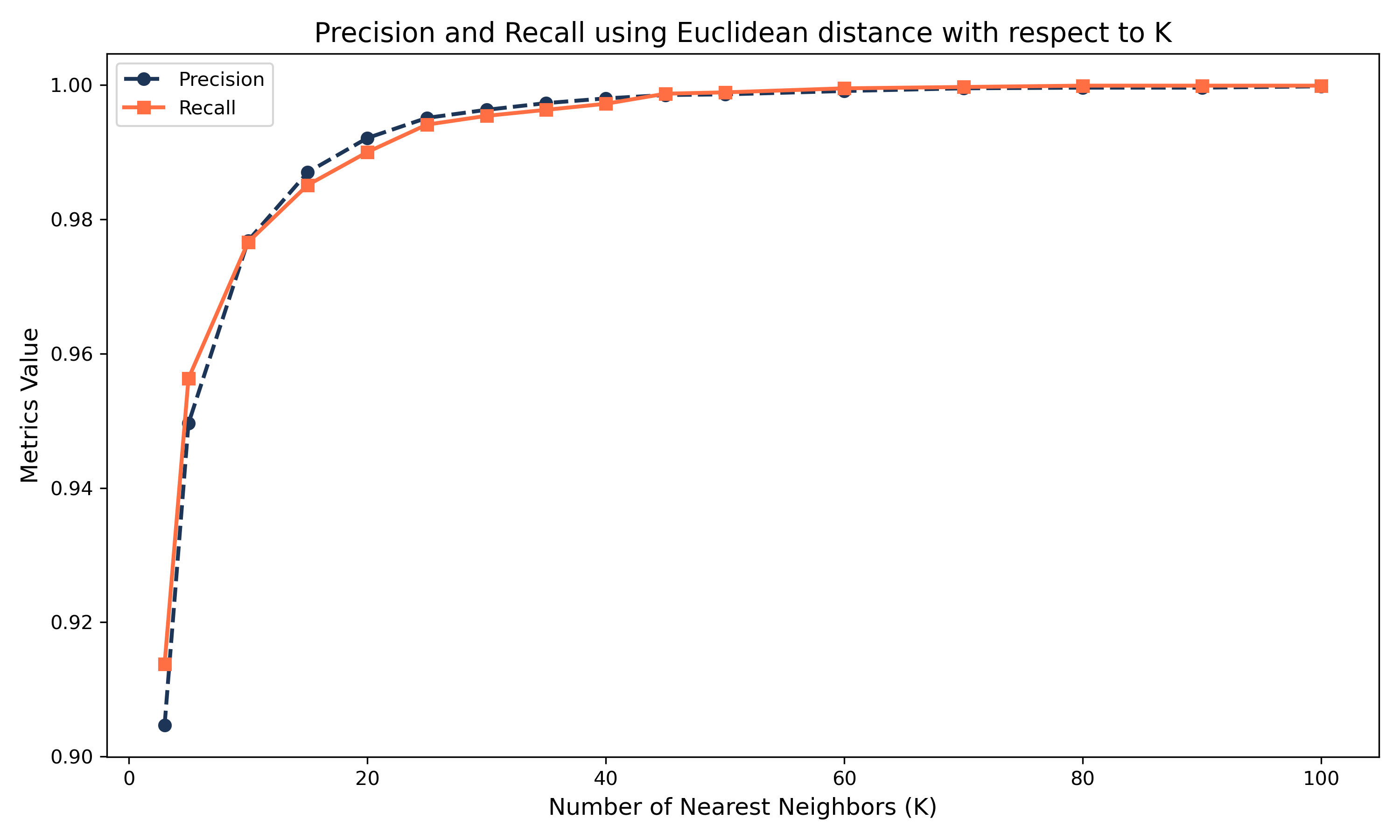}
        \subcaption{Euclidean Distance on Cow}\label{fig:euc_cow}
    \end{minipage}%
    \hfill
    \begin{minipage}{0.46\textwidth}
        \centering
        \includegraphics[width=\textwidth]{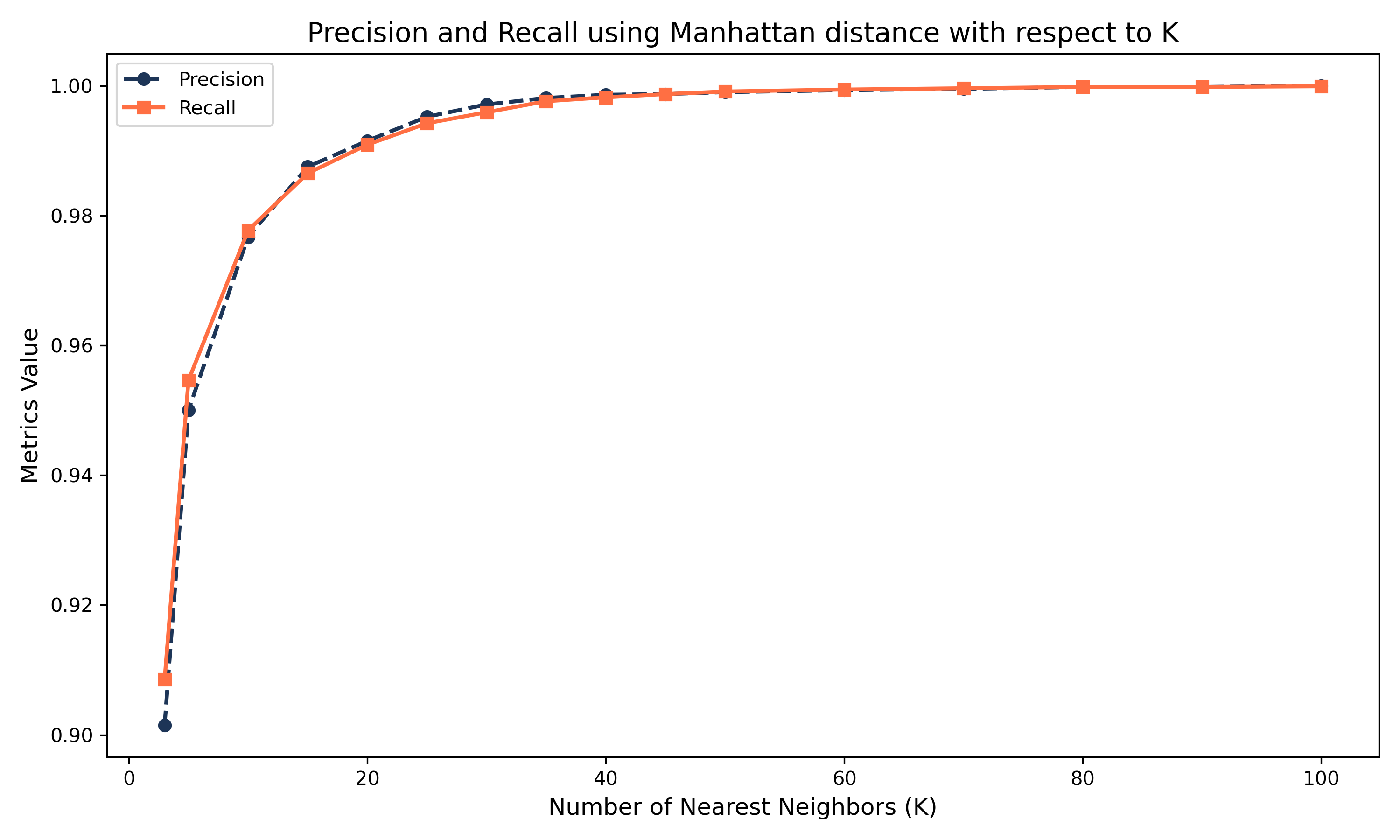}
        \subcaption{Manhattan distance on Cow}\label{fig:mht_cow}
    \end{minipage}\\[1ex]
    \begin{minipage}{0.46\textwidth}
        \centering
        \includegraphics[width=\textwidth]{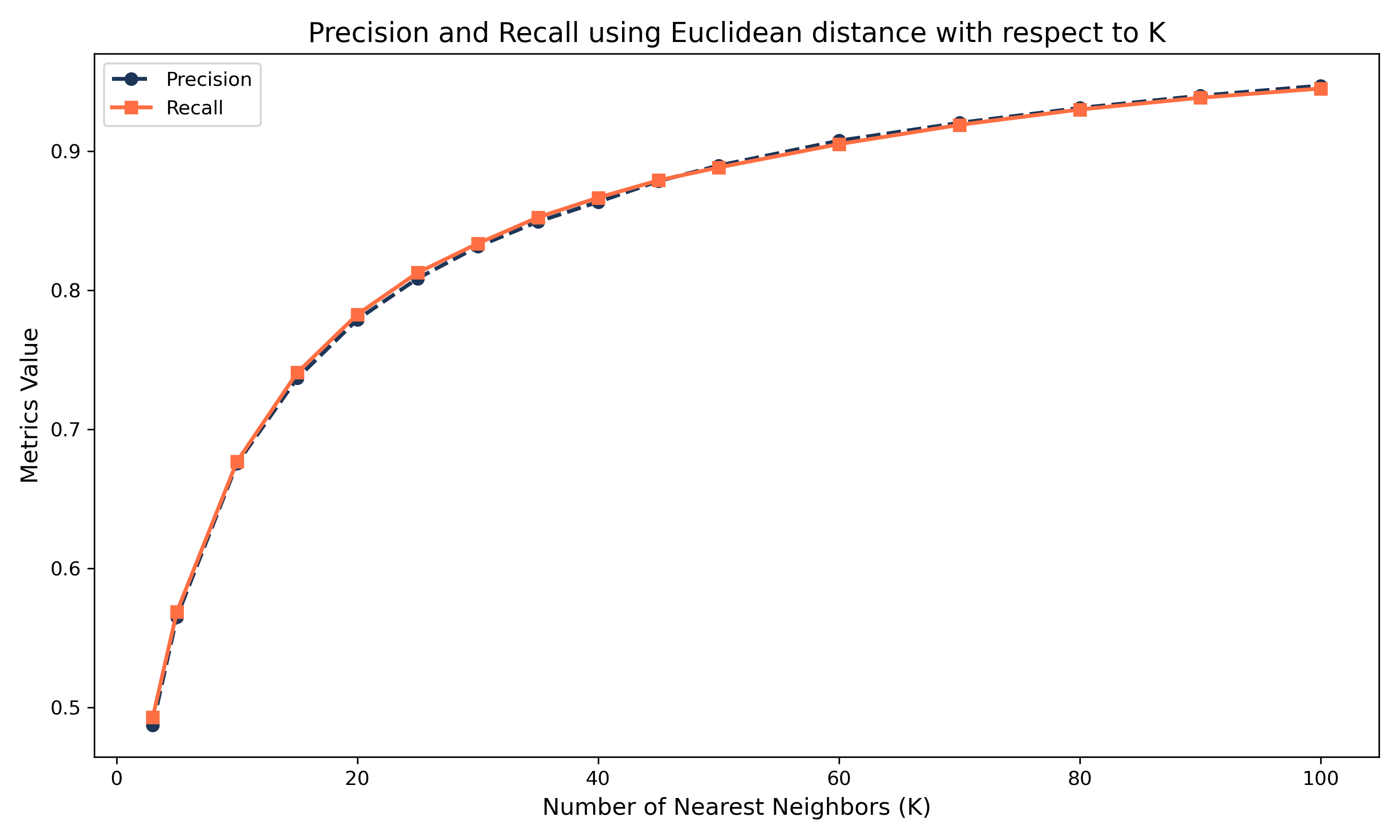}
        \subcaption{Euclidean Distance on Human}\label{fig:euc_ukb}
    \end{minipage}%
    \hfill
    \begin{minipage}{0.46\textwidth}
        \centering
        \includegraphics[width=\textwidth]{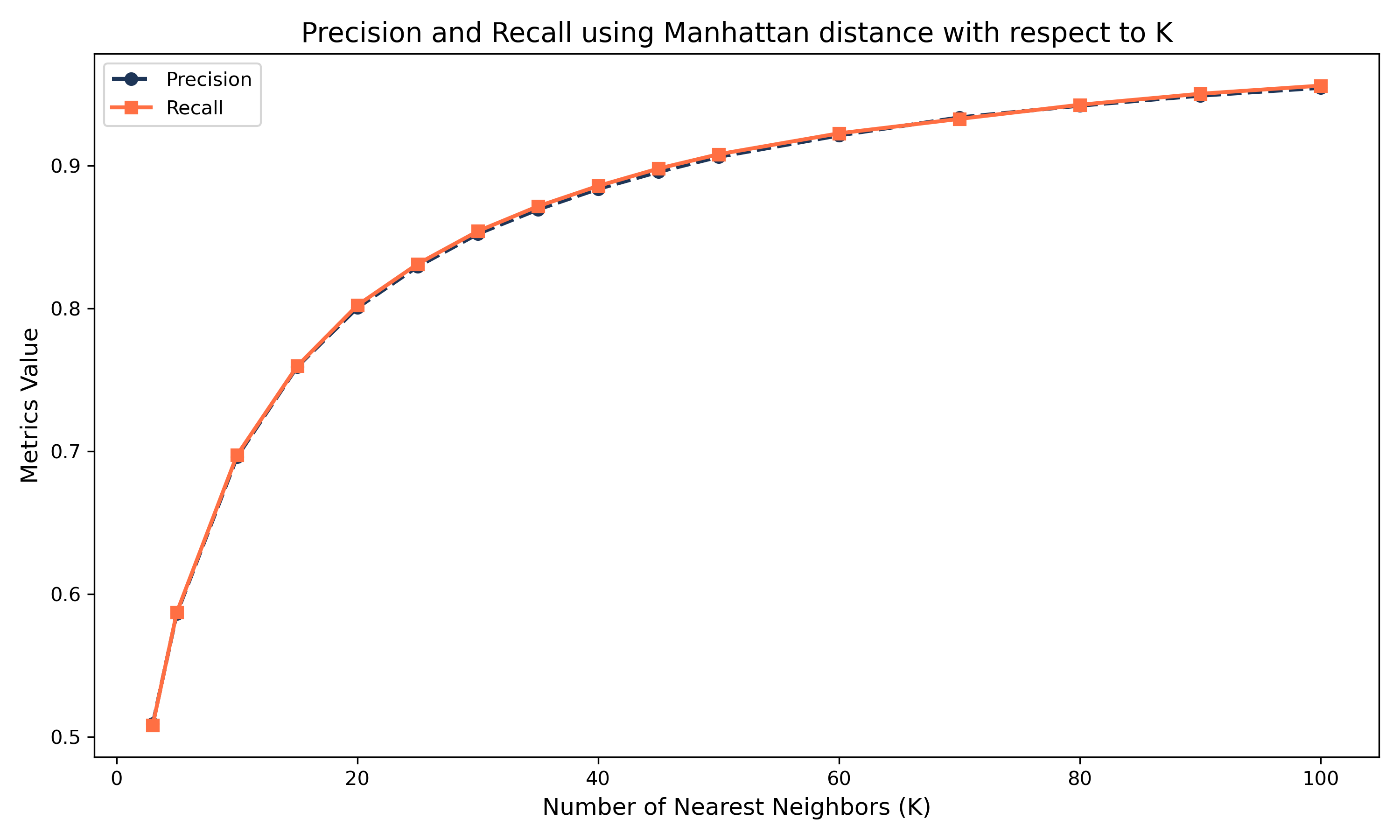}
        \subcaption{Manhattan Distance on Human}\label{fig:mht_ukb}
    \end{minipage}
    \caption{Evolution of Precision and Recall with respect to $k$ using Euclidean and Manhattan Distances on Cow and Human Datasets.}
\end{figure}

\newpage

\section{Method for Calculating Yield Deviation of Fat Content}
Fat content is measured by mid-infrared spectrometry on individual milk samples collected during monthly test days. For selection purpose, the trait analyzed is the average value over the entire lactation, with records repeated across lactations. To account for this, we use a multiplicative mixed model that adjusts for heterogeneous phenotypic variances within herd-year and region-year classes and accommodates multiple records per cow. The model accounts for fixed environmental effect, random permanent environmental effect, random genetic effect, and residual effect.

Let $i$ index cows, $j$ index repeated records for each cow, and $h$ index the level corresponding to variance heterogeneity. The $j^{th}$ record for cow $i$, associated with heterogeneity level $h$, is denoted by $y_{i,j,h}$. Formally, it is modeled as:
\begin{align*}
y_{i,j,h} =\ & ( \textcolor{blue}{\text{effect}_{\text{herd} \times \text{year}}} \\
& + \textcolor{blue}{\text{effect}_{\text{age at calving} \times \text{region} \times \text{year}}} \\
& + \textcolor{blue}{\text{effect}_{\text{calving year} \times \text{calving month} \times \text{region} \times \text{year}}} \hspace{2.35cm}  \textcolor{blue}{\text{fixed env effect}}\\
& + \textcolor{blue}{\text{effect}_{\text{dry period length} \times \text{region} \times \text{year}}} \\
& + \textcolor{violet}{b_{i}} \hspace{7.15cm} \textcolor{violet}{\text{permanent env effect}} \\
& + \textcolor{green}{a_{i}}  \hspace{8.35cm} \textcolor{green}{\text{genetic effect}}\\
& + \textcolor{orange}{e_{i}}) \hspace{8.1cm} \textcolor{orange}{\text{residual effect}}\\
& \times \textcolor{red}{e^{\gamma_{h}/2}} \hspace{3.45cm}  \textcolor{red}{\text{heterogeneous variance adjustment factor}}
\end{align*}

where we have:
\begin{itemize}
    \item \textcolor{blue}{\textbf{fixed env effect}}: fixed effect associated with different combinations of environmental factors.
    
    \item \textcolor{violet}{\textbf{permanent env effect}}: random effect accounting for the permanent environmental effect specific to cow $i$.
    
    \item \textcolor{green}{\textbf{genetic effect}}: additive genetic effect of cow $i$ such that $\text{Var}(a) = \mathbf{G} \sigma^2_a$, where $\mathbf{G}$ is the genomic relationship matrix and $\sigma^2_a$ is the additive genetic variance.
    
    \item \textcolor{orange}{\textbf{residual effect}}: random residual error term for cow $i$ such that $\text{Var}(e) = \mathbf{I} \sigma^2_e$, where $\sigma^2_e$ is the residual variance.
    
    \item \textcolor{red}{\textbf{heterogeneous variance adjustment factor}}: $\gamma_h$ is the parameter modeling heterogeneity in phenotypic variance, decided by [herd $\times$ year] and [region $\times$ year] \cite{hete_variance}. The factor $e^{\gamma_{h}/2}$ is applied to standardize variance across environments.
\end{itemize}

In our study, the estimated additive genetic variance $\sigma^2_a$ is $8.84$, the variance of the permanent environmental effect is $3.54$, and the residual variance $\sigma^2_e$ is $5.3$. Therefore, the heritability $h^2$ of the fat content trait is $8.84/(8.84 + 3.54 + 5.3) = 0.5$.

We first adjust each performance record $j$ of cow $i$ to account for heterogeneous variance, fixed environmental effects, and the permanent environmental effect. The adjusted record, denoted $y_{i,j}^{\text{adjusted}}$, is given by
\begin{align*}
    y_{i,j}^{\text{adjusted}} = y_{i,j,h} \times \textcolor{red}{e^{-\gamma_{h}/2}} 
    - \textcolor{blue}{\text{fixed env effect}} 
    - \textcolor{violet}{\text{permanent env effect}}
\end{align*}

Finally, the yield deviation $YD_i$ for cow $i$ is computed as a weighted average of all its adjusted records, given by
\begin{align*}
    YD_{i} = \frac{\sum_{j=1}^{3} w_{i,j} \times y_{i,j}^{\text{adjusted}}}{\sum_{j=1}^{3} w_{i,j}},
\end{align*}
where $w_{i,j}$ reflects the amount of information contained in the yield deviation, accounting for the number of elementary records per cow, the size of the environmental effect groups, the repeatability, and the heritability of the trait.

\section{Neural Network Architecture and Hyperparameter Selection for All Chromosomes of the Cow Dataset}
\subsection{VAE}

\subsubsection*{Model Architecture}
\begin{itemize}
  \item \textbf{Input dimension:} 150,483 features, corresponding to one-hot encoded genotypes for 50,161 SNPs.
  
  \item \textbf{Encoder:}
  \begin{itemize}
    \item Fully connected layers: 150,483 $\rightarrow$ 4,096 $\rightarrow$ 2,048 $\rightarrow$ 1,024
    \item One residual block at 1,024 units (2 linear layers with skip connection)
    \item Linear projection to 512, followed by two separate linear heads to infer:
    \begin{itemize}
      \item Mean vector $\mu \in \mathbf{R}^{256}$
      \item Log-variance vector $\log\sigma^2 \in \mathbf{R}^{256}$
    \end{itemize}
    \item All layers are followed by Batch Normalization to stabilize training and LeakyReLU activation (slope = 0.05) to avoid neuron inactivation.
  \end{itemize}

  \item \textbf{Latent space:} 256-dimensional latent vector

  \item \textbf{Decoder:}
  \begin{itemize}
    \item Fully connected layers: 256 $\rightarrow$ 512 $\rightarrow$ 1,024
    \item One residual block at 1,024 units (same structure as encoder)
    \item Continuation: 1,024 $\rightarrow$ 2,048 $\rightarrow$ 4,096 $\rightarrow$ 150,483 (The reverse one-hot decoding is handled during the loss computation, not as an explicit output layer)
    \item All hidden layers include Batch Normalization and LeakyReLU (slope = 0.05)
  \end{itemize}
\end{itemize}

\subsubsection*{Training Hyperparameters}
\begin{itemize}
  \item \textbf{Batch size:} 2,048
  \item \textbf{Optimizer:} Adam optimizer with $\beta_1 = 0.9$, $\beta_2 = 0.999$
  \item \textbf{Learning rate:} $5 \times 10^{-4}$
\end{itemize}

\subsection{GAN}
\subsubsection*{Model Architecture}
\begin{itemize}
  \item \textbf{Generator}
  \begin{itemize}
    \item \textbf{Input:} 256-dimensional latent vector
    \item \textbf{Initial transformation:} 
    \begin{itemize}
      \item Linear layer: 256 $\rightarrow$ 1,045
      \item Batch Normalization + LeakyReLU
    \end{itemize}
    \item \textbf{Residual blocks:} Two blocks with intermediate expansion:
    \begin{itemize}
      \item Block 1: 1,045 $\rightarrow$ 1,045 (residual) $\rightarrow$ 2,090
      \item Block 2: 2,090 $\rightarrow$ 2,090 (residual) $\rightarrow$ 4,180
      \item Each residual block consists of two fully connected layers with skip connection, Batch Normalization, and LeakyReLU
    \end{itemize}
    \item \textbf{Final projection:} 
    \begin{itemize}
    \item Linear: 4,180 $\rightarrow$ 150,483 (matches the one-hot encoded genotype size)
    \item Reshape to [batch size, 50161, 3]
    \item Apply Gumbel-Softmax to obtain discrete-like output while maintaining differentiability
    \end{itemize}
  \end{itemize}

  \item \textbf{Discriminator}
  \begin{itemize}
    \item \textbf{Input:} 150,483-dimensional vector
    \item \textbf{Initial transformation:}
    \begin{itemize}
      \item Linear layer: 150,483 $\rightarrow$ 4,180
      \item LeakyReLU activation
    \end{itemize}
    \item \textbf{Residual blocks:} Two blocks with progressive contraction:
    \begin{itemize}
      \item Block 1: 4,180 $\rightarrow$ 4,180 (residual) $\rightarrow$ 2,090
      \item Block 2: 2,090 $\rightarrow$ 2,090 (residual) $\rightarrow$ 1,045
      \item Each residual block includes two linear layers and LeakyReLU activations
    \end{itemize}
    \item \textbf{Final layer:} 
    \begin{itemize}
      \item Linear: 1,045 $\rightarrow$ 1
      \item Sigmoid activation for binary classification.
    \end{itemize}
  \end{itemize}

  \item \textbf{Activation functions:} All hidden layers use LeakyReLU (negative slope = 0.05)

  \item \textbf{Normalization:} Batch Normalization is used in the generator but omitted in the discriminator
\end{itemize}

\subsubsection*{Training Hyperparameters}
\begin{itemize}
  \item \textbf{Batch size:} 128
  \item \textbf{Optimizer:} Adam optimizer with $\beta_1 = 0.5$, $\beta_2 = 0.999$ for both generator and discriminator
  \item \textbf{Learning rate:} $1 \times 10^{-4}$ for both generator and discriminator
  \item \textbf{Gumbel-Softmax temperature:} Linearly annealed from 1.0 to 0.1 during training
\end{itemize}

\subsection{WGAN}
Here we present the conditional version, where phenotype information is provided during training. For the unconditional setting, it suffices to remove the phenotype-related tensors and adjust the tensor shapes accordingly.

\subsubsection*{Model Architecture}
\begin{itemize}
  \item \textbf{Generator:}
  \begin{itemize}
    \item \textbf{Input:} A 260-dimensional vector composed of:
    \begin{itemize}
      \item Latent noise vector $z$ of dimension 256
      \item Phenotype conditioning vector, repeated and concatenated of dimension 4
    \end{itemize}
    
    \item \textbf{Initial Layer:}
    \begin{itemize}
      \item Linear: 260 $\rightarrow$ 1,045
      \item Batch Normalization + LeakyReLU
    \end{itemize}
    
    \item \textbf{Residual Blocks:} Two sequential residual blocks:
    \begin{itemize}
      \item \textbf{Block 1:}
      \begin{itemize}
        \item ResNetBlock: A phenotype vector of dimension 4 is first injected, resulting in an input dimension of 1,049. The block consists of two linear layers with Batch Normalization and a skip connection, maintaining the dimensionality at 1,049. 
        \item Linear: 1,049 $\rightarrow$ 2,090 + LeakyReLU
      \end{itemize}
      \item \textbf{Block 2:}
      \begin{itemize}
        \item ResNetBlock: Similarly, a phenotype vector of dimension 4 is injected, resulting in a dimension of 2,094. The block maintains this dimensionality through two linear layers with Batch Normalization and a skip connection.
        \item Linear: 2,094 $\rightarrow$ 4,180 + LeakyReLU
      \end{itemize}
    \end{itemize}

    \item \textbf{Final projection:}
    \begin{itemize}
      \item Linear: inject phenotype vector of dimension 4 $\rightarrow$ 4,184 $\rightarrow$ 150,483
      \item Reshape to [batch size, 50161, 3]
      \item Apply Gumbel-Softmax
    \end{itemize}
  \end{itemize}

  \item \textbf{Critic:}
  \begin{itemize}
    \item \textbf{Input:} A flattened genotype sequence of dimension 150,483 is concatenated with a phenotype vector of dimension 4, resulting in a total input dimension of 150,487.

    \item \textbf{Initial Layer:}
    \begin{itemize}
      \item Linear: 150,487 $\rightarrow$ 4,180
      \item LeakyReLU
    \end{itemize}

    \item \textbf{Residual Blocks:}
    \begin{itemize}
      \item \textbf{Block 1:}
      \begin{itemize}
        \item ResNetBlock: inject phenotype vector of dimension 4 $\rightarrow$ 4,184 $\rightarrow$ 4,184
        \item Linear: 4,184 $\rightarrow$ 2,090 + LeakyReLU
      \end{itemize}
      \item \textbf{Block 2:}
      \begin{itemize}
        \item ResNetBlock: inject phenotype vector of dimension 4 $\rightarrow$ 2,094 $\rightarrow$ 2,094
        \item Linear: 2,094 $\rightarrow$ 1,045 + LeakyReLU
      \end{itemize}
    \end{itemize}

    \item \textbf{Final output:}
    \begin{itemize}
      \item Linear: 1,045 $\rightarrow$ 1 (critic score)
    \end{itemize}
  \end{itemize}

    \item \textbf{Activation functions:} All hidden layers use LeakyReLU (negative slope = 0.05)

  \item \textbf{Normalization:} Batch Normalization is used in the generator but omitted in the critic
\end{itemize}

\subsubsection*{Training Hyperparameters}
\begin{itemize}
  \item \textbf{Batch size:} 128
  \item \textbf{Optimizer:} Adam optimizer with $\beta_1 = 0.5$, $\beta_2 = 0.9$ for both generator and critic
  \item \textbf{Learning rate:} $1 \times 10^{-4}$ for both generator and critic
  \item \textbf{Critic updates per generator update:} 5
  \item \textbf{Gradient penalty coefficient ($\lambda$):} 10
  \item \textbf{Gumbel-Softmax temperature:} Linearly annealed from 1.0 to 0.1 during training
\end{itemize}

\subsection{DM}
Similarly, we provide the conditional version here. To switch to the unconditional setting, simply remove all phenotype-related tensors and adjust the tensor shapes accordingly.

\subsubsection*{Model Architecture}

\begin{itemize}
  \item \textbf{Input structure:}
  \begin{itemize}
    \item The input tensor is a concatenation of:
    \begin{itemize}
      \item Noisy data vector $x \in \mathbf{R}^{4819}$ (corresponding to the PCA-latent genotype data)
      \item Time embedding $t_{\text{emb}} \in \mathbf{R}^{256}$
      \item Phenotype embedding $pheno_{\text{emb}} \in \mathbf{R}^{64}$
    \end{itemize}
    \item Total input dimension: 5,139
  \end{itemize}

  \item \textbf{Time embedding:}
  \begin{itemize}
    \item Uses a sinusoidal positional encoding, similar to that in the original DDPM implementation.
    \item Embedding dimension is set to 256
  \end{itemize}

  \item \textbf{Phenotype embedding:}
  \begin{itemize}
    \item A continuous label (scalar) is projected to a higher-dimensional space using a linear layer
    \item Embedding dimension is set to 64
  \end{itemize}

  \item \textbf{Noise predictor architecture:}
  \begin{itemize}
    \item \texttt{fc1:} Input layer mapping from 5,139 $\rightarrow$ 8,192
    \item After the first hidden layer, time and phenotype embeddings are reinjected (8192 + 256 + 64 = 8512)
    \item \texttt{fc2:} 8,512 $\rightarrow$ 8,192
    \item \texttt{fc3:} 8,192 $\rightarrow$ 6,144
    \item \texttt{out:} Final output layer from 6,144 $\rightarrow$ 4,819
    \item \textbf{Residual connection:}
    \begin{itemize}
        \item \texttt{res:} A skip connection projects the input via a linear layer: 5,139 $\rightarrow$ 4,819
        \item The final output is computed as \texttt{out + res}
    \end{itemize}
    \item \textbf{Normalization:} Layer Normalization is applied after each internal fully connected layer
    \item \textbf{Activation:} ReLU is used after each normalization layer
\end{itemize}
\end{itemize}

\subsubsection*{Training Hyperparameters}
\begin{itemize}
\item \textbf{Batch size:} 4,086
\item \textbf{Diffusion process:}
  \begin{itemize}
    \item Number of diffusion steps: 1,500
    \item $\beta$ schedule: linear
  \end{itemize}

\item \textbf{Time Sampling Strategy:} Antithetic sampling

  \item \textbf{Optimizer:}
  \begin{itemize}
    \item Adam optimizer with $\beta_1 = 0.9$, $\beta_2 = 0.999$
    \item Learning rate: $3 \times 10^{-4}$
    \item Learning rate scheduler: Cosine Annealing 
    \item Minimum learning rate: $3 \times 10^{-6}$
    \item Warm-up:
    \begin{itemize}
      \item Strategy: Linear warm-up
      \item Period: first 1,000 steps
    \end{itemize}
\end{itemize}
\end{itemize}

\section{Impact of SNP Dependence on the Difficulty of Generative Modeling} \label{sec:data_vs_diff}
\begin{figure}[H]
    \centering
    \begin{minipage}{0.49\textwidth}
        \centering
        \includegraphics[width=\textwidth]{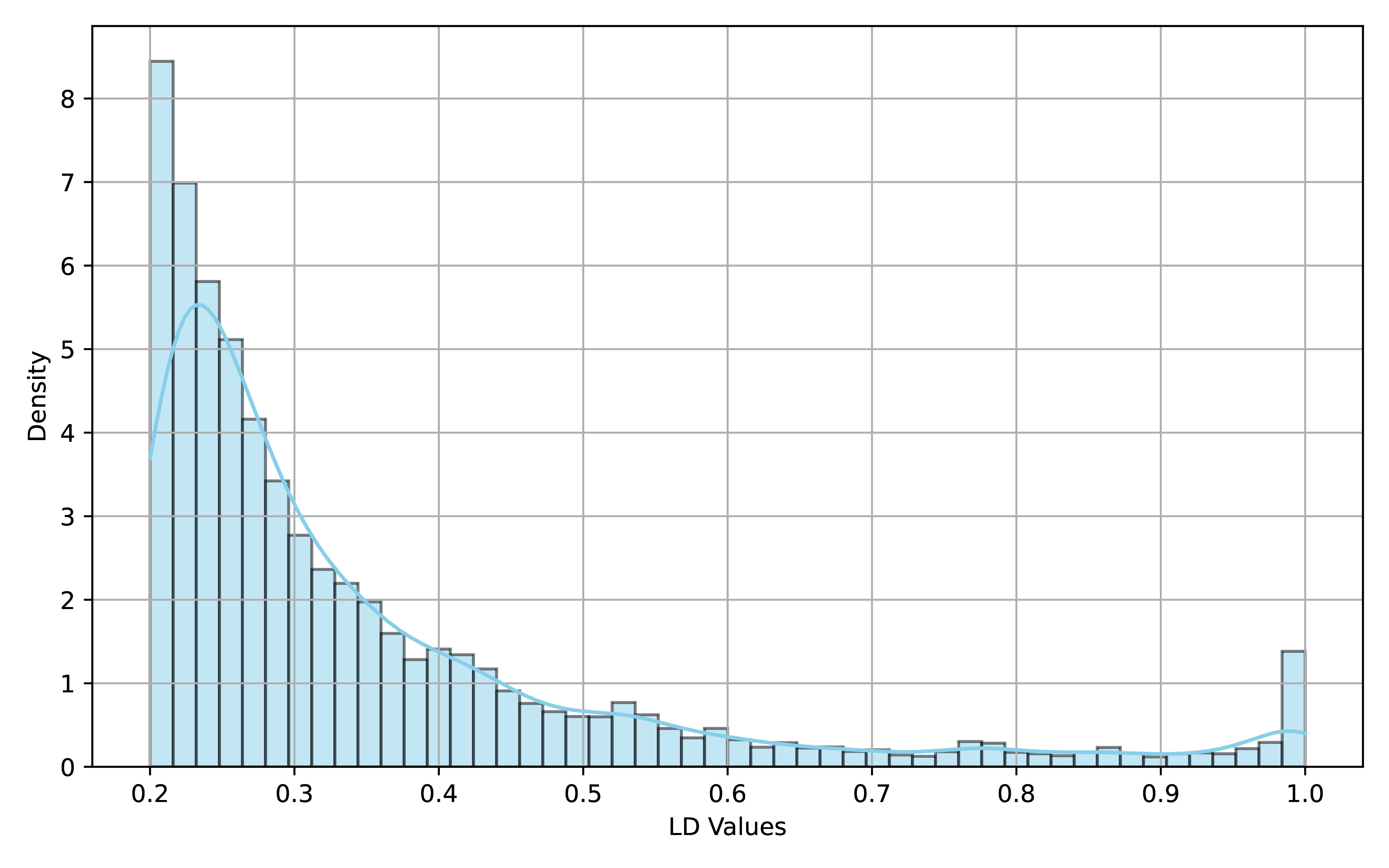}
        \subcaption{Cow CHR 5} \label{fig:ld_cow_ch5}
    \end{minipage}%
    \hfill
    \begin{minipage}{0.49\textwidth}
        \centering
        \includegraphics[width=\textwidth]{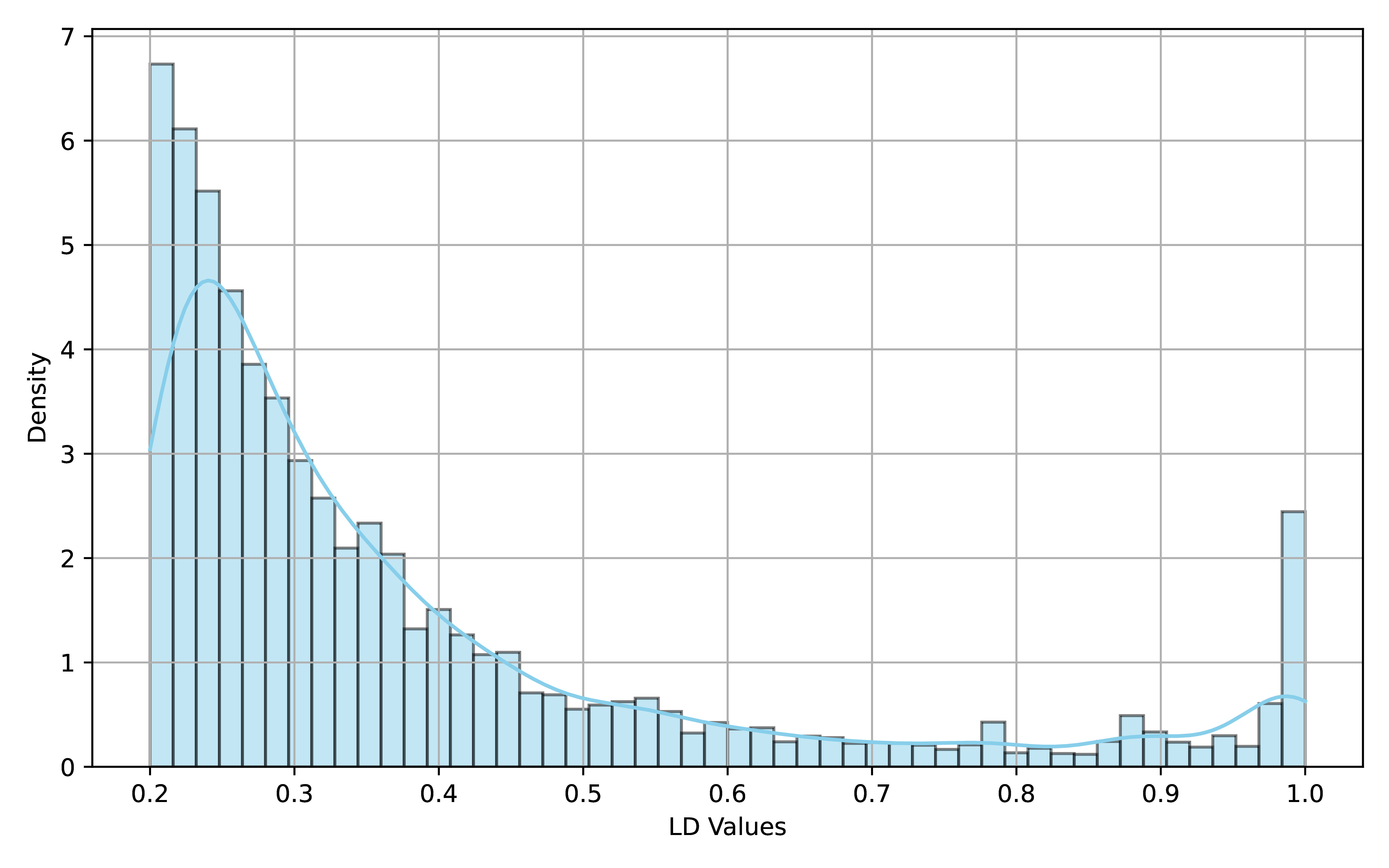}
        \subcaption{Cow CHR 14} \label{fig:ld_cow_ch14}
    \end{minipage}\\[1ex]
    \begin{minipage}{0.49\textwidth}
        \centering
        \includegraphics[width=\textwidth]{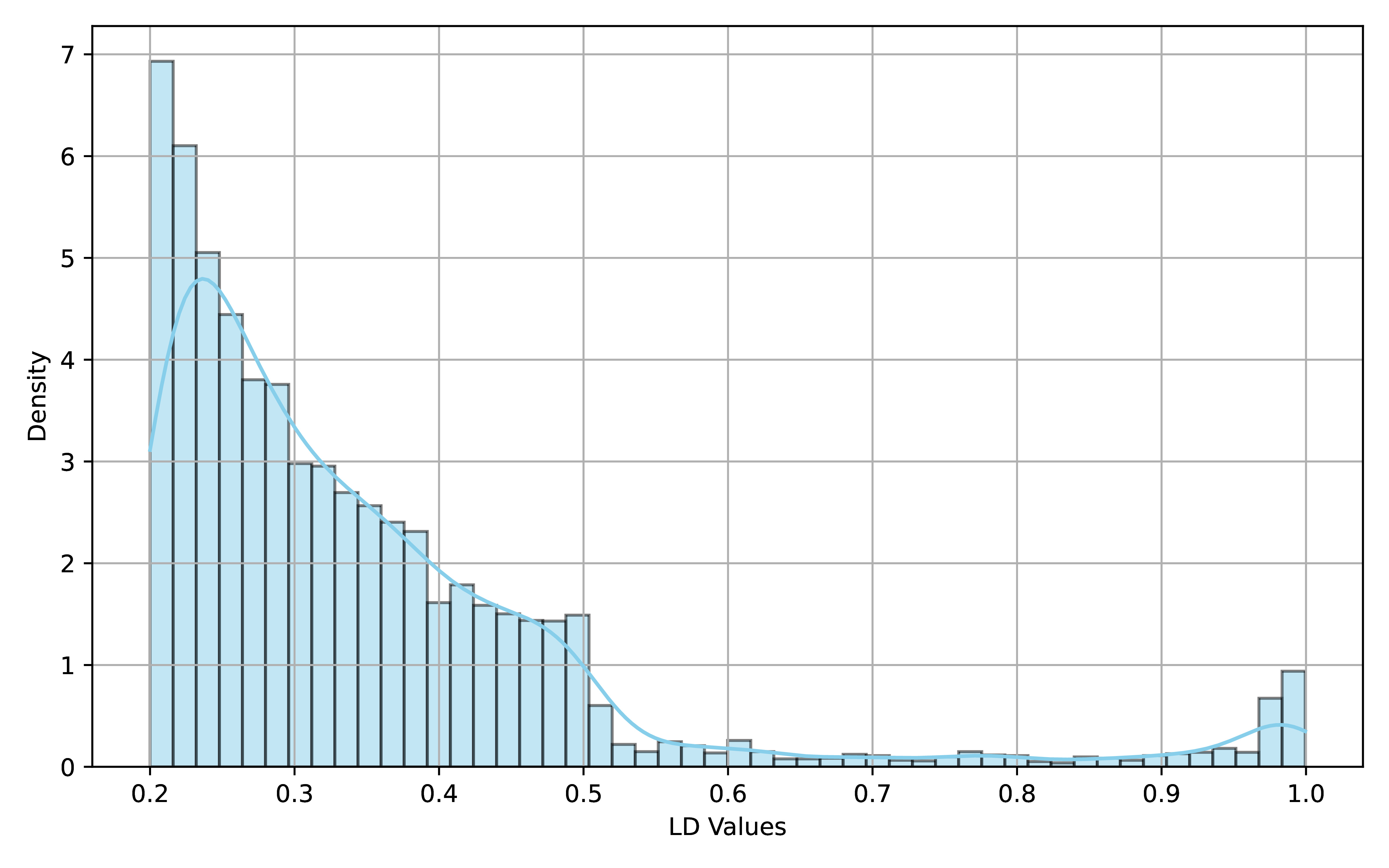}
        \subcaption{Human CHR 6} \label{fig:ld_ukb_ch6}
    \end{minipage}%
    \hfill
    \begin{minipage}{0.49\textwidth}
        \centering
        \includegraphics[width=\textwidth]{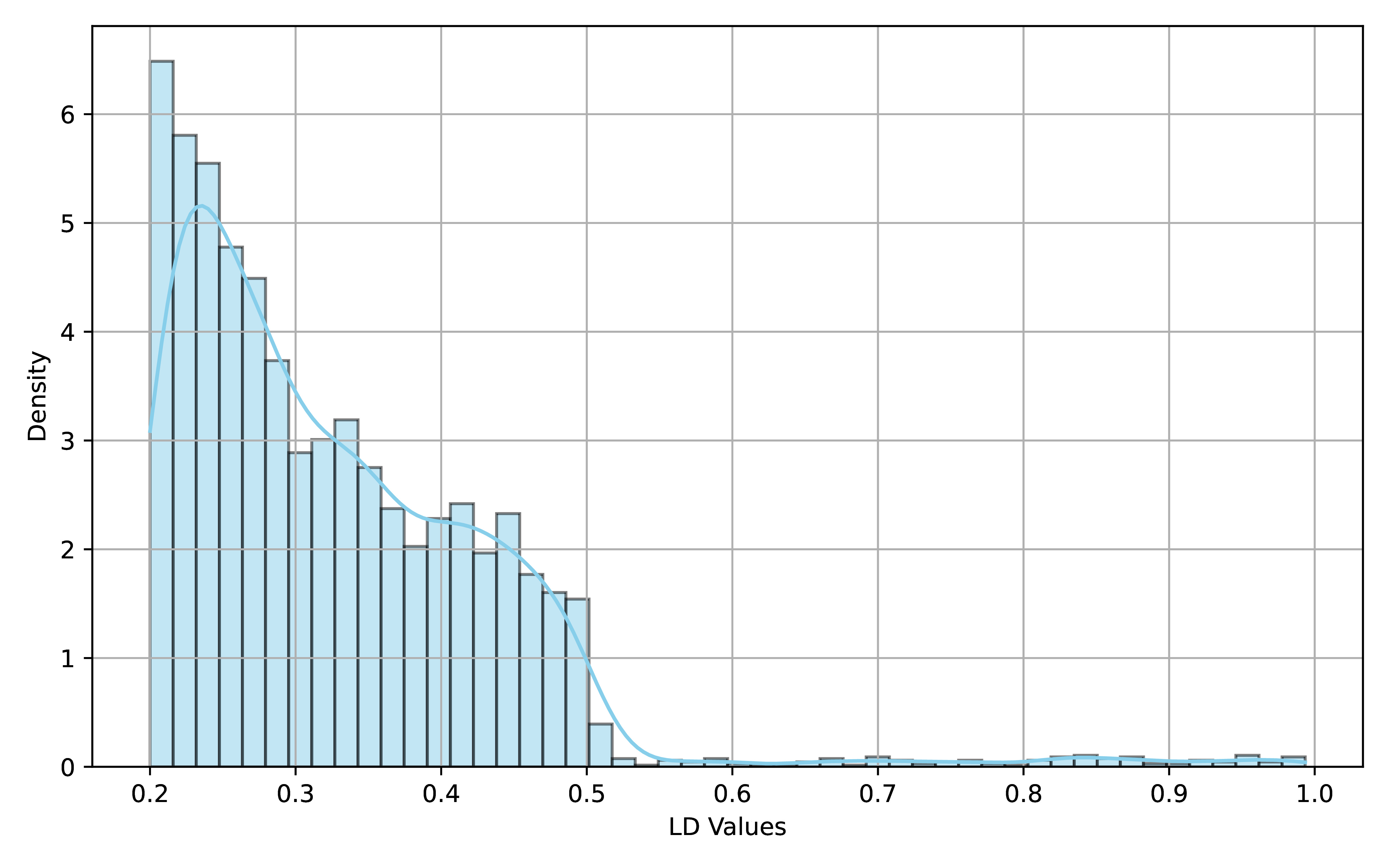}
        \subcaption{Human CHR 12} \label{fig:ld_ukb_ch12}
    \end{minipage}
    \caption{Distribution of LD values across different datasets, focusing on values $\geq 0.20$. In Cow dataset (a and b), SNPs exhibit stronger correlations compared to the Human dataset (c and d). In Human dataset, Chromosome 6 (c) contains more high-LD SNP pairs than Chromosome 12 (d). This makes it easier for generative models to learn Chromosome 6, despite its larger dimension compared to Chromosome 12.}
    \label{fig:ld_perf}
\end{figure}

\section{Limitation of AA Score as a Robust Metric}
\label{sec:limit_AA}
\begin{figure}[H]
    \centering
    \begin{minipage}{0.45\textwidth}
        \centering
        \includegraphics[width=\textwidth]{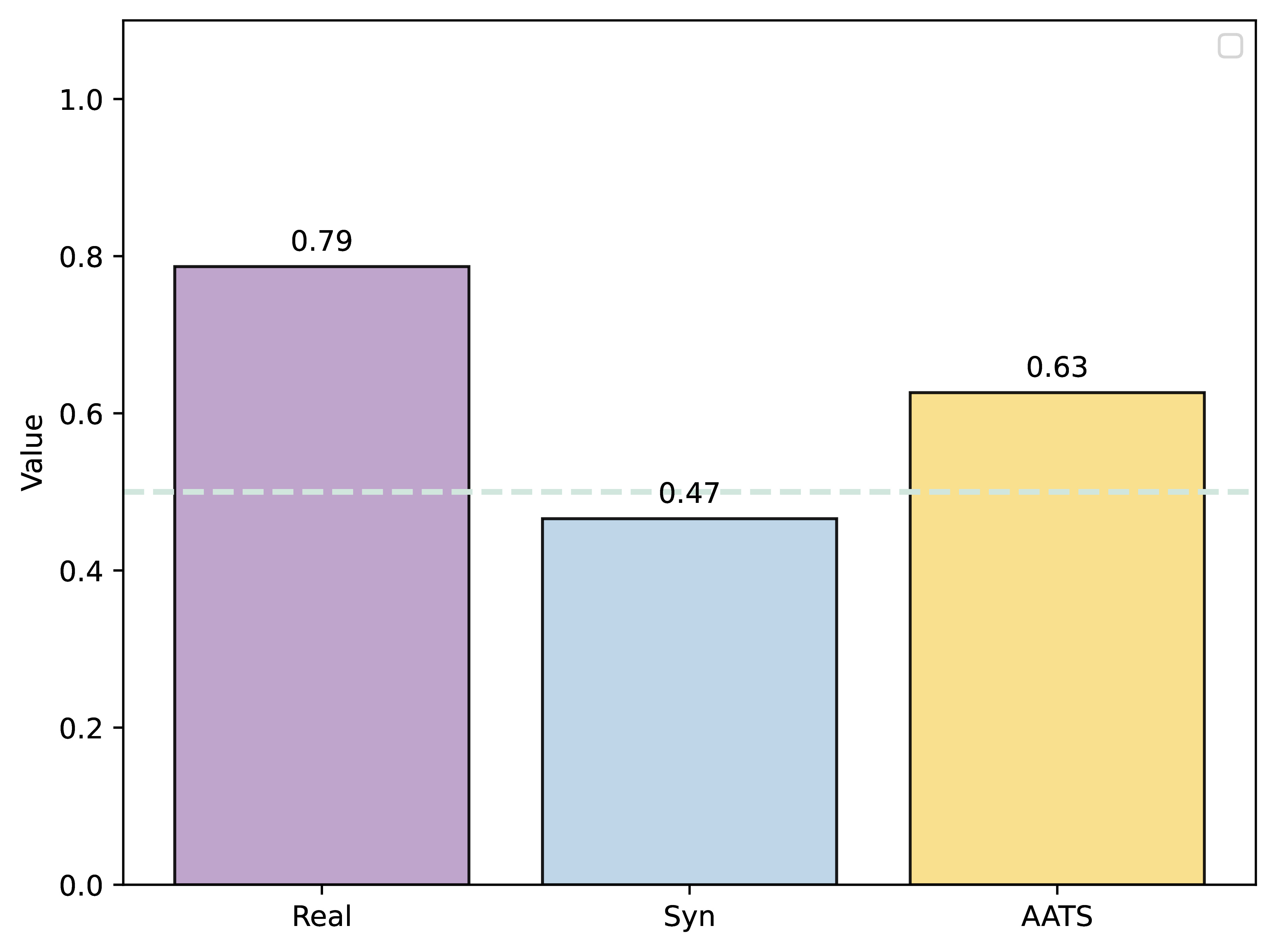}
        \subcaption{WGAN on Cow CHR 5} \label{fig:AA_cow_ch5}
    \end{minipage}%
    \hfill
    \begin{minipage}{0.45\textwidth}
        \centering
        \includegraphics[width=\textwidth]{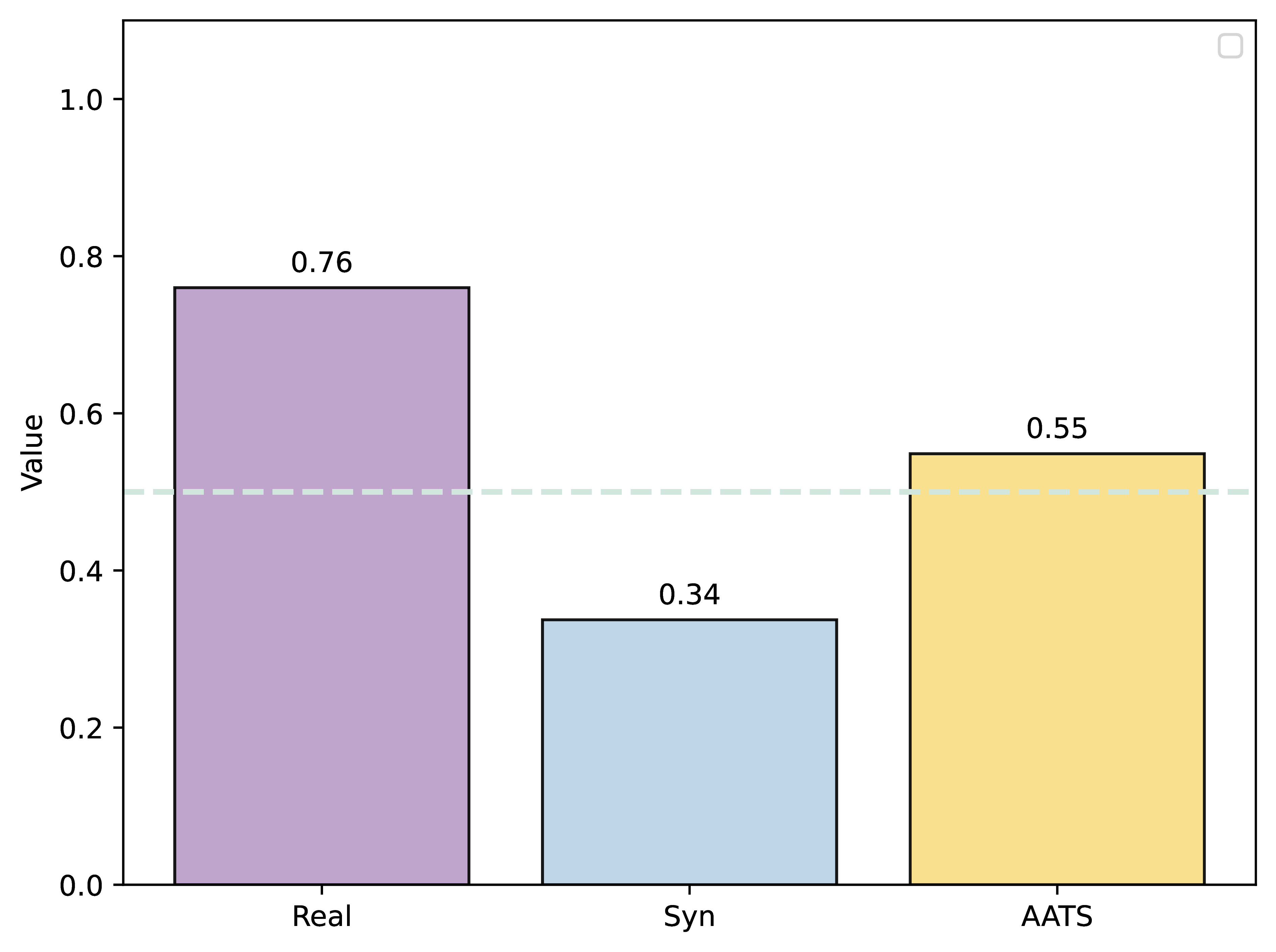}
        \subcaption{WGAN on Cow CHR 14} \label{fig:AA_cow_ch14}
    \end{minipage}\\[1ex]
    \begin{minipage}{0.45\textwidth}
        \centering
        \includegraphics[width=\textwidth]{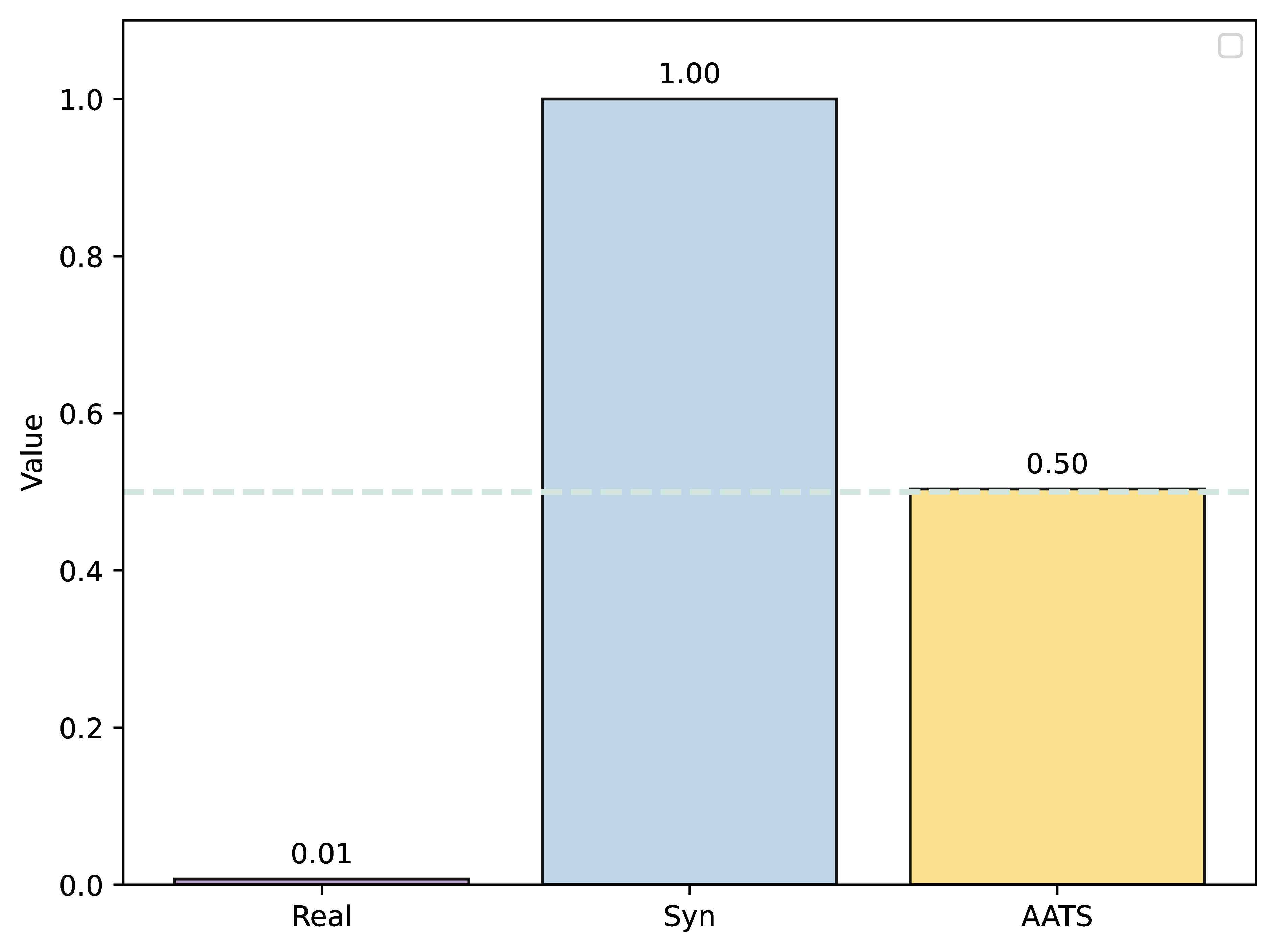}
        \subcaption{DM on Human CHR 6} \label{fig:AA_ukb_ch6}
    \end{minipage}%
    \hfill
    \begin{minipage}{0.45\textwidth}
        \centering
        \includegraphics[width=\textwidth]{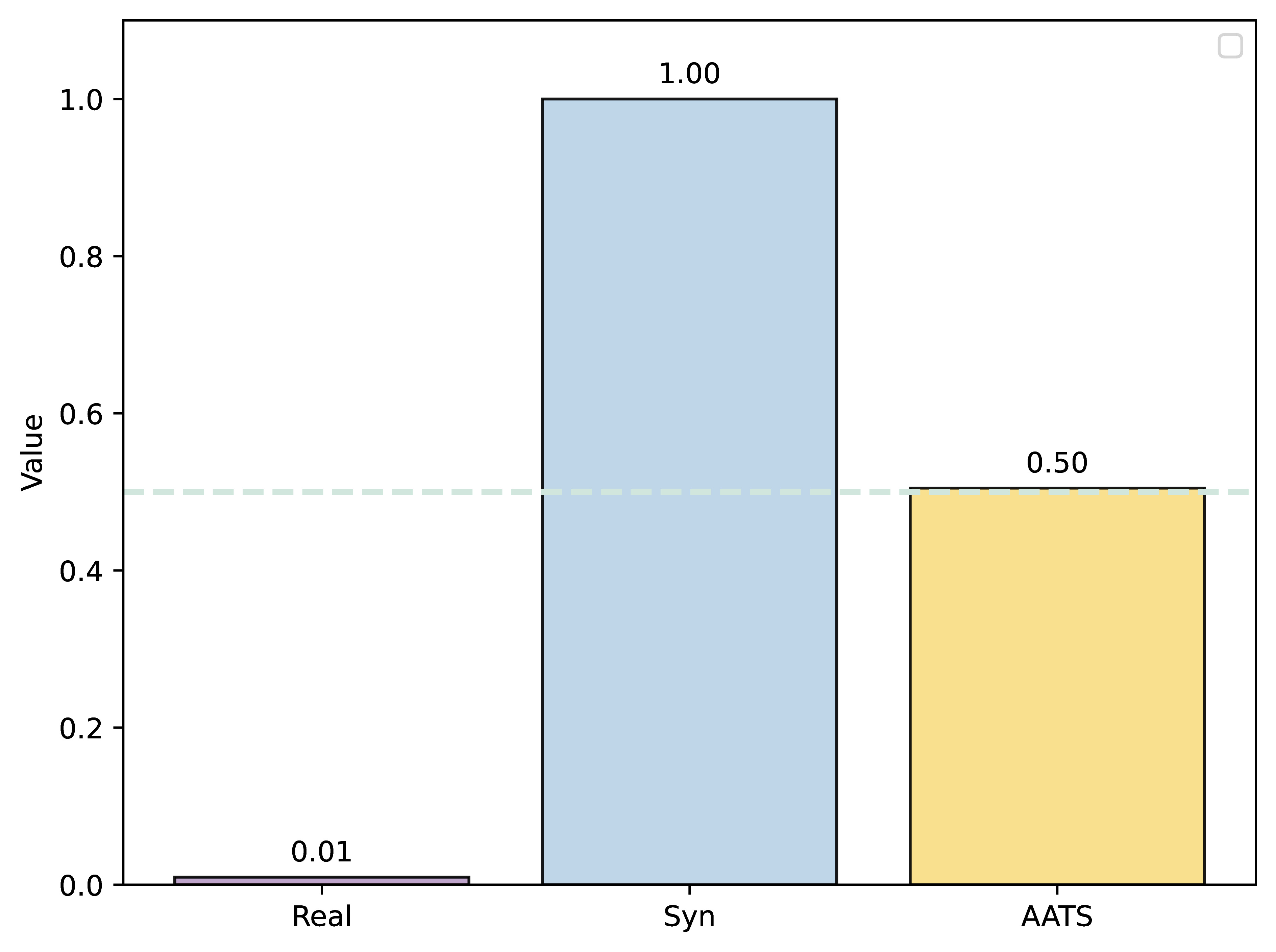}
        \subcaption{DM on Human CHR 12} \label{fig:AA_ukb_ch12}
    \end{minipage}
    \caption{A detailed investigation of the $AA$ score: (a) and (b) show cases where the $AA$ score serves as a good evaluation metric, with both $AA_{real}$ and $AA_{syn}$ yielding good scores, resulting in an $AA$ score around 0.50. (c) and (d) depict an anomalous scenario where $AA_{real} \approx 0$ and $AA_{syn} \approx 1$, yet the $AA$ score remains around $0.50$, highlighting a limitation of using $AA$ as a metric.}
    \label{fig:AA_incorrect}
\end{figure}

\begin{figure}[H]
  \centering
  \includegraphics[width=0.8\textwidth]{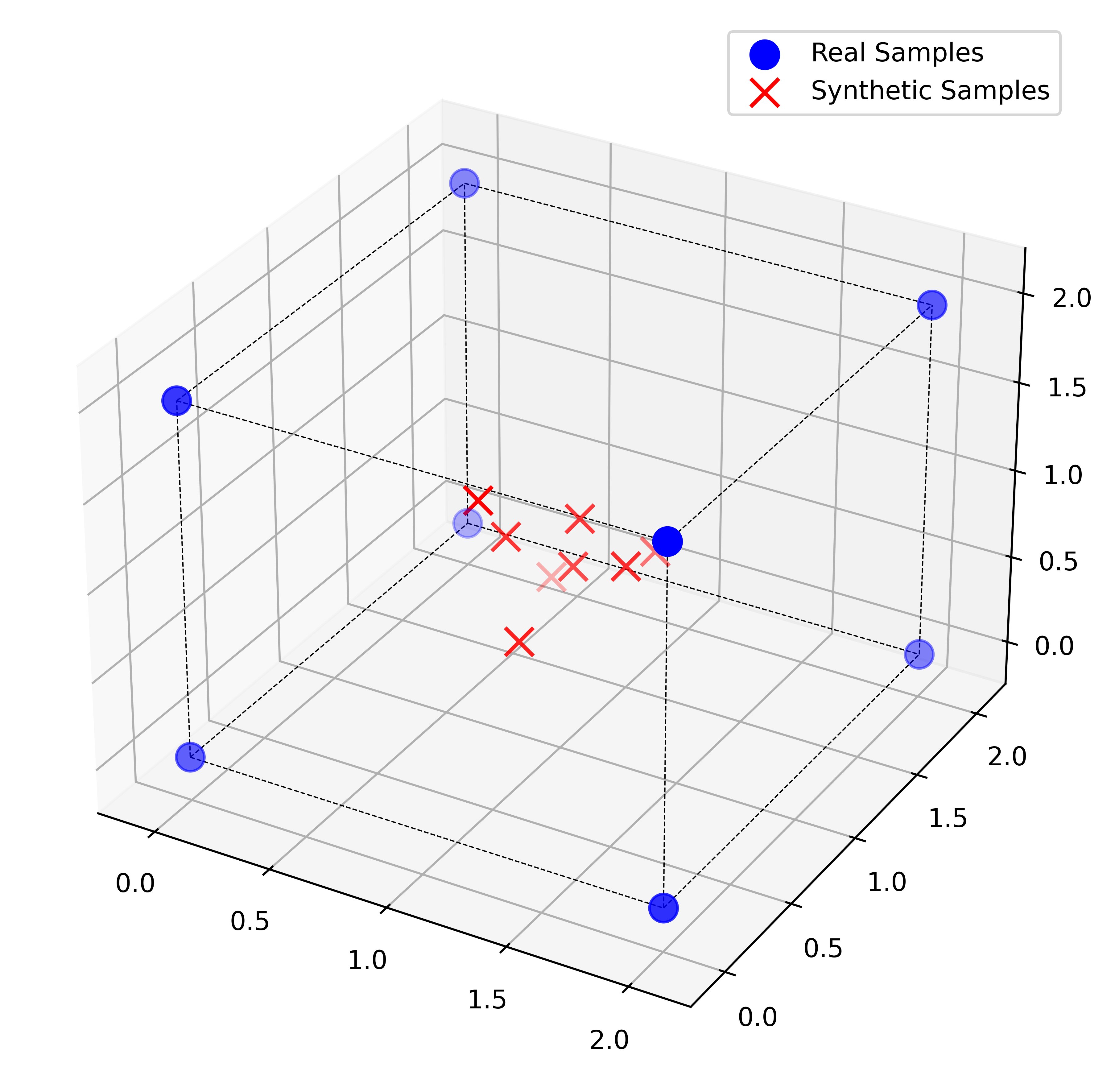}
  \caption{Geometric illustration of a scenario where $AA_{real} = 0$ and $AA_{syn}=1$: Eight real samples are positioned at the corners of a cube, with each pair of real samples' distance equal to 2 ($d_{RR} = 2$). Eight synthetic samples are tightly clustered at the center of the cube [1, 1, 1]. In this setup, each real sample is closer to a synthetic sample than to any other real samples ($d_{RS} < \sqrt{3} < d_{RR}$), yielding $AA_{real} = 0$. Conversely, each synthetic sample is closest to another synthetic sample in the cluster, yielding $AA_{syn} = 1$. If we relate this scenario to the precision and recall metrics with $\forall k$, we obtain a precision of $1$ because every synthetic sample falls within the support of a real sample. However, the recall is $0$, as none of the real samples fall within the support of any synthetic sample.}
  \label{fig:thre4}
\end{figure}

\printbibliography